\documentclass{aa}
\usepackage{graphicx}
\usepackage{braket}
\usepackage{txfonts}
\usepackage{subfig}
\usepackage{hyperref}
\usepackage{upgreek}
\bibpunct{(}{)}{;}{a}{}{,} 

\hypersetup{
  colorlinks   = true, 
  urlcolor     = blue, 
  linkcolor    = red, 
  citecolor    = blue 
}

\begin{document}

\title{First evidence of a connection between cluster-scale diffuse radio emission in cool-core galaxy clusters and sloshing features}

\author{N. Biava\inst{1,2,3}, A. Bonafede\inst{1,2}, F. Gastaldello\inst{4}, A. Botteon\inst{2}, M. Brienza\inst{5,1,2}, T. W. Shimwell \inst{6,7}, G. Brunetti\inst{2}, \\ L. Bruno\inst{1,2}, K. Rajpurohit\inst{8}, C. J. Riseley\inst{1,2,9}, R.J. van Weeren\inst{7}, M. Rossetti\inst{4}, R. Cassano\inst{2}, F. De Gasperin\inst{2,10}, \\ A. Drabent\inst{3}, H.J.A. Rottgering\inst{7}, A.C. Edge\inst{11}, C. Tasse\inst{12,13}}

\institute{Dipartimento di Fisica e Astronomia, Università di Bologna, via P. Gobetti 93/2, I-40129 Bologna, Italy
\and 
INAF - IRA, via P. Gobetti 101, I-40129 Bologna, Italy
\and 
Th\"uringer Landessternwarte, Sternwarte 5, D-07778 Tautenburg, Germany 
\\
\email{nbiava@tls-tautenburg.de}
\and 
INAF - IASF Milano, Via A. Corti 12, 20133 Milano, Italy
\and 
INAF – OAS, Via P. Gobetti 93/3, 40129 Bologna, Italy
\and 
ASTRON, Netherlands Institute for Radio Astronomy, Oude Hoogeveensedijk 4, 7991 PD, Dwingeloo, The Netherlands
\and 
Leiden Observatory, Leiden University, PO Box 9513, 2300 RA Leiden, The Netherlands
\and 
Harvard-Smithsonian Center for Astrophysics, 60 Garden Street, Cambridge, MA 02138, USA
\and 
CSIRO Space \& Astronomy, PO Box 1130, Bentley, WA 6102, Australia 
\and 
Hamburger Sternwarte, Universit\"at Hamburg, Gojenbergsweg 112, 21029, Hamburg, Germany 
\and 
Centre for Extragalactic Astronomy, Durham University, DH1 3LE, UK
\and 
GEPI \& ORN, Observatoire de Paris, Université PSL, CNRS, 5 Place Jules Janssen, 92190 Meudon, France 
\and 
Department of Physics \& Electronics, Rhodes University, PO Box 94, Grahamstown, 6140, South Africa\\
}

\date{Received / Accepted}

\abstract 
{Radio observations of a few cool-core galaxy clusters have revealed the presence of diffuse emission on cluster scales, similar to what was found in merging clusters in the form of radio halos. These sources might suggest that a minor merger, while not sufficiently energetic to disrupt the cool core, could still trigger particle acceleration in the intracluster medium on scales of hundreds of kiloparsecs.}
{We aim to verify the occurrence of cluster-scale diffuse radio emission in cool-core clusters and test the minor merger scenario.}
{With the  LOw Frequency ARray (LOFAR) at 144 MHz, we observed a sample of twelve cool-core galaxy clusters presenting some level of dynamical disturbances, according to X-ray data. We also performed a systematic search of cold fronts in these clusters, re-analysing archival \emph{Chandra} observations.}
{The clusters PSZ1G139.61+24, A1068 (new detection), MS 1455.0+2232, and RX J1720.1+2638 present diffuse radio emission on a cluster scale ($r\ge0.2R_{500}$). This emission is characterised by a double component: a central mini-halo confined by cold fronts and diffuse emission on larger scales, whose radio power at 144 MHz is comparable to that of radio halos detected in merging systems with the same cluster mass. The cold fronts in A1068 are a new detection. We also found a candidate plasma depletion layer in this cluster. No sloshing features are found in the other eight clusters. Two of them present a mini-halo, with diffuse radio emission confined to the cluster core. We also found a new candidate mini-halo. Whereas, for the remaining five clusters, we did not detect halo-like emission. For clusters without cluster-scale halos, we derived upper limits to the radio halo power.}
{We found that cluster-scale diffuse radio emission is not present in all cool-core clusters when observed at a low frequency, but it is correlated to the presence of cold fronts. The coexistence of cluster-scale diffuse radio emission and cold fronts in cool-core clusters requires a specific configuration of the merger and so it puts some constraints on the turbulence, which deserves to be investigated in the future with theoretical works.}

\keywords{galaxies: clusters: intracluster medium}

\titlerunning{Cluster-scale diffuse radio emission in cool-core clusters with sloshing features}
\authorrunning{N. Biava et al.}

   \maketitle


\section{Introduction}
Galaxy clusters are the most massive (with masses up to $10^{15}\rm{M_{\odot}}$) gravitationally bound objects in the Universe, which form through hierarchical aggregation of smaller structures \citep[for a review, see][]{Borgani2012}.
According to their dynamical stage, galaxy clusters are classified in relaxed or disturbed clusters. Disturbed, merging clusters are characterised by an irregular distribution of thermal gas emission, and they have high central entropy. Relaxed, cool-core clusters, instead, have a spherical distribution of gas, peaked X-ray emission, and a low central entropy. Cool-core clusters can also be perturbed by minor or off-axis mergers, causing sloshing of the cool gas in the central potential well, without disrupting the dense core. As a consequence of these encounters, cold fronts develop, which are arc-shaped X-ray surface brightness discontinuities, characterised by a jump in temperature and constant pressure across the front \citep[e.g.][]{Markevitch2007}. Sloshing lasts for several gigayears; therefore, the cluster can appear relatively relaxed albeit with cold fronts being present.

Mergers between clusters are expected to inject an amount of turbulence in the intracluster medium \citep[ICM, e.g.][]{Vazza2012} depending on the mass of the involved clusters and the merger configuration. Turbulence can then re-accelerate particles and amplify magnetic fields, generating large-scale diffuse synchrotron emission in the form of radio halos \citep[e.g.][]{Brunetti2001,Petrosian2001,Brunetti2016}.
Most of the massive merging clusters do indeed present megaparsec-sized diffuse radio emission, called a giant radio halo (RH), which follows the distribution of the thermal ICM and is characterised by a steep radio spectrum ($\alpha\ge 1$, with $S(\nu) \propto \nu^{-\alpha}$) \citep[for a review, see][]{vanWeeren2019}.
Diffuse radio emission has also been observed at the centre of relaxed cool-core clusters, surrounding the brightest cluster galaxy (BCG), with a smaller size of 0.1 - 0.5 Mpc, and referred to as a mini-halo \citep[MH][for a review]{vanWeeren2019}. Mini-halo emission is often bounded by X-ray cold fronts, suggesting particles are re-accelerated by turbulence connected with the sloshing of the cluster core \citep[e.g.][]{Mazzotta2008,ZuHone2013}.
Another possibility is that MHs are generated through the continuous injection of electrons by inelastic collisions of relativistic cosmic-ray protons with the cluster thermal proton population, according to the so-called hadronic model \citep[e.g.][]{Pfrommer2004}.

A statistical connection between RHs and cluster mergers has been found with a RH generally hosted by clusters showing signs of dynamical disturbance as found from X-ray and optical observations \citep[e.g.][]{Buote2001,Cassano2010,Wen2013,Cuciti2021,Cassano2023}. The RH found in the cool-core cluster CL1821+643 \citep{Bonafede2014} represents an outlier with respect to these observations. 
Recently, low-frequency observations ($\le200$ MHz), mainly with the LOw Frequency ARray \citep[LOFAR;][]{vanHaarlem2013}, have improved and complicated the present picture by revealing the presence of diffuse emission on larger scales, both in relaxed and merging clusters. Diffuse emission extending up to the periphery of clusters (at least up to $R_{200}$\footnote{The radius within which the mean mass over-density of the cluster is 200 times the cosmic critical density at the cluster redshift}), embedding the central RH, has been observed in merging systems \citep[e.g.][]{Botteon2022b,Cuciti2022}. In addition, LOFAR revealed the presence of diffuse emission extending beyond the cluster core in two cool-core clusters that host a central MH and show signs of minor dynamical disturbance \citep{Savini2018,Savini2019}. More recently, the analysis of the dynamical properties of \textit{Planck} clusters in the LOFAR-DR2 area has revealed the presence of RH in less disturbed systems and that the fraction of newly detected RHs by LOFAR increases from merging to more relaxed systems \citep{Cassano2023}. 

In this paper, we aim to investigate the presence of cluster-scale diffuse radio emission in cool-core clusters.
To explain this emission, it has been proposed that less energetic mergers could still inject enough turbulence in the ICM to trigger particle acceleration on scales of hundreds of kiloparsecs, without disrupting the cool core \citep{Savini2018}.
Less energetic mergers should generate lower levels of turbulence in the ICM and thus particles should be accelerated with a lower efficiency, producing steeper spectra radio emission with $\alpha\ge1.5$ \citep{Cassano2006,Brunetti2008}.
Low-frequency observations have revealed several ultra-steep spectrum radio halos (USSRHs) in less massive disturbed systems or in clusters presumably experiencing fewer energetic merger events \citep[e.g.][]{Edler2022,Biava2021b,Bruno2023b}, confirming the predictions of the turbulent re-acceleration model \citep[e.g.][]{Cassano2023}. 
Furthermore, merger events responsible for this emission may have left signatures in the thermal emission of the clusters, such as sloshing features, although these details are not directly predicted by semi-analytical models that are based on merger trees of dark matter halos \citep{Cassano2016}. 
In this scenario, cool-core clusters that show minor signs of dynamical interactions on scales larger than the core should commonly show ultra-steep spectrum diffuse radio emission on large scales, which is best detected at lower frequencies.

\begin{figure}
\centering
\subfloat{
\resizebox{\hsize}{!}{\includegraphics{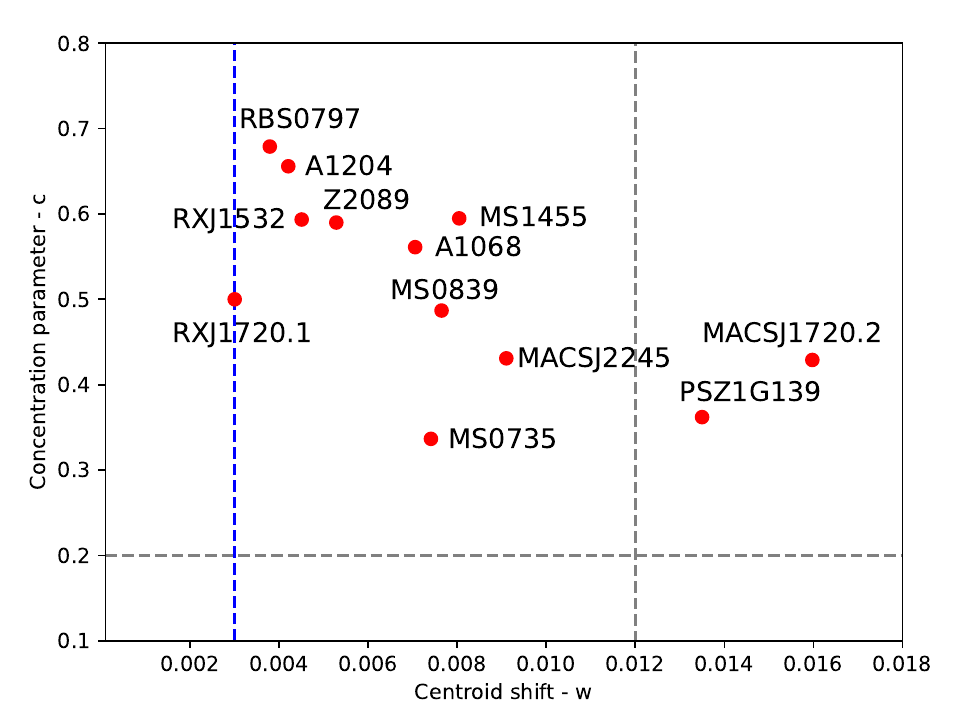}}}\\
\subfloat{
\resizebox{\hsize}{!}{\includegraphics{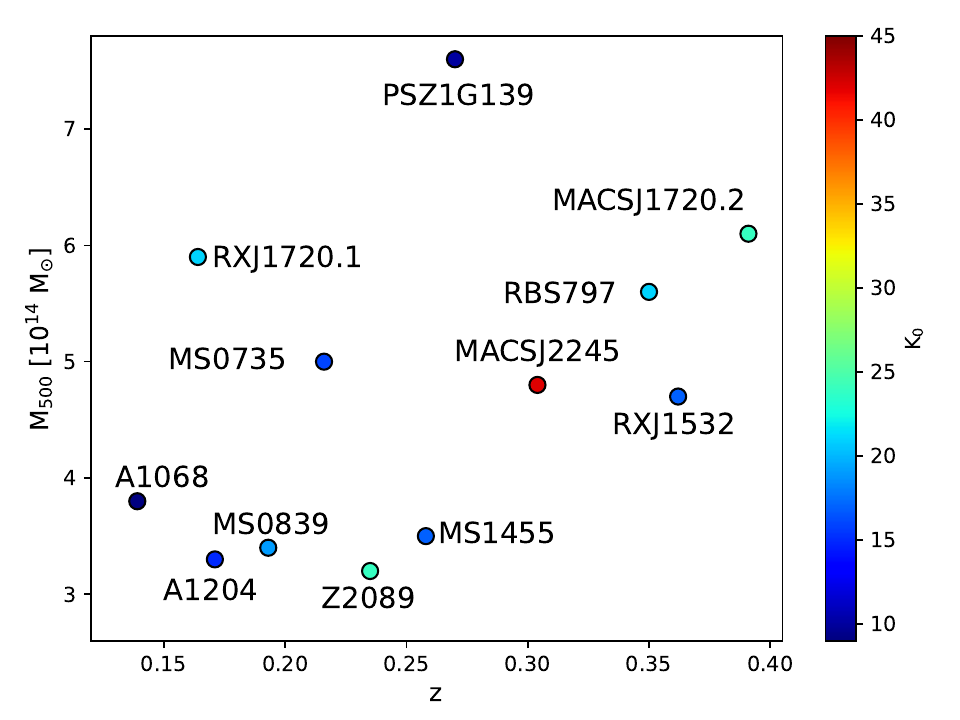}}}
\caption{Physical parameters of selected clusters. \emph{Top panel}: Morphological parameters' plot. The dashed blue line indicates the $w$ value used to select clusters in our sample. Dashed grey lines indicate the $c$ and $w$ values that separate clusters with a RH ($c\le0.2$ and $w\ge0.012$) from clusters without a RH, according to \cite{Cassano2010}. \emph{Bottom panel}: Redshift-mass distribution of clusters coloured according to their central entropy ($K_0$).}
\label{fig:c-w}
\end{figure}


\begin{table*}
    \caption{Properties of selected clusters.}
    \renewcommand\arraystretch{1.2}
    \centering
    \begin{tabular}{ccccccccc} \hline
Cluster name                &RA         &DEC            &z         &Scale   &$M_{500}$ &$K_0$ &c                    &    w                                  \\ 
&&&&[kpc/$\arcsec$] &[$10^{14}\ \rm{M_{\odot}}$] &[$\rm{keV\ cm^2}$]&&  \\\hline
PSZ1G139.61+24      &06:21:18.60        &+74:42:07.0      &0.270   &4.17   &7.6  &10 &0.36    &0.0135 \\
MS 0735.6+7421       &07:41:44.31       &+74:14:39.51      &0.216   &3.50   &5.0    &16 &0.34   &0.0074 \\ 
MS 0839.8+2938       &08:42:55.99       &+29:27:29.16      &0.193   &3.21   &3.4\tablefootmark{$*$} &19   &0.49   &0.0076  \\ 
Z2089               &09:00:36.85        &+20:53:40.62      &0.235   &3.67   &3.2\tablefootmark{$*$} &24   &0.59   &0.0045  \\ 
RBS 0797            &09:47:12.72        &+76:23:13.85      &0.350   &4.94   &5.6  &21  &0.68   &0.0038   \\
A1068               &10:40:44.53        &+39:57:11.94     &0.139   &2.45   &3.8  &9  &0.56   &0.0071   \\ 
A1204               &11:13:20.50        &+17:35:40.97      &0.171   &2.91   &3.3   &15 &0.66   &0.0042  \\ 
MS 1455.0+2232       &14:57:15.11       &+22:20:34.14      &0.258   &3.99   &3.5\tablefootmark{$*$}  &17  &0.59   &0.0080   \\ 
RX J1532.9+3021      &15:32:53.93       &+30:21:01.24      &0.362   &4.89   &4.7\tablefootmark{$*$} &17   &0.59   &0.0053 \\ 
RX J1720.1+2638     &17:20:10.10    &+26:37:29.50      &0.164   &2.81   &5.9  &21  &0.50      &0.0030        \\ 
MACS J1720.2+3536           &17:20:16.84        &+35:36:26.89      &0.391   &5.30   &6.1   &24 &0.43   &0.0159     \\ 
MACS J2245.0+2637    &22:45:04.58       &+26:38:04.46     &0.304   &4.49   &4.8\tablefootmark{$*$}  &42   &0.43   &0.0091        \\\hline
    \end{tabular}\\
    \label{tab:sample}
    \tablefoot{ Col. 1: Cluster name. Cols. 2 and 3: Coordinates. Col. 4: Redshift. Col. 5: Angular to linear scale conversion. Col. 6: Cluster mass from \cite{Planck2016}. Values marked with $*$ were estimated by \cite{Giacintucci2017} from the $M_{500}$ - $T_X$ relation of \cite{Vikhlinin2009}. Col. 7: Core entropy. Col. 8: Concentration parameter. Col. 9: Centroid shift.}
\end{table*}

To test this hypothesis and verify the occurrence of these sources, we observed with the LOFAR high band antennas (HBA; at 144 MHz) a sample of twelve cool-core clusters selected for a range of dynamical states. 
We classified the radio emission detected with LOFAR and compared it with observations at higher frequencies. We characterised the diffuse radio emission, analysing radial surface brightness profiles and estimating the radio power at 144 MHz. In the case of a non-detection, instead, we provided upper limits. 
Finally, we performed a systematic search for sloshing features in all the systems re-analysing \emph{Chandra} X-ray archival data, and we compared radio and X-ray properties of the clusters in our sample to try to understand under which conditions cluster-scale diffuse radio emission is generated in cool-core clusters.

The paper is organised as follows: in Section \ref{sec:sample} we present the selection criteria of the sample. In Section \ref{sec:data} we describe the data used in our analysis and their reduction. In Section \ref{sec:results} we present the morphology of the radio and X-ray emission for each object. In Section \ref{sec:analysis} we analyse and explain how we modelled the surface brightness profiles of the diffuse emission, estimated the associated power, and derived upper limits for non-detections. We discuss our results in Section \ref{sec:discussion} and we present our conclusions in Section \ref{sec:conclusions}. 
Throughout the paper we assumed a standard $\Lambda$CDM cosmological model with $\Omega_{\Lambda} = 0.7$, $\Omega_m = 0.3,$ and $H_0 = 70\ {\rm km\ s^{-1}\ Mpc^{-1}}$.

\section{The sample}\label{sec:sample}
To test the minor-merger scenario as a possible explanation for the presence of cluster-scale diffuse emission in relaxed clusters, we selected a sample of clusters that have a cool-core and show varying levels of minor dynamical disturbances. 
The dynamical state of clusters has been inferred from X-ray observations, through the computation of morphological parameters. In particular, the concentration parameter ($c$) and the centroid shift ($w$) can provide an effective evaluation of the cluster dynamical state \citep[][]{Lovisari2017}.
We describe these parameters as follows:
\begin{itemize}
    \item the concentration parameter, $c$, defined as the ratio of the central (within a radius $r=100$ kpc) over the ambient (within $r=500$ kpc) number of exposure-corrected counts, $S$ \citep{Santos2008}. 
    \begin{equation}
        c = \frac{S(r < 100 {\rm kpc})}{S(r < 500 {\rm kpc})};
    \end{equation}
    \item the centroid shift, $w$, defined as the standard
deviation of the projected separation $\Delta$ between the X-ray peak and
the cluster X-ray centroid computed within N circles of increasing radius, up to $R=500$ kpc \citep{Poole2006,Maughan2008}.
\begin{equation}
    w = \left[\frac{1}{N-1}\sum(\Delta_i - \Braket{\Delta} )^2\right]^{1/2} \times \frac{1}{R}.
\end{equation}
\end{itemize}
Large values of $c$ indicate a dense core, typical of cool-core clusters that have not undergone a recent major merger, while large values of $w$ indicate a distribution of gas that is not spherically symmetric, that is, a merging cluster.
\cite{Cassano2010} found that most of the clusters in their sample with $c$ < 0.2 and $w$ > 0.012 (i.e. merging) host a RH, while no cluster with $c$ > 0.2 and $w$ < 0.012 (i.e. relaxed clusters) showed RH detections at frequencies $\nu$ > 610 MHz.
However, more recently the statistical analysis of the \textit{Planck} clusters in the LOFAR DR2 has shown that the number of RHs in more dynamically relaxed clusters is not negligible at low radio frequency \citep{Cassano2023}. In particular, \cite{Savini2019} found that the most relaxed cluster with giant halo-like emission has $c = 0.3$ and $w = 0.003$.

\begin{table*}
    \caption{Observation details.}
    \renewcommand\arraystretch{1.2}
    \centering
    \begin{tabular}{ccccc} \hline
Cluster name  &LOFAR Project & LoTSS pointing &\emph{Chandra} ObsID &\emph{Chandra} exposure\\ 
&&&&[ks] \\\hline
PSZ1G139.61+24    &LC8\_022 &P098+74  &15139, 15297 &27.5 \\
MS 0735.6+7421    &LC10\_010 &-  &4197, 10468, 10469 &550\\ 
&&&10470, 10471, 10822 \\ &&&10918, 10922 \\
MS 0839.8+2938    &LT10\_010 &P129+29, P132+29 &2224 &29.8   \\ 
Z2089             &LC10\_010 &-   &10463 &40.6 \\ 
RBS 0797          &LC10\_010& -  &7902 &38.3  \\
A1068             &LC4\_034 &P157+40, P160+42, P161+40  &1652 & 26.8  \\ 
A1204             &LC10\_010 &-  &2205 &23.6  \\ 
MS 1455.0+2232    &LT10\_010 &P225+22 &4192 &91.9     \\ 
RX J1532.9+3021   &LC10\_010 &- &1649, 1665, 14009 &108.2\\ 
RX J1720.1+2638   &LC7\_024, LT10\_010 &P260+28, P261+25 &3224, 4361, 1453 &57.3   \\ 
MACS J1720.2+3536         &LC10\_010&-  &3280, 7718, 6107 &61.7  \\ 
MACS J2245.0+2637 &LC6\_015  &P341+26, P342+28 &3287 &16.9       \\\hline
    \end{tabular}
    \label{tab:obs}
\end{table*}

To search for cool-core clusters, we chose our targets from two X-ray samples of galaxy clusters: the Archive of \emph{Chandra} Cluster Entropy Profile Tables (ACCEPT) sample \citep{Cavagnolo2009} and the ME-MACS sample \citep{Mann2012,Rossetti2017}. 
We selected all ME-MACS clusters that are classified as cool-core by \cite{Rossetti2017}, according to the concentration parameter, and all ACCEPT clusters that are classified as cool-core by \cite{Cavagnolo2009} and \cite{Giacintucci2017}, according to the core entropy ($K_0 \le 50\ \rm{keV\ cm^2}$).
We then selected those clusters presenting minor signs of dynamical interactions on large scales, using the centroid shift as an indication of that, requiring $w$ > 0.003, i.e. larger than the $w$ of the cluster RX J1720.1+2638, in which diffuse emission outside the cluster core has been observed by \cite{Savini2019}. 
We computed the morphological parameters of the selected clusters from \emph{Chandra} archival observations that we have reprocessed as described in section \ref{sec:Chandra}, excluding those clusters without \emph{Chandra} observations covering the central $\rm{Mpc^2}$.
In addition, we considered only clusters that could easily be observed with LOFAR, i.e. with declination greater than $10^{\circ}$, and with a redshift z < 0.4 to resolve emission on scales of 500 kpc at the resolution of $20\arcsec$.
The final sample consists of 12 clusters listed in Table \ref{tab:sample} and shown in Fig. \ref{fig:c-w} in the $c-w$ plot. 
We also report in Table \ref{tab:sample} the core entropy values estimated by \cite{Cavagnolo2009} and \cite{Giacintucci2017} for the selected clusters. We note that all the clusters in the sample have a core entropy $K_0 \le 50\ \rm{keV\ cm^2}$, and so are cool-core clusters according to \cite{Giacintucci2017}.


\section{Observations and data reduction}\label{sec:data}

\subsection{LOFAR data}\label{sec:HBA_cal}
The selected clusters were observed with LOFAR HBA in the frequency range 120--168 MHz. Half of these were observed by LoTSS \citep[LOFAR Two-metre Sky Survey;][]{Shimwell2017}, multiple observations are used if the cluster falls between different survey pointings to achieve comparable sensitivity to those where the target is in the pointing centre. The rest of the sample was observed as part of a dedicated proposal (project LC10\_010, P.I. Bonafede), and so the observations are centred on the clusters.
The observational details are reported in Table \ref{tab:obs}. For the clusters observed by LoTSS, we indicate both the project ID and the pointings used.
Each observation consists of an 8 hr time-on-source book-ended by 10 min scans on the flux density calibrator using HBA stations in the \verb|HBA_DUAL_INNER| mode, implying that only central antennas are used for the remote stations to mimic the size of the core stations.

All the data were processed with the standard
Surveys Key Science Project pipeline \footnote{https://github.com/mhardcastle/ddf-pipeline/} \citep[see][]{Shimwell2019,Tasse2021}. 
Here we report only the main steps of data reduction, and we refer to \cite{Shimwell2022} for more details.
The data were firstly corrected for the direction-independent (DI) effects \citep[for a description of the procedure][see]{vanWeeren2016,Williams2016,deGasperin2021} and then for the direction-dependent (DD) effects through the pipeline that uses \verb|killMS| and \verb|DDF| \citep{Tasse2014a,Tasse2014b,SmirnovTasse2015,Tasse2018}. 
Finally, to improve the calibration on the target region we followed the extraction and self-calibration procedure described in \cite{vanWeeren2021}. Here we subtracted all the sources outside a small region (typically radius r = 0.3-0.6 deg) centred on the cluster, using the DD gains, and then performed several cycles of phase and amplitude self-calibration in the extracted region using \verb|DPPP| and \verb|WSclean| \citep{DPPP2018,Offringa2014}. When multiple observations of a cluster are present we combined them during the self-calibration imaging whilst accounting for the different beam attenuation.
The flux density scale was set according to \cite{ScaifeHeald2012}, and subsequently aligned with LoTSS-DR2 data release, where the flux calibration uncertainty is estimated to be 10 percent \citep[][]{Hardcastle2021,Shimwell2022}.

The calibrated data were then imaged at different resolutions using \verb|WSClean| \citep{Offringa2014}.
The high-resolution images (resolution $\sim6\arcsec$) were produced using a Briggs weighting scheme with robust $= -0.5$, applying an inner $uv$-cut at $80\lambda$ and using a pixel-scale of 1$\arcsec$. For the low-resolution images (resolution $\sim20\arcsec$) instead we used a pixel-scale of 4$\arcsec$, a $uv$-taper of 20$\arcsec$ and multi-scale deconvolution to improve the recovery and modelling of diffuse emission. Finally, we performed a subtraction of the central source and of surrounding discrete sources which could influence the study of diffuse emission. The source subtraction was performed in the visibility plane, using a model from high-resolution images. When diffuse emission is present, we applied a inner $uv$-cut to exclude short baselines and filtered out the emission on that scales. It is well known that the subtraction procedure is not a trivial task and may leave residual artefacts that contaminate the diffuse emission \citep[see a detailed discussion in][]{Bruno2023a}. In the case of non-negligible subtraction residuals, we masked the emission of extended sources during the analysis. A more detailed case-by-case description is provided in Section \ref{sec:results}. For each cluster, in Figs. \ref{fig:PSZ1G139} - \ref{fig:RXJ1720_2} we show the high-resolution radio image and the source-subtracted low-resolution radio contours overlaid onto the \emph{Chandra} X-ray image. 

\subsection{Chandra data}\label{sec:Chandra}
We investigated the thermal emission of our clusters using \emph{Chandra} X-ray observations to search for small-scale surface brightness discontinuities with high angular resolution.
We reprocessed \emph{Chandra} archival data, listed in Table \ref{tab:obs}, using CIAO 4.14 (\emph{Chandra} Interactive Analysis of Observation), starting from the \verb|level = 1| event file.
We searched and removed time periods affected by soft proton flares by inspecting the light curve extracted in the 0.5–7.0 keV band.
When multiple observations of a single cluster were present, we combined them to improve the exposure time. 
Then we replaced point sources, identified with the \verb|wavdetect| task and confirmed by eye, with the Poissonian noise of a surrounding region, using the CIAO task \verb|dmfilth|. Exposure-corrected images have been created in the energy band 0.5–7\,keV, and binned to a common resolution of $0.98\arcsec$.
A background image was produced by re-projecting the blank sky event files to the corresponding event files for every ObsID and adding the background files of multiple observations, when present, using the CIAO tasks \verb|blanksky| and \verb|blankskyimage|. 

Our goal is to understand if the selected clusters have undergone a minor merger. We therefore carried out a systematic search for cold fronts in all the galaxy clusters of the sample, regardless of the presence of diffuse radio emission.
To enhance the detection of possible surface brightness discontinuities in the X-ray images we produced X-ray surface brightness gradient maps using the adaptively smoothed
Gaussian Gradient Magnitude (GGM) filter introduced by \cite{Sanders2016}. We also inspected residual images obtained by subtracting a double circular or elliptical double $\beta$-model, depending on the case, from the original image to highlight deviations from a relaxed cool-core.

Then using \verb|pyproffit|\footnote{https://pyproffit.readthedocs.io} \citep{Eckert2020} we extracted and fitted surface brightness radial profiles on background-subtracted exposure-corrected images along the direction of candidate discontinuities and in adjacent sectors to cover the whole thermal emission. 
We established quantitative criteria to affirm the presence of surface brightness discontinuities, fitting the radial profiles with a double $\beta$-model and considering as candidate discontinuities those profiles with $\chi^2_{\rm red}\ge1.5$. We then further investigated these discontinuities modelling the region of interest with a broken power-law projected along the line of sight, of the form:
\begin{equation}
    n(r) = 
    \begin{cases}
    C(r/r_j)^{\alpha_1} & \text{if $r\le r_j$} \\
    (r/r_J)^{\alpha_2} & \text{if $r > r_j$},
    \end{cases}
\end{equation}
where $\alpha_1$ and $\alpha_2$ are the power-law slopes and $C$ is the surface brightness jump at the discontinuity radius $r_j$. 

Finally, we checked the nature of surface brightness discontinuities that are well modelled by a broken power-law, by performing spectral analysis.
We extracted and fitted spectra (using all the ObsIDs available
for each cluster, when multiple observations are present) in the 0.7–10.0 keV band using the package \verb|XSPEC| v12.12.1 with the \cite{Anders1989} abundances table. Cash statistic \citep{Cash1979} was adopted. 
We used as background files those extracted from the blank sky events, as we are analysing spectra in the bright central regions of the clusters where the background is not critical. We modelled the ICM emission with a thermal model taking into account the Galactic absorption in the direction of the clusters (phabs * apec), adopting the values from \cite{HI4PI2016}. Temperature, abundance, and normalisation are free to vary. 
A cold front is characterised by a jump in temperature with higher temperature in the outer region, while a shock front presents a reversed temperature jump. We therefore looked at trends in the projected temperature profiles (a conservative approach since projection smooths gradients) across the surface brightness discontinuities to characterise their nature.

\begin{table*}
    \caption{LOFAR classification and X-ray features.}
    \renewcommand\arraystretch{1.2}
    \centering
    \begin{tabular}{ccccccc} \hline
Cluster name  &$S_{6\arcsec}$ &$S_{\rm excess}$ &$R_{\rm max}$ &$0.2R_{500}$ &Radio &X-ray\\
&[mJy] &[mJy] &[kpc] &[kpc] &classification &features \\\hline
PSZ1G139.61+24 &$22\pm2$ &$2.1\pm1.2$ &584 &240 &MH+RH  &Elongated emission + CF \\
MS 0735.6+7421 &4$500\pm450$ &0 &284 &  &BCG+lobes &Multiple cavities + shock  \\ 
MS 0839.8+2938  &$170\pm17$ &0 &67 &200  &BCG+lobes & -    \\ 
Z2089     &$100\pm10$ &0 &57 &192         &BCG & -    \\ 
RBS 0797   &$110\pm11$ &0 &124 &232       &MH &Multiple cavities + multiple shocks   \\
A1068       &$36\pm4$ &$1.3\pm0.7$ &233 &206      &MH+RH\tablefootmark{$*$} &3 CF\tablefootmark{$*$}   \\ 
A1204            &$14\pm2$ &$0.4\pm0.9$ &73 &178 &BCG  &Cavities  \\ 
MS 1455.0+2232   &$155\pm16$ &$1.0\pm1.6$ &335 &196 &MH+RH  &2 CF  \\ 
RX J1532.9+3021  &$92\pm9$ &$0.1\pm0.8$ &122 &208 &MH &Cavities \\ 
RX J1720.1+2638  &$850\pm85$ &$2.0\pm2.0$ &343 &248 &MH+RH &2 CF\\ 
MACS J1720.2+3536       &$160\pm16$ &0 &80 &224  &BCG  & Cavities        \\ 
MACS J2245.0+2637 &$44\pm4$ &$0.2\pm0.9$ &125 &214 &cMH\tablefootmark{$*$} &Elongated emission      \\\hline
    \end{tabular} \\
    \tablefoot{ Col. 1: Cluster name. Col 2: Radio flux density at 6\arcsec resolution. Col. 3: Radio flux density excess at 20\arcsec resolution, with respect to 6\arcsec resolution. Col. 4: Maximum radius measured on $3\sigma$ radio contours. Col. 5: Comparing radius for the classification of the radio emission. Col. 6: Radio classification: MH = mini-halo, RH = giant radio halo, cMH = candidate mini-halo, BCG = brightest cluster galaxy. Col. 7: X-ray features: CF = cold front. $^*$ indicates new detections.} 
    \label{tab:results_sample}
\end{table*}

\section{Radio and X-ray morphology}\label{sec:results}
In the following subsections, we briefly discuss each cluster reporting information from the literature and our new results based on LOFAR observations and re-analysis of \emph{Chandra} data.
In Table \ref{tab:results_sample} we summarise the radio and X-ray characteristics of these sources.
To classify the radio emission in our clusters we took into account the MH definition provided by \cite{Giacintucci2017}, according to which the MH emission is not directly connected to the central AGN, and has an extension of $50\rm{kpc}\le R_{\rm radio}\le 0.2R_{500}$. As most of the sources in the sample exhibit an irregular and elongated morphology, to quantify the extension of the radio emission we considered the maximum radii measured on $3\sigma$ surface brightness contours. We then proceed in the following way:
\begin{itemize}
    \item First, we checked the high-resolution radio images to identify the presence of radio emission associated with the central AGN, like lobes or tails. Sometimes this step required the inspection of data at higher resolution from the literature and optical and X-ray counterpart identification.
    \item Then, we produced radio images at lower resolution with an $uv$-taper of 20\arcsec, to improve the detection of diffuse emission. Decreasing the resolution, however, does not necessarily imply collecting more emission, sometimes we are simply spreading the radio emission on a larger area. Therefore, we compared the radio flux density within a region encompassed by the $3\sigma$ contour in the high and low-resolution radio images (with discrete sources subtracted), to check for the presence of an excess of radio emission. 
    \item Finally, we measured the extension of the radio emission. If the flux density measured from the low resolution image is larger than the flux density measured from high resolution image, we have used the former to estimate the size of the diffuse emission. In the other cases, we have estimated the size of the diffuse emission from the high resolution image. 
\end{itemize}
Following the above procedure we classified a source as: 'without diffuse emission' if its emission is clearly or possibly associated with the central AGN and no more emission is measured in the lower resolution images; 'with a mini-halo' if there is a clear presence of diffuse emission and its extension is $50\rm{kpc}\le R_{\rm radio}\le 0.2R_{500}$; 'with cluster-scale diffuse emission' if the diffuse emission extension is $R_{\rm radio}\ge 0.2R_{500}$.
For the sake of the clarity of this section, we anticipate here our results on the radio and X-ray morphology of the clusters of our sample.
LOFAR observations at 144 MHz revealed the presence of cluster-scale diffuse radio emission in the clusters PSZ1G139.61+24, RX J1720.1+2638, MS 1455.0+2232 and A1068. 
We also detected two MHs in the clusters RX J1532.9+3021 and RBS 0797, and a candidate MH in the cluster MACS J2245.0+2637. In the remaining six clusters we have not detected halo-like diffuse emission.
X-ray analysis shows evident signatures of sloshing in those clusters presenting cluster-scale diffuse emission. 
Systems of cavities and sometimes also shock fronts are instead detected in clusters presenting a powerful central galaxy \citep{Schindler2001,Hlavacek2013,Vantyghem2014,Ubertosi2021,Ubertosi2023}. 
Most of these features were already known in the literature, while we detected for the first time a pair of cold fronts in the cluster A1068.
The only discrepancy with the literature was found for the cluster RX J1532.9+3021, for which we are not able to confirm the cold front reported in \cite{Hlavacek2013}.


\subsection{Clusters with cluster-scale diffuse emission}

\subsubsection{RX J1720.1+2638}
\begin{figure*}
\centering
\includegraphics[width=17cm]{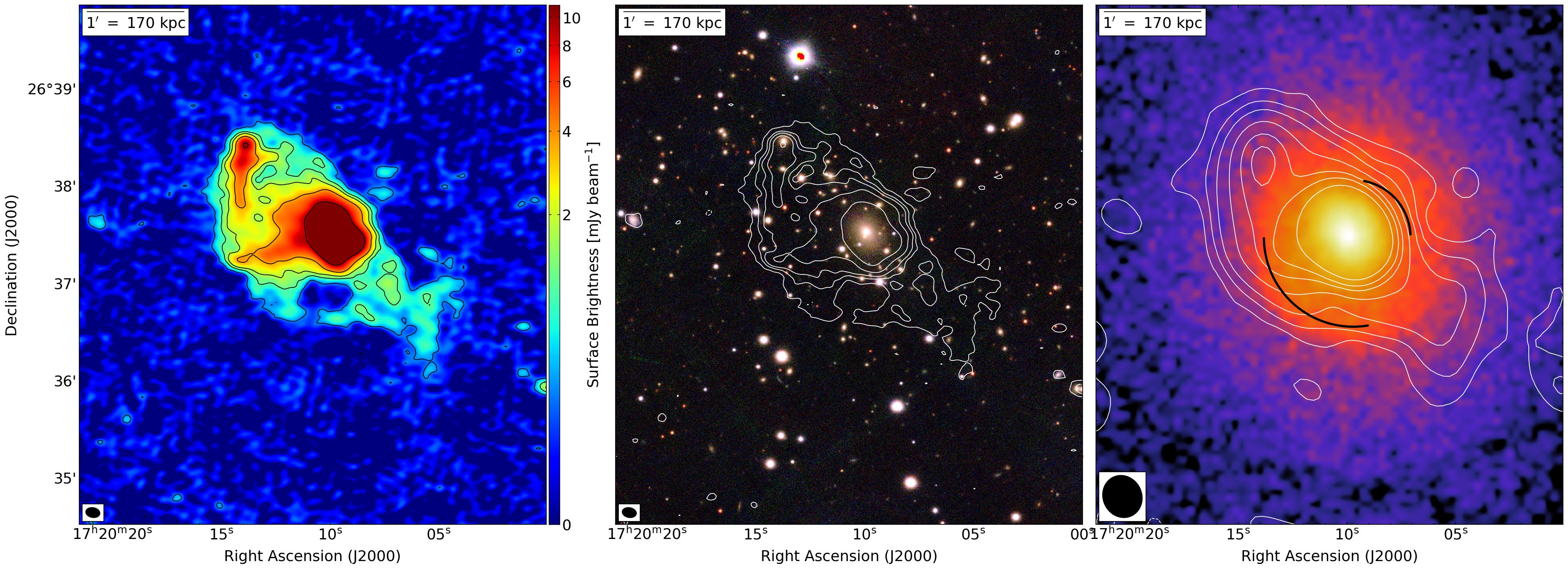}
\caption{Multi-wavelength images of RX J1720.1+2638. \emph{Left panel}: High-resolution 144 MHz LOFAR image. Levels:[-3, 3, 9, 15, 30, 80]$\times\sigma$ (where $\sigma=0.12$ mJy $\rm{beam}^{-1}$). The beam is $8.8\arcsec\times6.1\arcsec$ and is shown in the bottom left corner of the image. 
\emph{Central panel}: Optical \emph{Pan-STARRS} RGB image with high-resolution LOFAR contours overlaid. \emph{Right panel}: \emph{Chandra} X-ray image smoothed on a scale of $3\arcsec$ with overlaid low-resolution LOFAR contours (Levels=[-3, 3, 9, 20, 50, 100, 150, 250]$\times\sigma$, where $\sigma=0.22$ mJy $\rm{beam}^{-1}$ and the beam is $26.5\arcsec\times24.0\arcsec$). The black arcs indicate the position of the cold fronts found by \protect\cite{Mazzotta2008}. All the panels depict the same region of the sky.}
\label{fig:RXJ1720_1}
\end{figure*}

RX J1720.1+2638 (hereafter RXJ1720.1) is a cool-core galaxy cluster located at z=0.164, which hosts a bright central MH with a size of 70 kpc and a spectral index of $\alpha_{237}^{4850}=1.0\pm0.1$, and a lower surface brightness extension to the east \citep[][]{Giacintucci2014a}.
Two cold fronts detected with \emph{Chandra} on opposite sides with respect to the cluster centre, tracing the spiral pattern of the sloshing motion, appear to confine the MH \citep[][]{Mazzotta2001, Mazzotta2008}.
In the cluster field, there is also a head-tail radio galaxy, associated with a cluster member galaxy \citep[][]{Owers2011}, located north-east of the cluster core. The nature of this source is confirmed also by radio spectral analysis, presenting a flat core and  steepening along the tail \citep{Giacintucci2014a,Biava2021b}.
LOFAR HBA observations revealed the presence of diffuse emission on larger scales, which extends in the north-east--south-west direction up to $D_{\rm max}\sim560$ kpc \citep{Savini2019}.
Here, we re-analysed LOFAR HBA data combining two LoTSS observations of this source (Fig. \ref{fig:RXJ1720_1}), resulting in a 15\% improvement in sensitivity with respect to the \cite{Savini2019} images.
The cluster-scale diffuse emission is clearly visible south-west of the cluster core, even in the high-resolution LOFAR image (left panel of Fig. \ref{fig:RXJ1720_1}).
This emission is characterised by an ultra-steep spectrum with $\alpha\sim3$ in the frequency range 54 - 144 MHz \citep[][]{Biava2021b}. Radio and X-ray analysis of this source performed by \cite{Biava2021b}, suggests the cluster-scale diffuse emission resembles a RH, generated by the re-acceleration of particles after a minor merger.

\subsubsection{MS 1455.0+2232}
\begin{figure*}
\centering
\includegraphics[width=17cm]{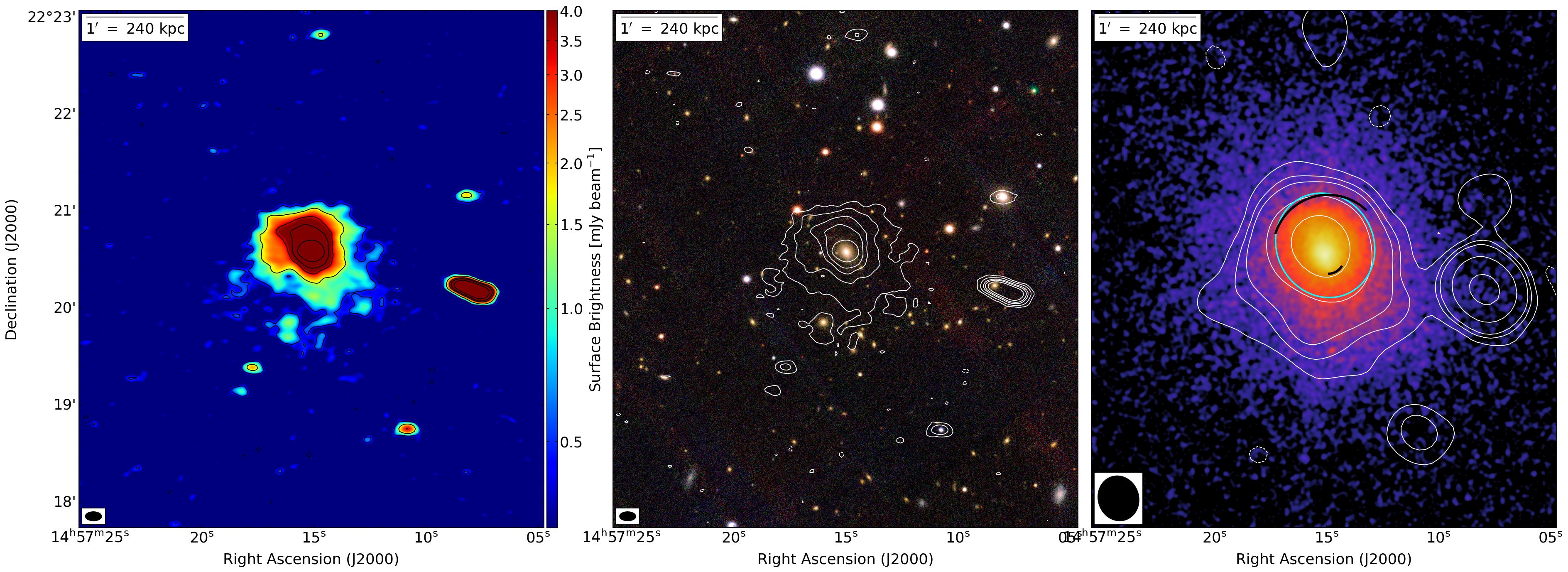}
\caption{Multi-wavelength images of MS 1455.0+2232. \emph{Left panel}: High-resolution 144 MHz LOFAR image. Levels:[-3, 3, 9, 20, 30, 50]$\times\sigma$ (where $\sigma=0.17$ mJy $\rm{beam}^{-1}$). The beam is $10.1\arcsec\times5.6\arcsec$ and is shown in the bottom left corner of the image. 
\emph{Central panel}: Optical \emph{Pan-STARRS} RGB image with high-resolution LOFAR contours overlaid. \emph{Right panel}: \emph{Chandra} X-ray image smoothed on a scale of $1\arcsec$ with overlaid low-resolution LOFAR contours after the subtraction of the central point source (Levels=[-3, 3, 9, 15, 50, 150]$\times\sigma$, where $\sigma=0.22$ mJy $\rm{beam}^{-1}$ and the beam is $28.1\arcsec\times25.1\arcsec$). The black arcs indicate the position of the cold fronts found by \protect\cite{Mazzotta2008}, while the cyan circle indicates the boundaries of the sloshing region pointed out in \protect\cite{Riseley2022}. All the panels depict the same region of the sky.}
\label{fig:MS1455}
\end{figure*}

GMRT (610 MHz) and VLA (1400 MHz) observations of MS 1455.0+2232 (hereafter MS1455, z=0.258) indicate the presence of a central MH with a linear size of $D_{\rm max}\sim350$ kpc and a spectral index of $\alpha=1.46\pm0.22$ \citep{Venturi2008,Giacintucci2019}. A pair of cold fronts, indicated with black arcs in the bottom right panel of Fig. \ref{fig:MS1455}, surrounds the MH emission \citep{Mazzotta2008}.
A re-analysis of the \emph{Chandra} data performed by \cite{Riseley2022} points out that the sloshing region is larger than previously reported by \cite{Mazzotta2008}. The cyan circle in the bottom right panel of Fig. \ref{fig:MS1455} delimits that region. While this article was in preparation, \citep{Giacintucci2024} reported the detection of a further cold front southern of the cluster centre at $r\sim114\arcsec\cong455$ kpc.
The diffuse emission is well visible with LOFAR (Fig. \ref{fig:MS1455}) both in the high and low resolution image (obtained subtracting the central compact source, modelled applying an inner $uv$-cut of $5k\lambda\cong160$ kpc), where it extends up to $D_{\rm max}\sim470$ kpc. The radio emission thus extends beyond the central cold fronts, while it is contained within the large-scale sloshing recently detected by \cite{Giacintucci2024}.
This source was also observed with MeerKAT at 1.28 GHz, detecting even more emission than LOFAR \citep[total extension at 1.28 GHz of $D_{\rm max}\sim586$ kpc][]{Riseley2022}, thanks to the very high sensitivity of these data ($8\ \rm{\mu Jy\  beam^{-1}}$ for MeerKAT data compared to $159\ \rm{\mu Jy\  beam^{-1}}$ for LOFAR data at same resolution of $15\arcsec$).
The diffuse emission has a mean spectral index between 144 and 1283 MHz of $\alpha=1.1\pm0.1$, with a slightly larger dispersion in the regions outside the sloshing boundaries ($\sigma=0.08$, respect to $\sigma=0.05$ inside the sloshing region).

\subsubsection{PSZ1G139.61+24}
\begin{figure*}
\centering
\includegraphics[width=17cm]{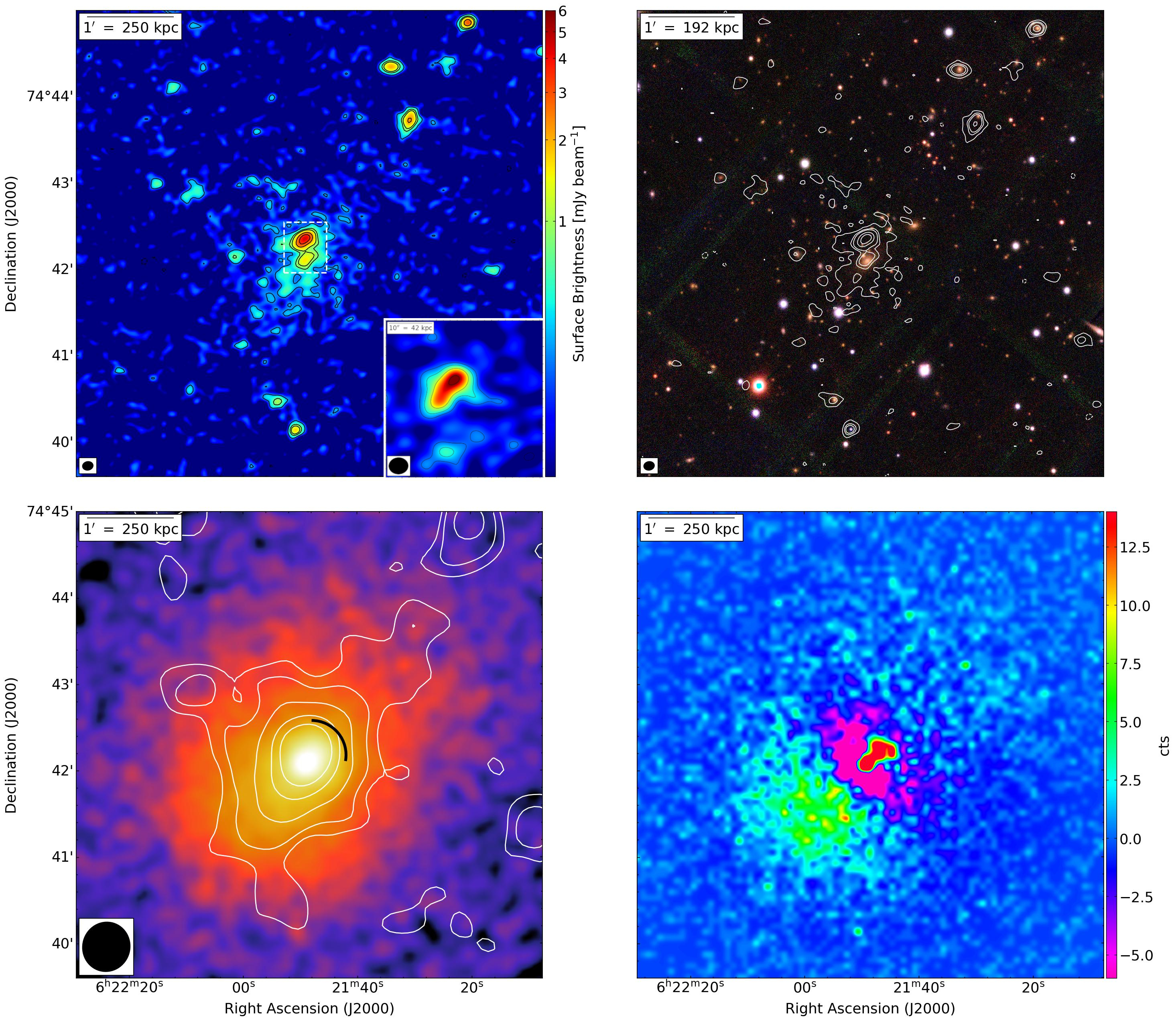}
\caption{Multi-wavelength images of PSZ1G139.61+24. \emph{Upper left panel}: High-resolution 144 MHz LOFAR image. Levels:[-3, 3, 7, 10, 20, 30]$\times\sigma$ (where $\sigma=0.11$ mJy $\rm{beam}^{-1}$). The beam is $7.4\arcsec\times6.0\arcsec$ and is shown in the bottom left corner of the image. The insert box shows the image of the two central sources at $4.5\arcsec\times3.8\arcsec$ resolution, obtained with Briggs weighting robust=-1.25. 
\emph{Upper right panel}: Optical \emph{Pan-STARRS} RGB image with high-resolution LOFAR contours overlaid. \emph{Lower left panel}: \emph{Chandra} X-ray image smoothed on a scale of $1\arcsec$ with overlaid low-resolution source-subtracted LOFAR contours (Levels=[-3, 3, 5, 9, 15, 20]$\times\sigma$, where $\sigma=0.3$ mJy $\rm{beam}^{-1}$ and the beam is $34.6\arcsec\times33\arcsec$). The black arc indicates the position of the cold front found by \protect\cite{Savini2018}. 
\emph{Lower right panel:} Residual X-ray image after the subtraction of a circular double $\beta$-model.
All the panels depict the same region of the sky.}
\label{fig:PSZ1G139}
\end{figure*}

The cluster PSZ1G139.61+24 (hereafter PSZ1G139), located at z=0.27, presents elongated X-ray emission in the north-west--south-east direction with a cold front $\sim100$ kpc to the north-west \citep{Giacintucci2017,Savini2018}. We note that the thermal emission is mostly elongated towards the south-east and also the residual map presents an excess of emission in that direction (Fig. \ref{fig:PSZ1G139}, lower panels). This feature is a further indication of sloshing. According to sloshing simulations, the classical spiral morphology is observed when the line-of-sight is perpendicular to the orbital plane of the merger or at low inclinations of the orbital plane, becoming difficult to recognise over 70 degrees \citep{Roediger2011,Gastaldello2013}. The morphology observed in the residual map of PSZ1G139 is, instead, displayed in simulations with an orbital plane parallel to the line of sight.

In the radio band the cluster was observed by \cite{Savini2018} with the Giant Metrewave Radio Telescope (GMRT) at 610 MHz and LOFAR at 144 MHz. Both telescopes detect a central MH with an extension of $\sim200$ kpc, whilst LOFAR maps also show additional diffuse radio emission beyond the cluster core, reaching a total extension of $D_{\rm max}\sim500$ kpc. In this frequency range, the central MH has a spectral index of $\alpha\sim1.3$, while the diffuse emission on a larger scale has a steeper spectrum with a $2\sigma$ lower-limit spectral index of $\alpha\ge1.7$ \citep{Savini2018,Savini2019}.

We have reprocessed the LOFAR data of \cite{Savini2018,Savini2019} using the procedure described in Section \ref{sec:HBA_cal}, obtaining significant improvements in dynamic range and image fidelity \citep[Fig. \ref{fig:PSZ1G139}, $\sigma = 300\ \rm{\mu Jy\ beam^{-1}}$ at $34.6\arcsec\times33\arcsec$ resolution, against $\sigma = 500\ \rm{\mu Jy\ beam^{-1}}$ at $35\arcsec\times35\arcsec$ resolution in][]{Savini2019}.
At high resolution there are two peaks of radio emission (upper left panel of Fig. \ref{fig:PSZ1G139}): the faintest one is associated with the BCG and coincides with the X-ray peak, while the brightest one on the north does not have an obvious optical counterpart, and could be a background radio galaxy. The two central sources are surrounded by diffuse emission.
To improve the detection of this emission, we subtracted the central sources and compact near by sources, producing a model with an inner $uv$-cut of $5k\lambda\cong170$ kpc at cluster redshift, and re-imaged the resulting data with a $uv$-taper of 20\arcsec. The radio contours of this image are reported onto the X-ray image in lower left panel of Fig. \ref{fig:PSZ1G139}.
We detected radio emission extending even beyond the cold front, reaching a total extension of $D_{\rm max}\sim 760$ kpc.

\subsubsection{A1068}\label{A1068}
\begin{figure*}
\centering
\includegraphics[width=17cm]{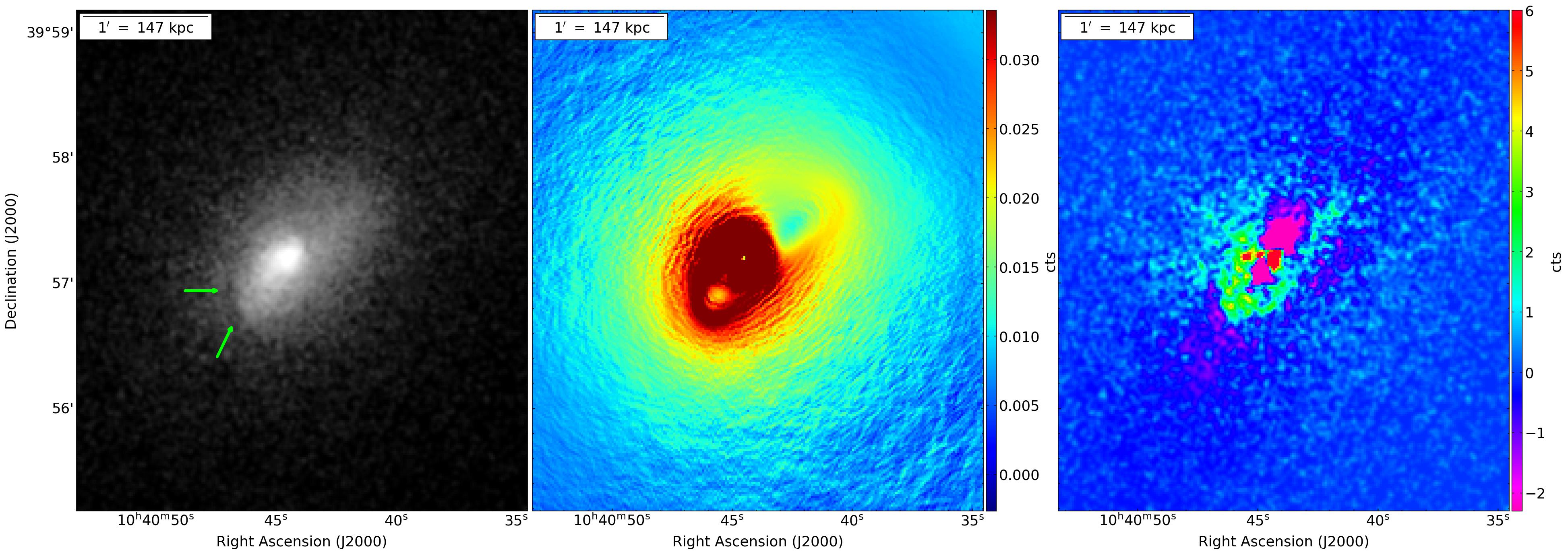}
\caption{X-ray images of A1068. \emph{Left panel:} \emph{Chandra} 0.5–7.0 keV image. Green arrows indicate the position of PDL. \emph{Central panel:} X-ray surface brightness gradient map. \emph{Right panel:} Residual image after the subtraction of an elliptical double $\beta$-model.}
\label{fig:A1068_Xmaps}
\end{figure*}

The X-ray emission of A1068 (z=0.139) shows a highly elliptical morphology, with the major axis along the north-west south-east direction (Fig. \ref{fig:A1068_Xmaps}, left panel).
An elongated morphology is also reflected in the abundance map of the elements, where there is an enrichment along the major axis, suggesting the central gas is driven to large distances by turbulent motion \citep{Wise2004,McNamara2004}.
The GGM map (see Fig. \ref{fig:A1068_Xmaps}, central panel) shows a positive gradient of X-ray surface brightness in the south-west region of the cluster, suggesting the presence of a surface brightness discontinuity. A discontinuity could also be present on the north of the core, where there is a net jump in the surface gradient map.
North-west and south-east of the cluster centre there are instead two regions of low gradient, resembling holes, which indicate regions where the gas is denser and elongated by the sloshing. 
An excess of X-ray surface brightness is also visible in the residual map (Fig. \ref{fig:A1068_Xmaps}, right panel), obtained by subtracting an elliptical double $\beta$-model, similar to the excess of surface brightness gradient in the GGM map. 
 
\begin{figure*}[h!]
\centering
\subfloat{ \qquad \qquad
\includegraphics[width=6.5cm]{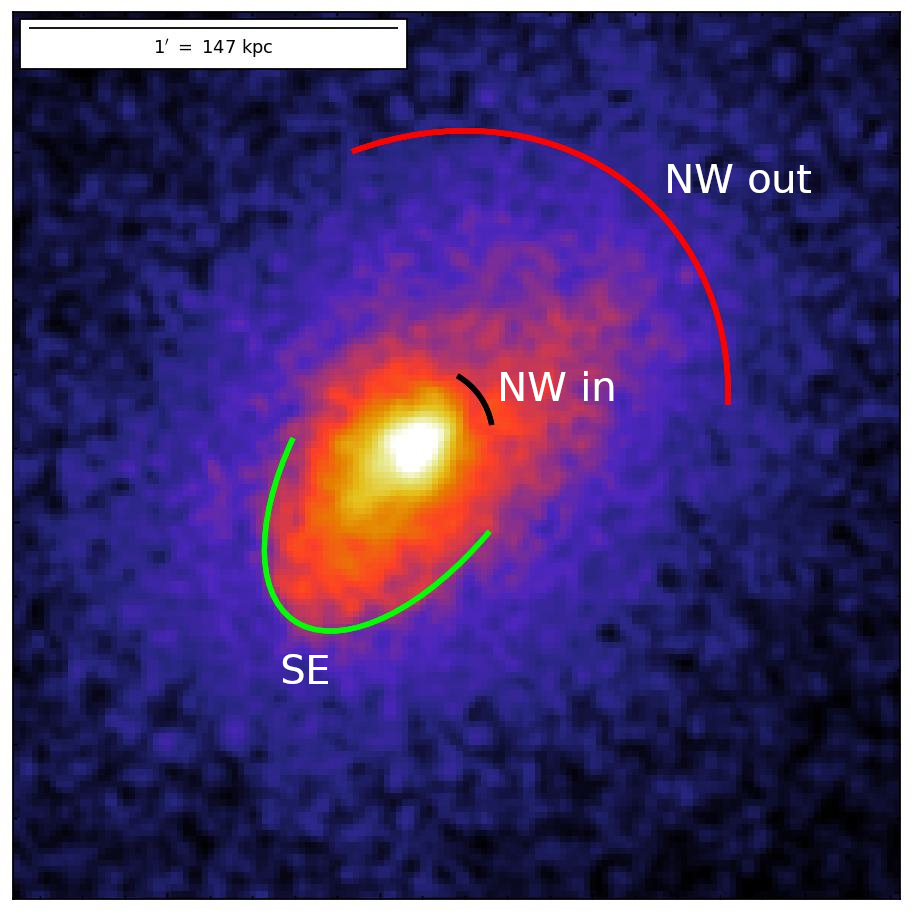} \qquad
\includegraphics[width=8.5cm]{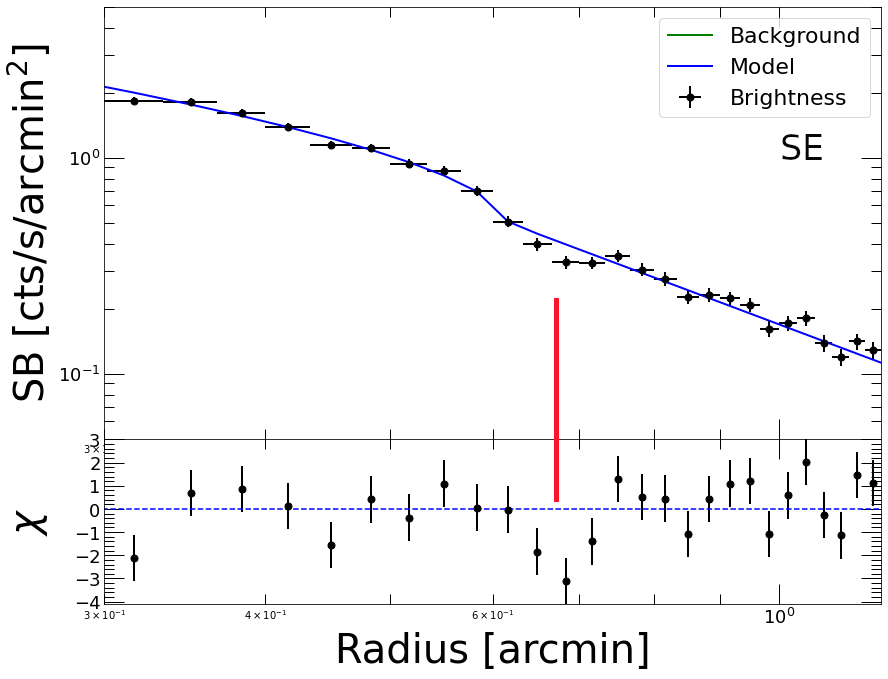}} \\
\subfloat{
\includegraphics[width=8.5cm]{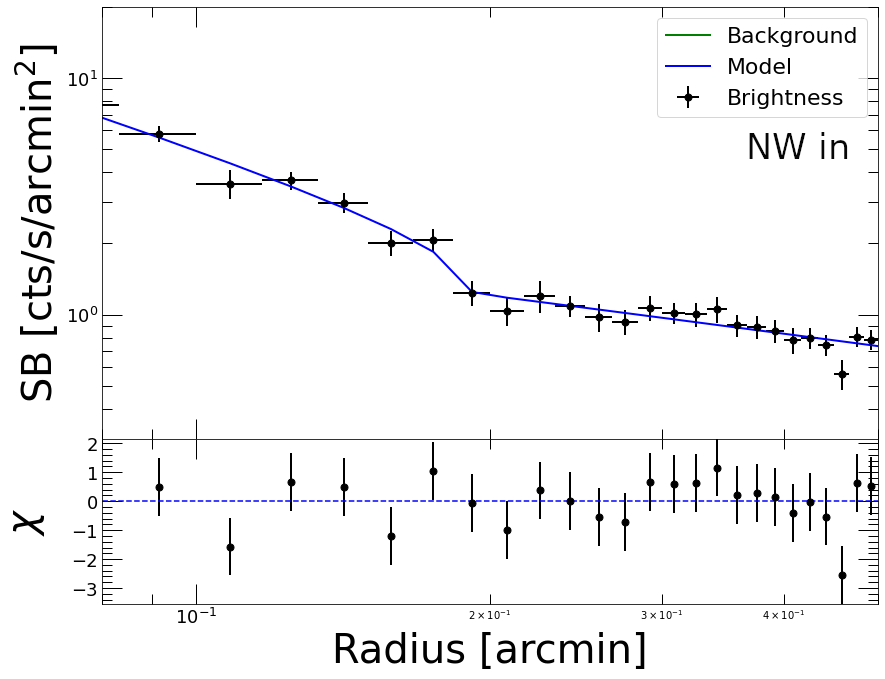}
\includegraphics[width=8.5cm]{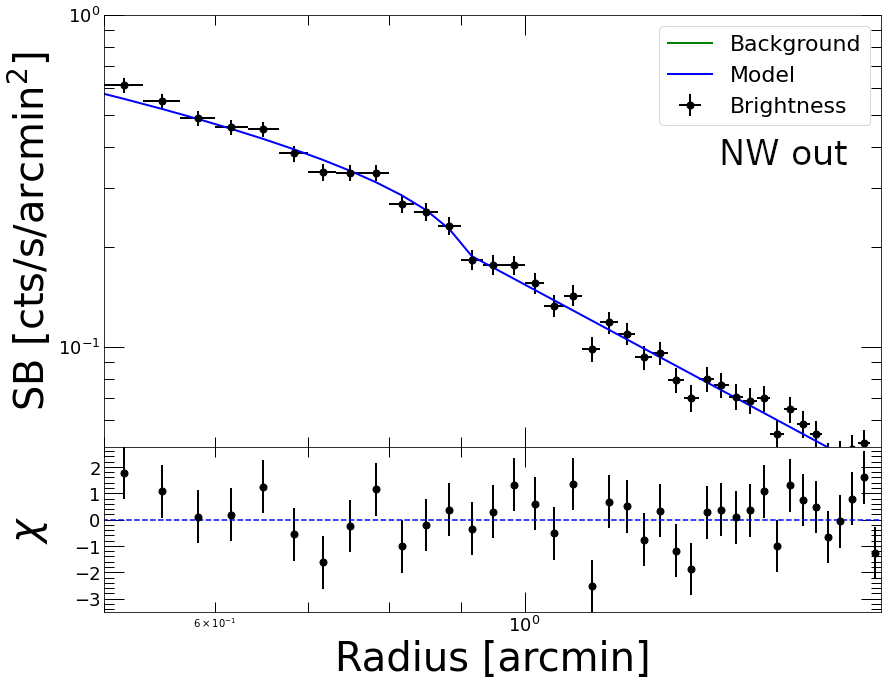}} \\
\subfloat{
\includegraphics[width=17cm]{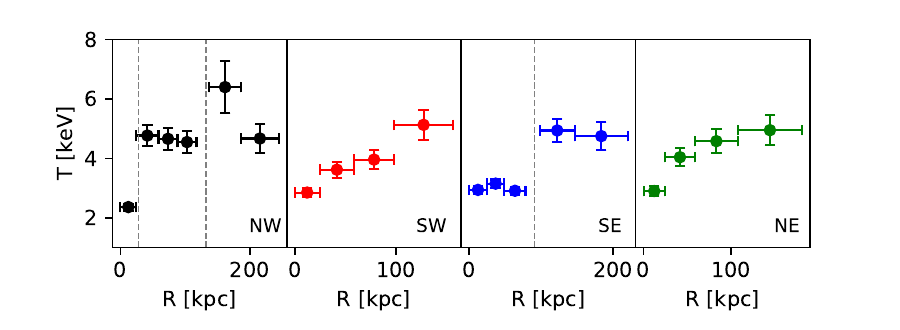}
}
\caption{A1068 cold fronts analysis. \emph{Upper left panel:} X-ray \emph{Chandra} image with indicated the positions of cold fronts: red arc for the NW outer CF, black arc for the NW inner CF and green arc for the SE CF. \emph{Upper right panel:} best-fitting broken power-law model (blue line) with associated residuals of the SB discontinuity in the south-east profile. The red line indicates the position of the putative PDL. \emph{Middle left panel:} same as before for the inner SB discontinuity in the north-west profile. \emph{Middle right panel:} same as before for the outer SB discontinuity in the north-west profile. \emph{Lower panels:} Temperature profiles across the north-west (black), south-west (red), south-east (blue) and north-east (green) sectors. The dashed grey lines indicate the position of cold fronts.}
\label{fig:A1068_CF}
\end{figure*}

We investigated this feature by extracting a radial profile in the south-east direction, using elliptical annuli optimised to follow the X-ray morphology, with a bin size of $2\arcsec$. The profile, reported in the upper right panel of Fig. \ref{fig:A1068_CF}, shows a clear discontinuity, which is well fitted ($\chi^2_{\rm red}=1.3$) by a broken power-law (in blue) with a jump in projected density of $1.34\pm0.04$ at a distance of 90 kpc. We indicate this feature with a green arc in the upper left panel of Fig. \ref{fig:A1068_CF}.

We also note a depression in the surface brightness profile in the proximity of the surface brightness discontinuity, at a distance of 0.6 - 0.8\arcmin{} from the cluster centre (indicated with a red line in the SE profile of Fig. \ref{fig:A1068_CF}), which appears as a dip channel in the residuals of the broken power-law fit. 
This is possible evidence of a Plasma Depletion Layer (PDL), a feature observed in simulations when the magnetic field in the ICM is stretched and amplified by a plasma flow, as in the proximity of cold fronts \citep[][]{Markevitch2007,ZuHone2011,Markevitch2018}. The magnetic pressure then rises and the gas is squeezed out of the region to keep pressure equilibrium. If enough gas is squeezed out of the region, a deficit of X-ray emission is visible, as in the case of A2142 \citep[][]{Markevitch2018}. An X-ray deficit is visible in the X-ray image of A1068, indicated with green arrows, and in the residual image (left and right panels of Fig. \ref{fig:A1068_Xmaps}, respectively).
Deeper \emph{Chandra} observations (PI Markevitch), that are in progress, will provide a more accurate profile of that feature.

We also searched for cold fronts in other sectors, using elliptical annuli adjusted to follow the X-ray morphology. 
We found two discontinuities in the north-west direction. The inner one (black arc in Fig. \ref{fig:A1068_CF}), modelled with an elliptical sector with angular aperture between 20-60 deg, is located at a distance of 28 kpc from the cluster centre and has a jump in surface brightness of 1.87, while the outer one (red arc in Fig. \ref{fig:A1068_CF}) is highlighted with a larger angular aperture between 10-100 deg and has a jump in the projected density of $1.26\pm0.04$ at a distance of 130 kpc from the cluster centre. The best-fitting broken power-law models and associated residuals of these discontinuities are represented in the central panels of Fig. \ref{fig:A1068_CF} (inner NW cold front on the left and outer NW cold front on the right).


To constrain the nature of the detected surface brightness discontinuities, we computed temperature profiles along the two directions containing discontinuities and the perpendicular ones.
The temperature is obtained by the fit of spectra extracted in concentric regions with a size adapted to contain a minimum of 2500 counts in the 0.7-10 keV band, after background subtraction. The results are shown in the bottom panels of Fig. \ref{fig:A1068_CF}. In all directions there is an increase in temperature at large radii, with a net jump in the sectors containing discontinuities, while the trend is smoother in the perpendicular sectors. In particular, in the north-west sector, there is a double jump in the temperature profile at radii corresponding to the surface brightness discontinuities, characterised by $ kT_{\rm in} = 2.36 \pm 0.12\ \rm{keV}$ and $ kT_{\rm out} = 4.77 \pm 0.35\ \rm{keV}$ for the inner jump and $ kT_{\rm in} = 4.6 \pm 0.4\ \rm{keV}$ and $ kT_{\rm out} = 6.4 \pm 0.9\ \rm{keV}$ for the outer jump; while in the south-east sector $ kT_{\rm in} = 2.9 \pm 0.1\ \rm{keV}$ and $ kT_{\rm out} = 4.9 \pm 0.4\ \rm{keV}$.
The observed discontinuities are therefore cold fronts, as the gas in the inner regions has a colder temperature.


\begin{figure*}
\centering
\includegraphics[width=17cm]{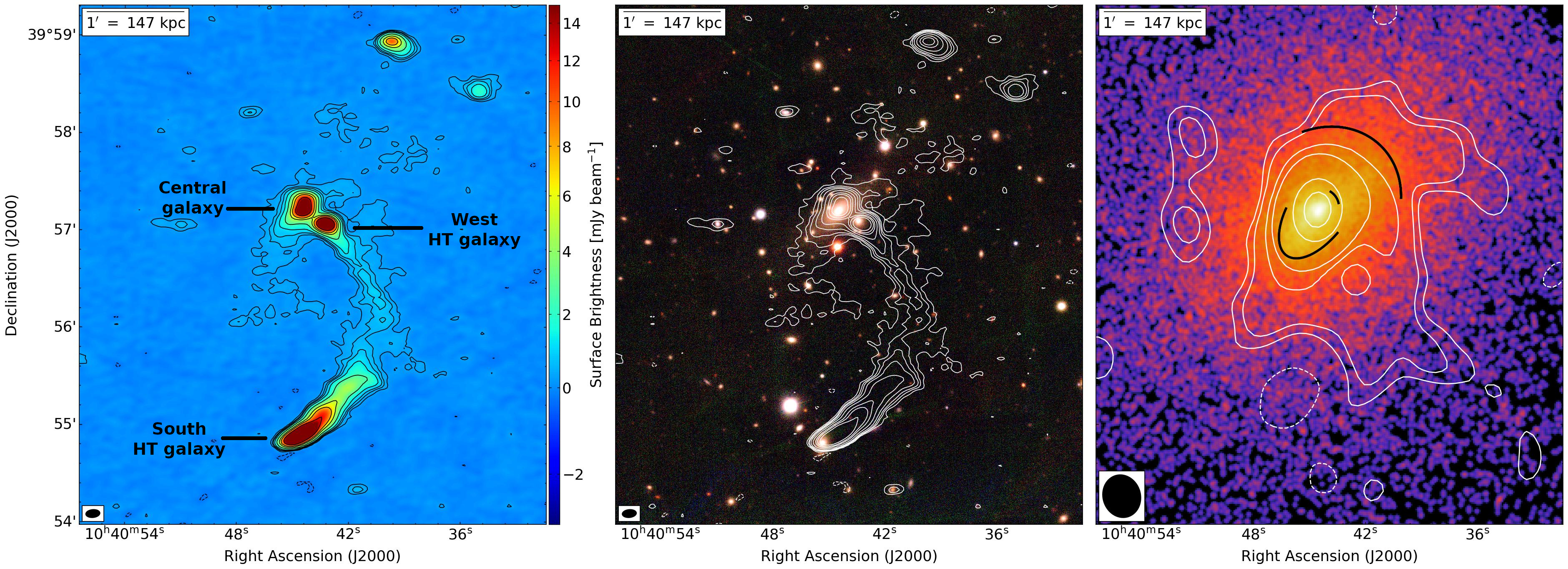}
\caption{Multi-wavelength images of A1068. \emph{Left panel}: High-resolution 144 MHz LOFAR image. Levels:[-3, 3, 6, 9, 15, 30, 60, 90, 150]$\times\sigma$ (where $\sigma=0.08$ mJy $\rm{beam}^{-1}$). The beam is $8.8\arcsec\times5.0\arcsec$ and is shown in the bottom left corner of the image.  \emph{Central panel}: Optical \emph{Pan-STARRS} RGB image with high-resolution LOFAR contours overlaid. \emph{Right panel}: \emph{Chandra} X-ray image smoothed on a scale of $1\arcsec$ with overlaid low-resolution source-subtracted LOFAR contours (Levels=[-3, 3, 5, 12, 20, 90, 150]$\times\sigma$, where $\sigma=0.15$ mJy $\rm{beam}^{-1}$ and the beam is $32.2\arcsec\times25.3\arcsec$). Black arcs indicate the position of cold fronts. All the panels depict the same region of the sky. }
\label{fig:A1068}
\end{figure*}

From a non-thermal point of view, the source was observed in the radio band with VLA at 1.4 GHz \citep{Govoni2009}, detecting hints of diffuse emission surrounding the central bright source and its companion.
In our high-resolution LOFAR image at 144\,MHz (left panel of Fig. \ref{fig:A1068}), we can clearly distinguish the central galaxy and the companion one, west of the centre, which presents a long tail. Another head-tail galaxy (z = 0.134, a likely cluster member) is located to the south.
The central galaxy is surrounded by diffuse emission, which extends in the north-west--south-east direction. 
\cite{Botteon2022a}, studying all the \textit{Planck} clusters in the LoTSS-DR2 region, catalogued the diffuse emission in A1068 as a radio halo, indicating in this category all clusters with central diffuse emission co-spatial with the ICM, without making a distinction between MH and RH. The source subtraction, necessary to separate diffuse emission from contamination of other sources and clearly classify it, is critical. Here we improved the subtraction of the discrete sources, which was performed in two steps: we subtracted the central compact source and the companion one applying an inner $uv$-cut of $10k\lambda\cong50$kpc; we then subtracted the contribution of the two tails.
At lower resolution, after the source-subtraction, the diffuse emission is elongated in the north-west south-east direction, following the X-ray emission, and extends beyond the cold fronts, with an overall size of $D_{\rm max}\sim400$ kpc (right panel of Fig. \ref{fig:A1068}).
The radio emission on the south, misaligned from the X-ray emission, is instead a residual of the subtraction of the west head-tail galaxy.


\subsection{Clusters with a mini-halo or a candidate mini-halo}

\begin{figure*}
\centering
\includegraphics[width=17cm]{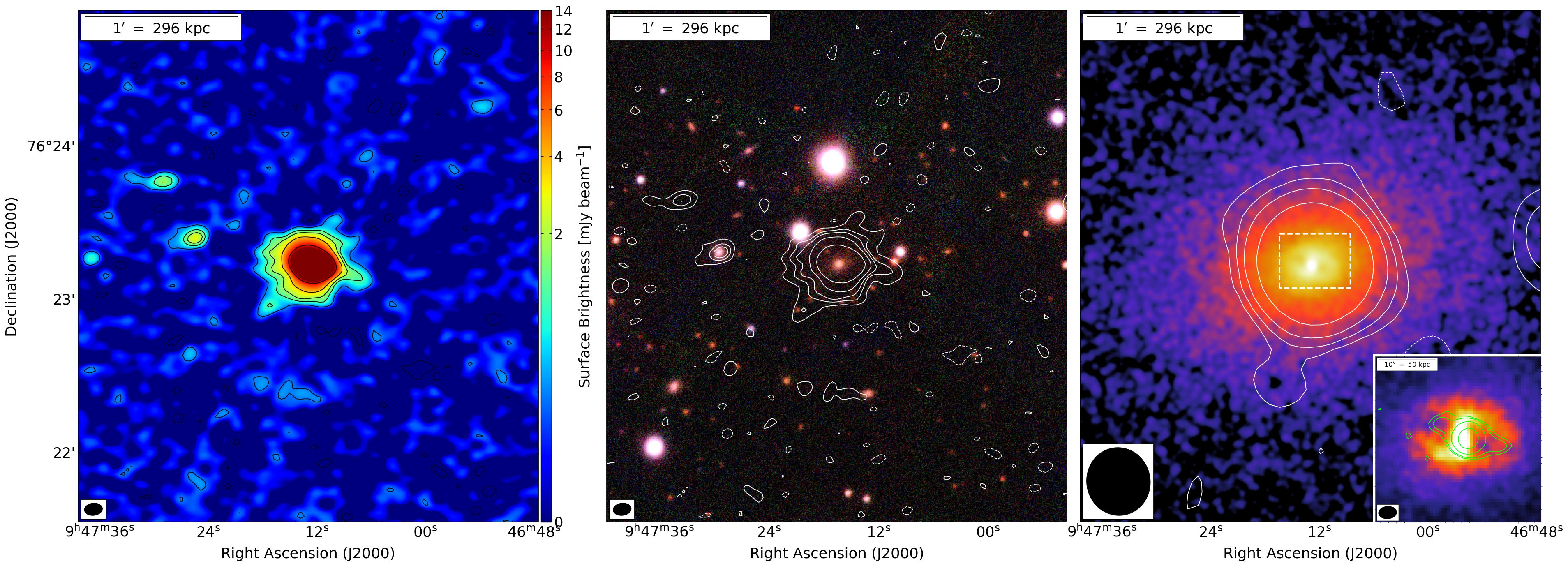}
\caption{Multi-wavelength images of RBS797. \emph{Left panel}: High-resolution 144 MHz LOFAR image. Levels:[-3, 3, 9, 15, 30, 80]$\times\sigma$ (where $\sigma=0.12$ mJy $\rm{beam}^{-1}$). The beam is $7.0\arcsec\times4.7\arcsec$ and is shown in the bottom left corner of the image.  
\emph{Central panel}: Optical \emph{Pan-STARRS} RGB image with high-resolution LOFAR contours overlaid. \emph{Right panel}: \emph{Chandra} X-ray image smoothed on a scale of $1\arcsec$ with overlaid low-resolution source-subtracted LOFAR contours (Levels=[-3, 3, 7, 15, 40]$\times\sigma$, where $\sigma=0.33$ mJy $\rm{beam}^{-1}$ and the beam is $26.5\arcsec\times24.8\arcsec$). The insert box shows a zoom of the X-ray image with overlaid LOFAR contours at $3.2\arcsec\times2.2\arcsec$ resolution, obtained with natural weighting and inner $uv$-cut. All the panels depict the same region of the sky.}
\label{fig:RBS797}
\end{figure*}

\subsubsection{RBS 0797}
The cluster RBS 0797, at z=0.35, was studied by \cite{Gitti2006} and \cite{Doria2012} who performed VLA radio observations at different frequencies (1.4$-$8.4 GHz) and resolutions (0.4$-$6.3 arcsec). They found radio emission at different scales: a MH with an extension of $D_{\rm max}\sim 220$ kpc, extended emission on the north-east south-west direction filling X-ray cavities located at a distance of $r\sim24$ kpc from the cluster centre \citep{Schindler2001}, and inner jets on scales of $r\sim13$ kpc pointing north-south, roughly perpendicular to the X-ray cavities. 
\cite{Ubertosi2021} recently detected a second pair of X-ray cavities in the north-south direction, at roughly the same radial distance as the north-east--south-west cavities.
Deeper \emph{Chandra} observations revealed the presence of three pairs of shock fronts at different distances from the cluster centre, presumably generated by different phases of AGN activity \citep{Ubertosi2023}.
With LOFAR (Fig. \ref{fig:RBS797}), we detected the central MH with an extension similar to that observed at higher frequency. At lower resolution, no further diffuse emission is recovered.
To search for emission from the lobes, we re-imaged the LOFAR data cutting the baselines shorter than $10\ k\lambda\cong100$kpc to filter out the MH emission. We achieved a resolution of $3.2\arcsec\times2.2\arcsec$, which allows us to detect the north-east south-west radio lobes (insert box in the bottom right panel of Fig. \ref{fig:RBS797}). The low-resolution radio contours presented in the right panel of Fig. \ref{fig:RBS797}, are obtained after the subtraction of the model obtained with an inner $uv$-cut of $10k\lambda\cong100$ kpc. A detailed analysis of LOFAR data of this source is reported in a dedicated paper \citep{Bonafede2023_RBS797}.

\begin{figure*}
\centering
\subfloat{
\includegraphics[width=10.5cm]{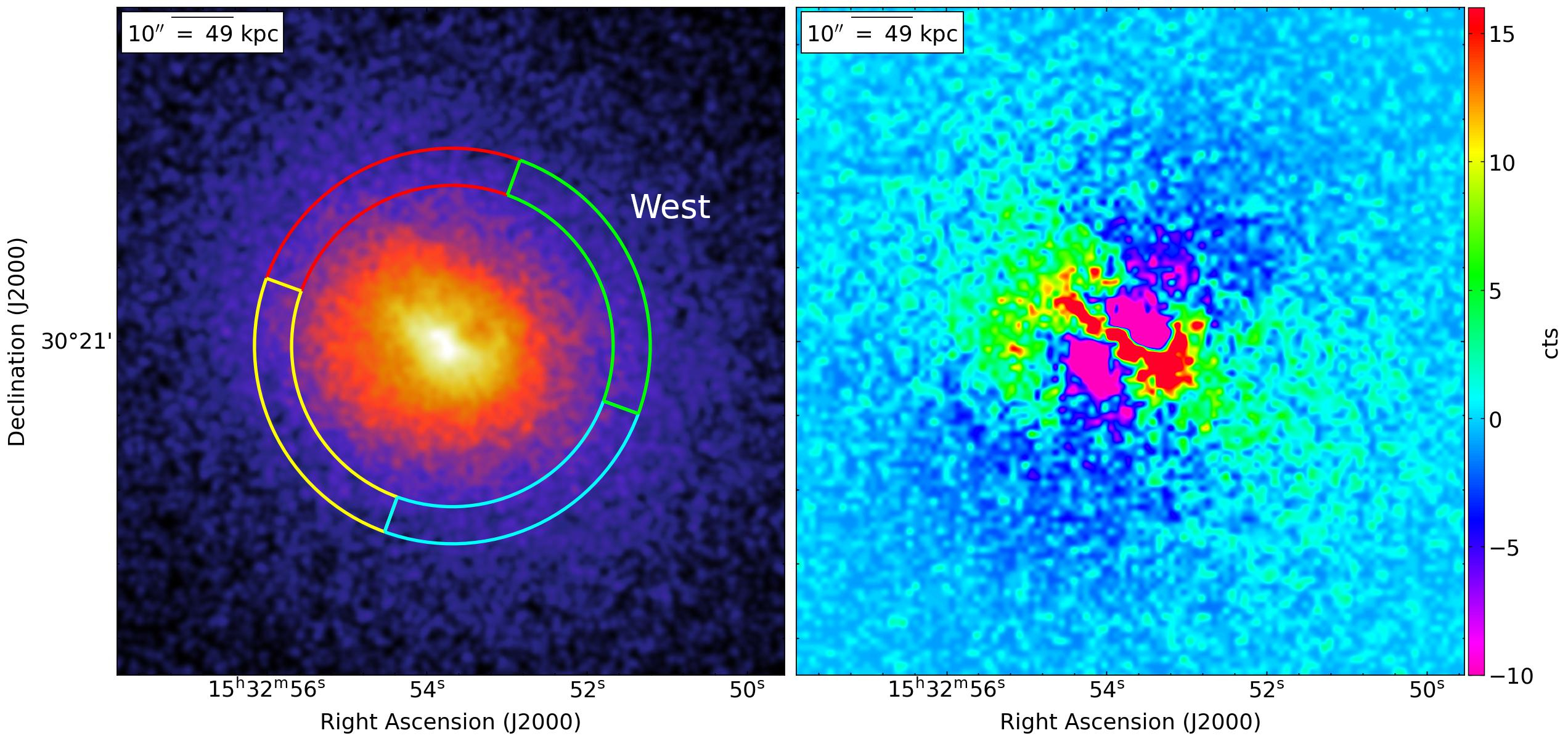}
\includegraphics[width=6.5cm]{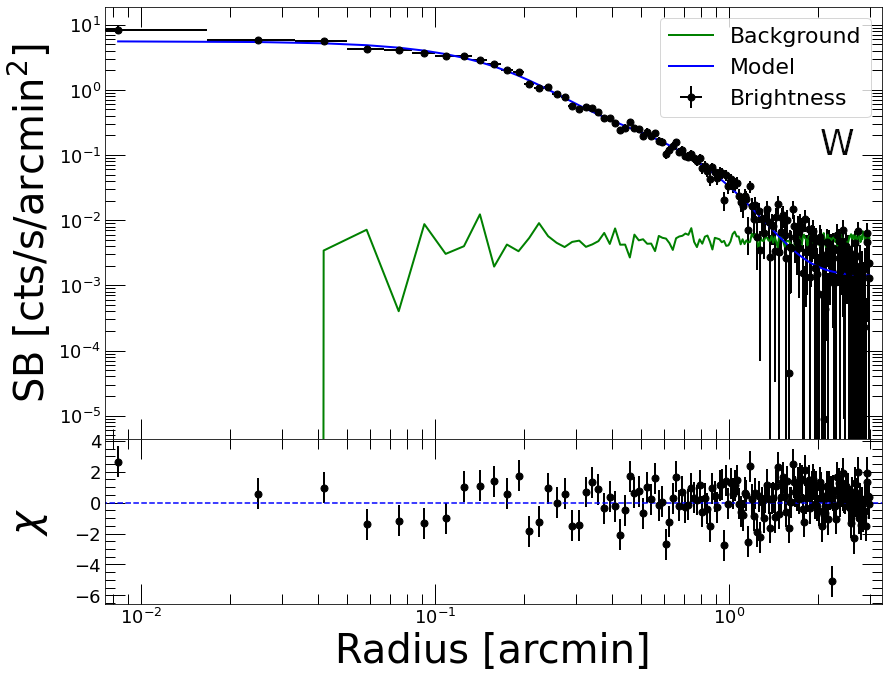}}
\caption{X-ray analysis of RX J1532.9+3021. \emph{Left panel:} \emph{Chandra} 0.5–7.0 keV image with a portion of the extraction regions superimposed. \emph{Central panel:} Residual image obtained subtracting a double $\beta$-model. \emph{Right panel:} Radial surface brightness profile extracted along the West sector and fitted with a double $\beta$-model (blue line), with associated residuals.}
\label{fig:RXJ1532_Xmaps}
\end{figure*}

\begin{figure*}
\centering
\includegraphics[width=17cm]{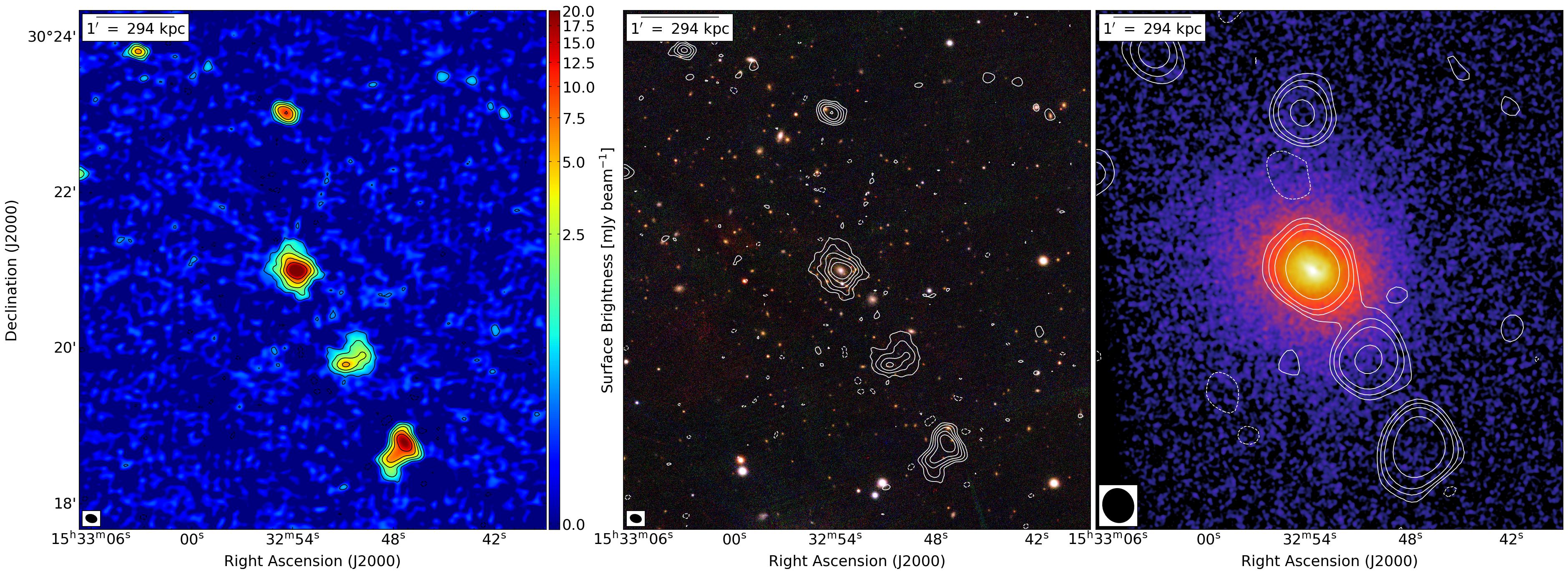}
\caption{Multi-wavelength images of RX J1532.9+3021. \emph{Left panel}: High-resolution 144 MHz LOFAR image. Levels:[-3, 3, 9, 20, 40,80]$\times\sigma$ (where $\sigma=0.13$ mJy $\rm{beam}^{-1}$). The beam is $8.8\arcsec\times6.2\arcsec$ and is shown in the bottom left corner of the image. 
\emph{Central panel}: Optical \emph{Pan-STARRS} RGB image with high-resolution LOFAR contours overlaid. \emph{Right panel}: \emph{Chandra} X-ray image with overlaid low-resolution LOFAR contours after the subtraction of the central source (Levels=[-3, 3, 7, 15, 50,100]$\times\sigma$, where $\sigma=0.24$ mJy $\rm{beam}^{-1}$ and the beam is $26.6\arcsec\times23.9\arcsec$). All the panels depict the same region of the sky.}
\label{fig:RXJ1532}
\end{figure*}

\subsubsection{RX J1532.9+3021}
RX J1532.9+3021 (hereafter RXJ1532) is a luminous X-ray cool-core cluster at z=0.362, which has two X-ray cavities detected west and east of the core \citep{Hlavacek2013} (Fig. \ref{fig:RXJ1532_Xmaps}, left panel). \cite{Hlavacek2013} reported the presence of a cold front 65 kpc $\equiv$ 0.2\arcmin{} west of the cluster centre, identifying by eye a break in the averaged surface brightness profile and a jump in the projected temperature profile in the West sector. However, the proximity of this feature to the edge of the western cavity complicates the detection. 
Re-analysing \emph{Chandra} X-ray data we were not able to confirm this feature. We inspected radial profiles using the sectors reported in the literature \citep{Hlavacek2013} and represented in the left panel of Fig. \ref{fig:RXJ1532_Xmaps}. We have not found statistical evidence of a jump in surface brightness in the West sector (Fig. \ref{fig:RXJ1532_Xmaps}, right panel), neither in the other sectors (see Appendix \ref{sec:appendix}).

\begin{figure*}
\centering
\includegraphics[width=17cm]{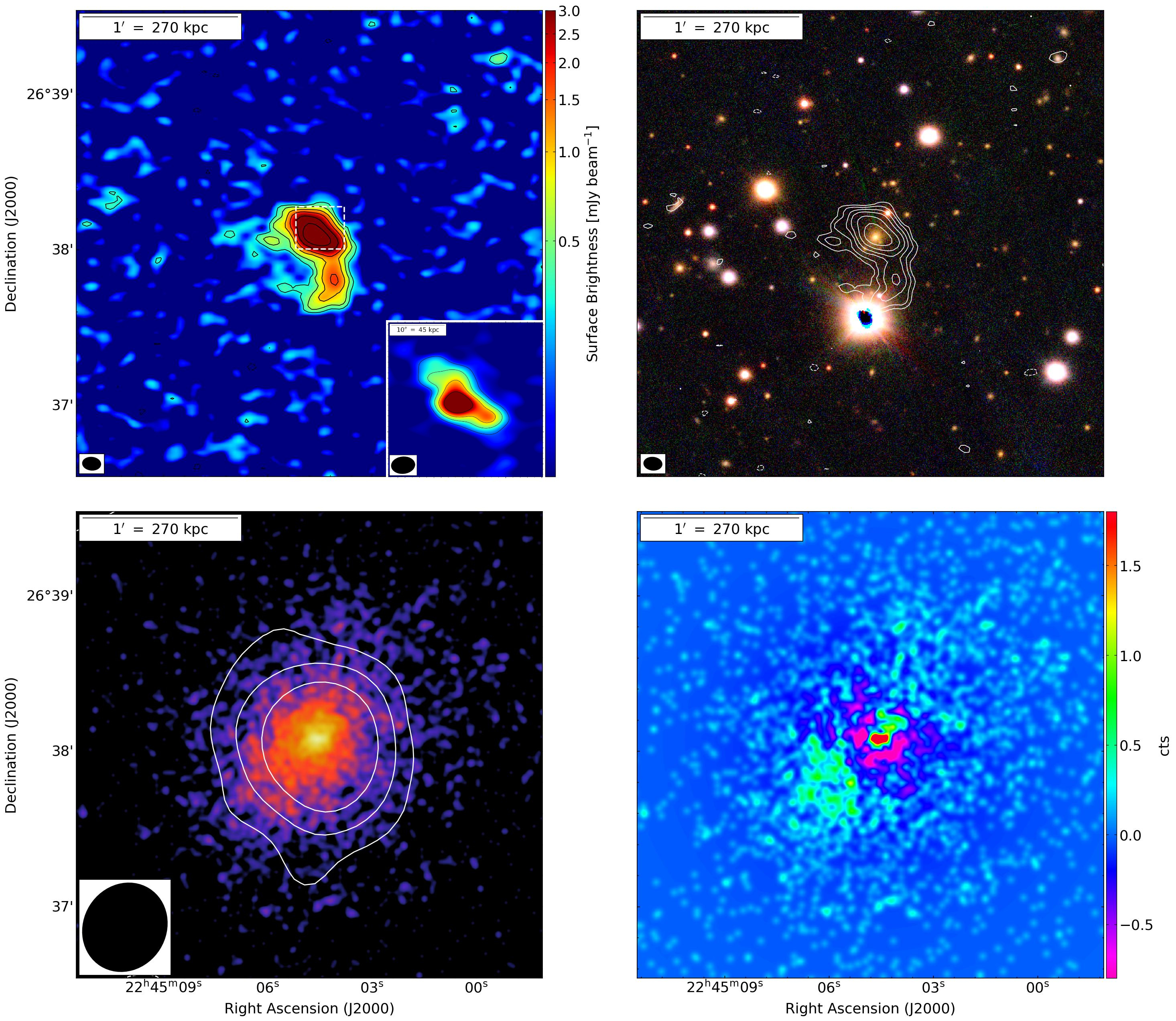}
\caption{Multi-wavelength images of MACS J2245.0+2637. \emph{Upper left panel}: High-resolution 144 MHz LOFAR image. Levels:[-3, 3, 5, 9, 15, 30, 60]$\times\sigma$ (where $\sigma=0.11$ mJy $\rm{beam}^{-1}$). The beam is $7.0\arcsec\times5.0\arcsec$ and is shown in the bottom left corner of the image. The insert box shows the image of the central source at $4.5\times3.2\arcsec$ resolution obtained with Briggs weighting robust = $-1.25$ and inner $uv$-cut. 
\emph{Upper right panel}: Optical \emph{Pan-STARRS} RGB image with high-resolution LOFAR contours overlaid. \emph{Lower left panel}: \emph{Chandra} X-ray image smoothed on a scale of $3\arcsec$ with overlaid low-resolution source-subtracted LOFAR contours (Levels=[-3, 3, 9, 20, 60]$\times\sigma$, where $\sigma=0.32$ mJy $\rm{beam}^{-1}$ and the beam is $35.1\arcsec\times31.7\arcsec$). \emph{Lower right panel}: Residual X-ray image after the subtraction of a circular double $\beta$-model. All the panels depict the same region of the sky.}
\label{fig:MACSJ2245}
\end{figure*}

In the radio band, the cluster was observed with the VLA and GMRT telescopes at 325 MHz, 610 MHz and 1.4 GHz, revealing a prominent MH in the cluster core, with a total extension of $D_{\rm max}\sim100$ kpc, surrounding the BCG \citep{Hlavacek2013,Giacintucci2014b}. The MH has a total spectral index of $\alpha = 1.20\pm0.07$. 
A second radio galaxy is observed at  $\sim2.6^{\prime}$ ($\sim$ 800 kpc) projected distance south-west of the cluster centre. In the same position, a cluster has been detected \citep[GMBCG J233.19725+30.31626;][]{Hao2010}, located at z=0.358, hence compatible with the redshift of RXJ1532.
Our LOFAR image (Fig. \ref{fig:RXJ1532}, left panel) confirms the presence of a central MH with a total extension of $D_{\rm max}\sim 240$ kpc. No more diffuse emission is detected at larger scales. The low-resolution radio contours are obtained after the subtraction of the central compact source, which model has been obtained performing an inner $uv$-cut at $10k\lambda\cong100$ kpc, and applying a taper of 20\arcsec on the residual visibilities. Recently, \cite{Timmerman2022} has imaged the LOFAR data of this source obtained with international baselines at sub-arcsecond resolution, detecting only a faint compact radio component.

\subsubsection{MACS J2245.0+2637}
The cluster MACS J2245.0+2637 (hereafter MACSJ2245), at z=0.304, presents a pronounced cool-core \citep{Ebeling2010} and a small potential X-ray cavity \citep{Hlavacek2012}. The thermal emission is elongated towards the south-east and there is an excess of emission in that direction in the residual map (Fig. \ref{fig:MACSJ2245}, lower panels), similar to the morphology of the cluster PSZ1G139 (see Fig. \ref{fig:PSZ1G139}), suggesting that also this system is not completely relaxed. 
However, the short exposure time of \emph{Chandra} archival observations prevents us from performing a more detailed analysis of this intermediate-redshift cluster.
\cite{Giacintucci2017} reported the absence of diffuse radio emission in this cluster at 610 MHz. 
With LOFAR (Fig. \ref{fig:MACSJ2245}, upper left panel), we detected the central radio galaxy and radio emission extending towards the south of uncertain interpretation. At a higher resolution of $4.5\times3.2\arcsec$ (insert box in the upper left panel of Fig. \ref{fig:MACSJ2245}), we found the central source presents two putative radio lobes that are not aligned. A hint of diffuse emission with low surface brightness surrounding the central galaxy is noticed. However, the flux density of this component is not significant even at lower resolution, where the surface brightness sensitivity to diffuse emission is higher. The low-resolution radio contours reported in the lower left panel of Fig. \ref{fig:MACSJ2245} are residuals of the subtraction of the central source and inner lobes, which model is obtained applying an inner $uv$-cut of $5k\lambda\cong180$ kpc and then imaged with an $uv$-taper of 20\arcsec. We classify this source as a candidate MH. Deeper observations are needed to better characterise the radio emission in the cluster.


\subsection{Clusters without diffuse emission}

\begin{figure*}
\centering
\includegraphics[width=17cm]{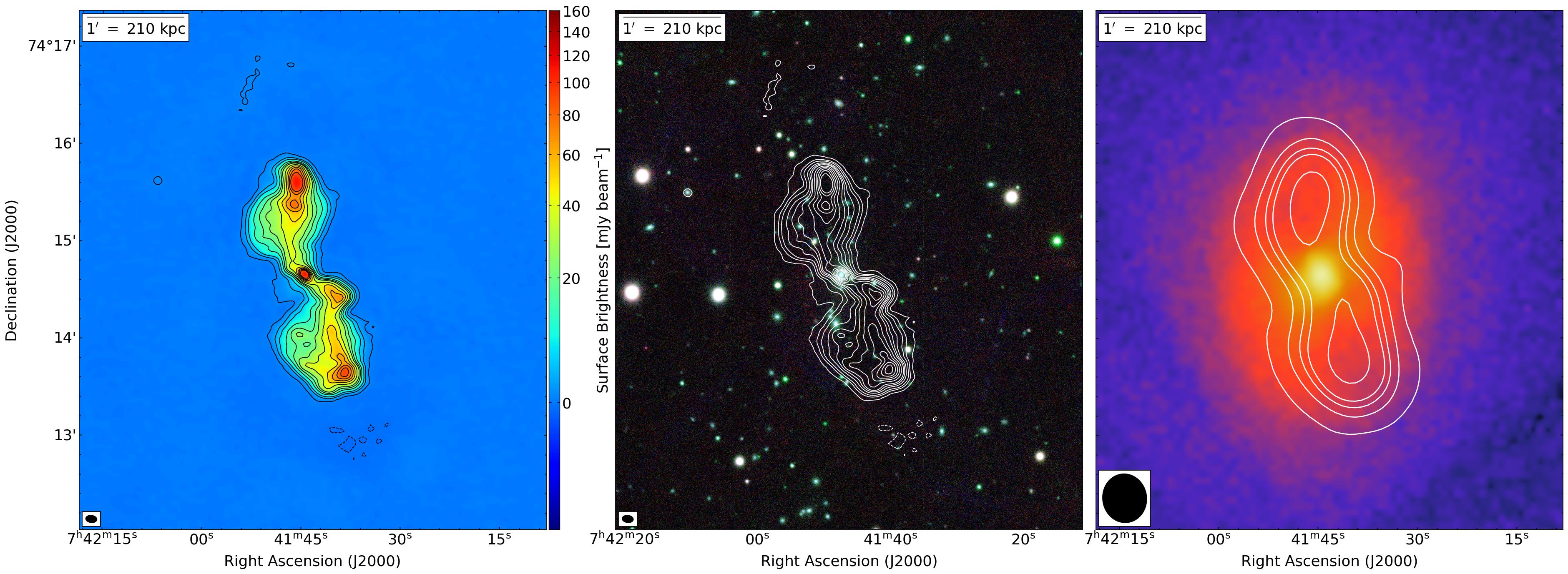}
\caption{Multi-wavelength images of MS 0735.6+7421. \emph{Left panel}: High-resolution 144 MHz LOFAR image. Levels:[-3, 3, 15, 30, 60, 90, 120, 180, 240, 300, 360, 420]$\times\sigma$ (where $\sigma=0.2$ mJy $\rm{beam}^{-1}$). The beam is $6.9\arcsec\times4.5\arcsec$ and is shown in the bottom left corner of the image. 
\emph{Central panel}: Optical \emph{Pan-STARRS} RGB image with high-resolution LOFAR contours overlaid. \emph{Right panel}: \emph{Chandra} X-ray image with low-resolution LOFAR contours overlaid (Levels=[-3, 3, 20, 40, 60, 120]$\times\sigma$, where $\sigma=5$ mJy $\rm{beam}^{-1}$ and the beam is $30.1\arcsec\times27.1\arcsec$). All the panels depict the same region of the sky.}
\label{fig:MS0735}
\end{figure*}

\begin{figure*}
\centering
\includegraphics[width=17cm]{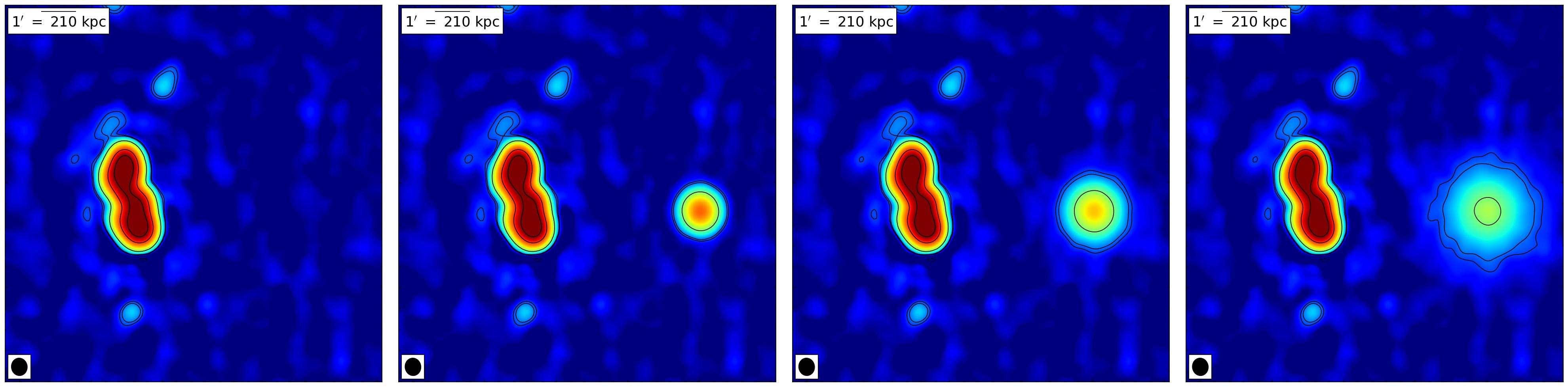}

\includegraphics[width=17cm]{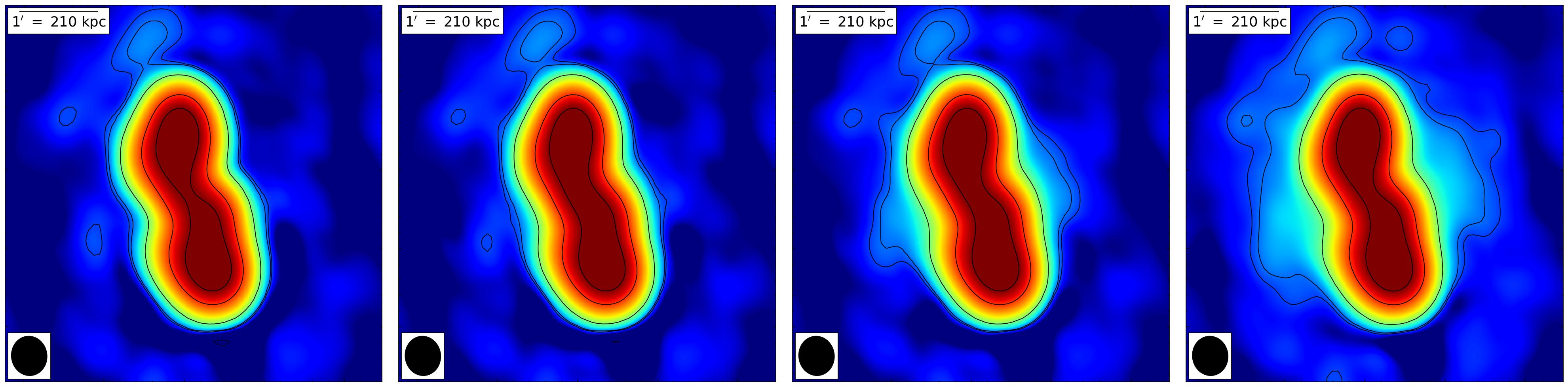}
\caption{Mini-halo injection in MS 0735.6+7421. \emph{Upper panels:} MH injection in a close-by region void of sources. \emph{Lower panels:} MH injection at cluster centre. \emph{First column:} LOFAR image before the injection. \emph{Second column:} Injection with $r_e$= 25 kpc. \emph{Third column:} Injection with $r_e$= 50 kpc. \emph{Last column:} Injection with $r_e$= 100 kpc.}
\label{fig:MS0735_inj}
\end{figure*}

\subsubsection{MS 0735.6+7421}
The cluster MS 0735.6+7421 (hereafter MS0735), at z=0.216, hosts in its centre a powerful radio galaxy with a well-defined core and radio lobes. 
The LOFAR data and multi-frequency analysis of this source are presented in \cite{Biava2021a}.  
The radio lobes at LOFAR frequency (Fig. \ref{fig:MS0735}) are wider than previously found with the Very Large Array (VLA) at higher frequencies \citep[][]{McNamara2005,Cohen2005} and now completely fill the X-ray outer cavities, each roughly 200 kpc in diameter \citep{McNamara2005}. Another pair of smaller cavities are located in the inner 20 kpc of the radio jet, indicating the central AGN experienced different phases of jet activity \citep[][]{Vantyghem2014}.
An intermediate phase of jet activity has been recently discovered and associated with the radio lobe located south-west of the core \citep{Biava2021a}. The position of this lobe corresponds to a new cavity discovered re-analysing \emph{Chandra} X-ray archival data \citep[][]{Biava2021a}. A shock front surrounds
the outer pair of cavities \citep{McNamara2005}.

\begin{figure*}
\centering
\includegraphics[width=17cm]{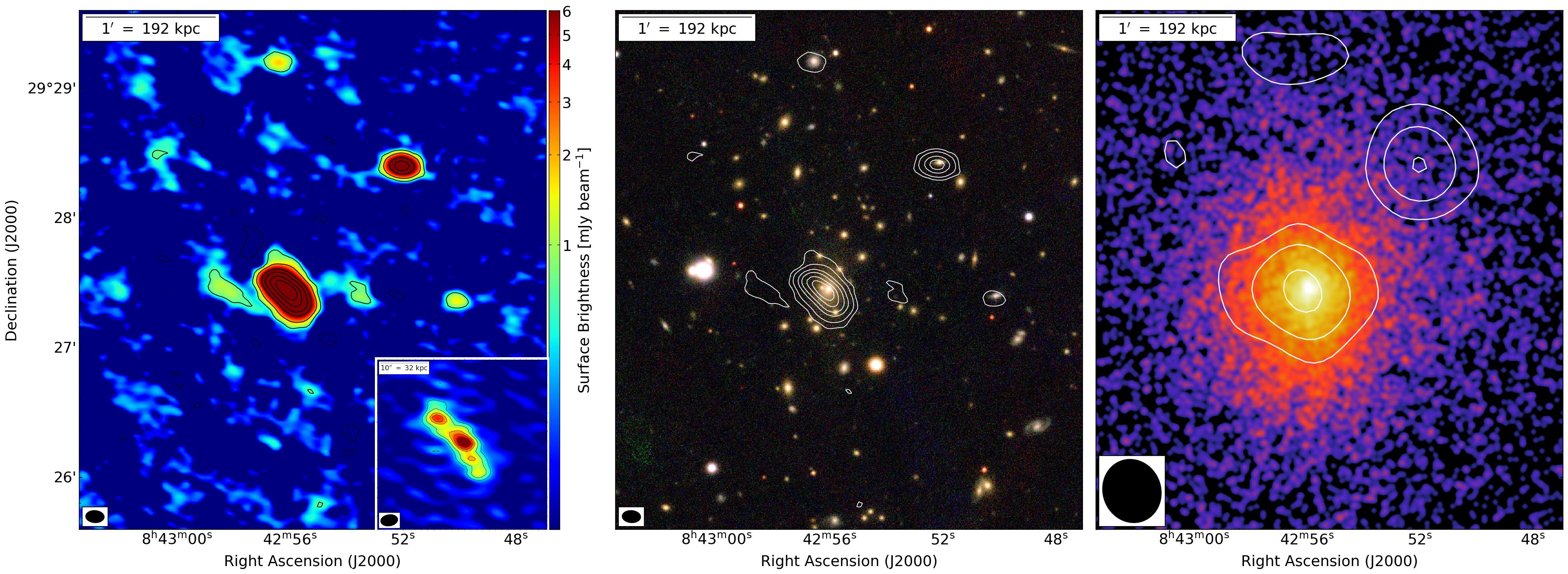}
\caption{Multi-wavelength images of MS 0839.8+2938. \emph{Left panel}: High-resolution 144 MHz LOFAR image. Levels:[-3, 3, 9, 30, 60, 120, 180]$\times\sigma$ (where $\sigma=0.22$ mJy $\rm{beam}^{-1}$). The beam is $8.5\arcsec\times5.6\arcsec$ and is shown in the bottom left corner of the image. The insert box shows the image of the central source at resolution of $3.7\arcsec\times2.5\arcsec$ obtained with uniform weightings. 
\emph{Central panel}: Optical \emph{Pan-STARRS} RGB image with high-resolution LOFAR contours overlaid. \emph{Right panel}: \emph{Chandra} X-ray image smoothed on a scale of $1\arcsec$ with overlaid low-resolution source-subtracted LOFAR contours (Levels=[-3, 3, 15, 40]$\times\sigma$, where $\sigma=0.55$ mJy $\rm{beam}^{-1}$ and the beam is $29.6\arcsec\times26.8\arcsec$). All the panels depict the same region of the sky.}
\label{fig:MS0839}
\end{figure*}

Recently, \cite{Begin2023} reported the possible detection of a MH in MS0735 using new VLA P-band (224–480 MHz) observations. They speculated that the detected excess of radio emission in VLA P-band array C image compared to VLA P-band array A image, could be associated with MH emission as an alternative to jet-related emission. As MH emission is typically characterised by a steep radio spectrum \citep[$\alpha\sim1$][]{Giacintucci2019}, it should be brighter at lower frequencies. However, with LOFAR we have not detected diffuse emission in the form of RH or MH in this cluster. 
As a further confirmation, we checked if the LOFAR sensitivity is enough to detect the excess of radio emission found by \cite{Begin2023}. To do so, we scaled the radio power associated with the MH at 144 MHz, considering a typical spectral index of $\alpha=1$, finding a power of $P_{\rm excess,\ 144\ MHz}=(3.9\pm2)\times10^{25}\ \rm{W\ Hz^{-1}}$. 
We then injected a mock halo in the \emph{uv}-plane of the cluster, modelled according to an exponential profile $I=I_{0} e^{-r/r_e}$, where $I_{0}=S/2\pi r_e^2$ (see Section \ref{sec:UL} for more details on this technique). We derived the flux density ($S=260$ mJy) from the expected MH power at 144 MHz, while for the $e$-folding radius we used a range of values, as no firm correlations with physical quantities of the cluster have been reported in literature  \citep{Bruno2023a}. Taking as reference a MH sample studied by \cite{Murgia2009}, the $e$-folding radius of MHs ranges between 25-100 kpc, so we decided to inject halos with $r_e=25,50,100$ kpc. At first, we injected the halo in a region free of point sources, to verify that it is detectable with $D_{2\sigma}\ge2\times r_e$. The detection criteria are satisfied for all the $e$-folding radii considered (Fig. \ref{fig:MS0735_inj}, upper panels). Then we injected the halo at the cluster centre (Fig. \ref{fig:MS0735_inj}, lower panels). We note that a halo with $r_e=25$ kpc is not distinguishable by eye from the lobes emission, but there is an increase of flux density in the region of injection corresponding to the value injected. Increasing the $e$-folding radius, the MH emission is spread on a larger volume, becoming visible. Therefore, a MH with the expected power derived by \cite{Begin2023} could certainly be detectable at LOFAR sensitivity. However, we have not found evidence of diffuse emission in our high-resolution map and even after decreasing the resolution and using an $uv$-taper to improve the recovery of diffuse emission, the flux density in the $3\sigma$ region remains the same. In conclusion, is more likely that the excess of radio emission detected by \cite{Begin2023} is associated with the wider outer radio lobes we detected with LOFAR \citep{Biava2021a}, rather than a MH.

\subsubsection{MS 0839.8+2938}
The cluster MS 0839.8+2938 (hereafter MS0839), at z=0.193, was observed with VLA at 1.4 GHz, detecting a central radio galaxy with small radio lobes, which extends up to $D_{\rm max}\sim80$ kpc \citep[][]{Giacintucci2017}.
A LOFAR image of this cluster observed with LoTSS was presented by \cite{Birzan2020}, confirming the presence of a central radio source with small radio lobes.
Our image (Fig. \ref{fig:MS0839}) is made by combining two LoTSS pointings and the data are calibrated with more advanced techniques, meaning we achieved an enhanced signal-to-noise ratio \citep[Fig. \ref{fig:MS0839}, $\sigma = 220\ \rm{\mu Jy\ beam^{-1}}$ at $8.5\arcsec\times5.6\arcsec$ resolution, against $\sigma = 331\ \rm{\mu Jy\ beam^{-1}}$ at $10.5\arcsec\times5.4\arcsec$ resolution in][]{Birzan2020}. The quality of the image, however, is still affected by the presence of a bright radio source in the north not visible in the image. In the higher resolution ($3.7\arcsec\times2.5\arcsec$) zoom box in the upper left panel, obtained with uniform $uv$-weighting, we can distinguish the central compact source and the associated radio lobes. 
At lower resolution and after the central source subtraction (the model of the central source with radio lobes obtained using uniform weighting), we detected some residual emission coincident with the cluster, but given the high level of contamination from the radio lobes and the strong calibration artefacts, we do not consider the emission as a reliable detection of diffuse radio emission form the ICM. Further investigation with higher-quality data is necessary.

\begin{figure*}
\centering
\includegraphics[width=17cm]{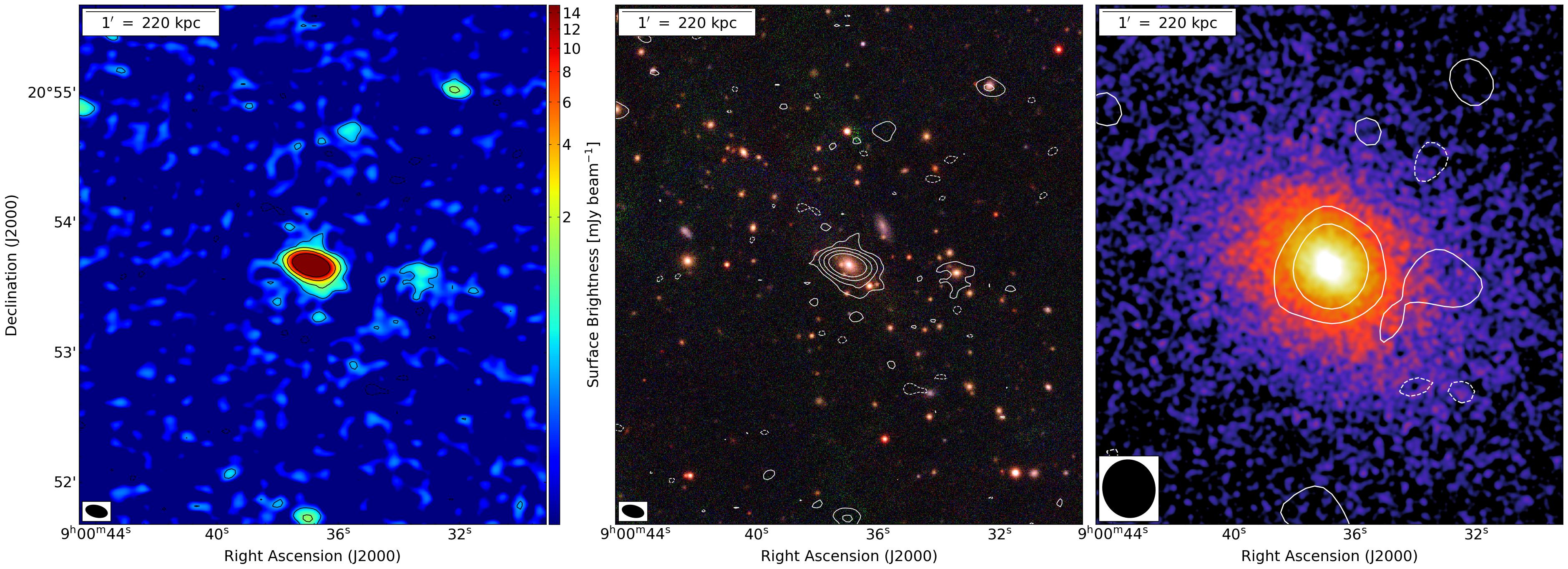}
\caption{Multi-wavelength images of Z2089. \emph{Left panel}: High-resolution 144 MHz LOFAR image. Levels:[-3, 3, 9, 30, 90]$\times\sigma$ (where $\sigma=0.14$ mJy $\rm{beam}^{-1}$). The beam is $10.2\arcsec\times5.5\arcsec$ and is shown in the bottom left corner of the image. 
\emph{Central panel}: Optical \emph{Pan-STARRS} RGB image with high-resolution LOFAR contours overlaid. \emph{Right panel}: \emph{Chandra} X-ray image smoothed on a scale of $1\arcsec$ with overlaid low-resolution source-subtracted LOFAR contours (Levels=[-3, 3, 9]$\times\sigma$, where $\sigma=0.28$ mJy $\rm{beam}^{-1}$ and the beam is $26.8\arcsec\times24.2\arcsec$). All the panels depict the same region of the sky.}
\label{fig:Z2089}
\end{figure*}

\subsubsection{Z2089}
The cluster Z2089, at z=0.235, was observed with the GMRT at 610 MHz, and no diffuse radio emission was detected \citep[][]{Venturi2008}. 
The central source is detected by the FIRST survey \citep{Becker1995}, with an integrated flux density of $S_{\rm 1.4~GHz}= 8.6$ mJy.
In our LOFAR observations (Fig. \ref{fig:Z2089}), we detected a central discrete source with a small extension ($D_{\rm max}\sim 100$ kpc) in the north-south direction at modest signal-to-noise ratio.
The radio emission associated with the compact central source has a flux density of $S_{\rm 144~MHz}=95$ mJy.
Therefore it has a steep spectral index of $\alpha\sim1$ between 144$-$1400 MHz. This relatively steep spectrum suggests the central source is not resolved in the LOFAR image and the flux density could be dominated by steep radio lobes.
At lower resolution, no more emission is collected. The subtraction of the central point source, modelled with uniform weighting, has left some residuals that are enhanced at lower resolution \citep[see][for a discussion on this effect]{Bruno2023a}. The small emission around the central source could be then associated with the central AGN. Further investigation is necessary to confirm this case.

\begin{figure*}
\centering
\includegraphics[width=17cm]{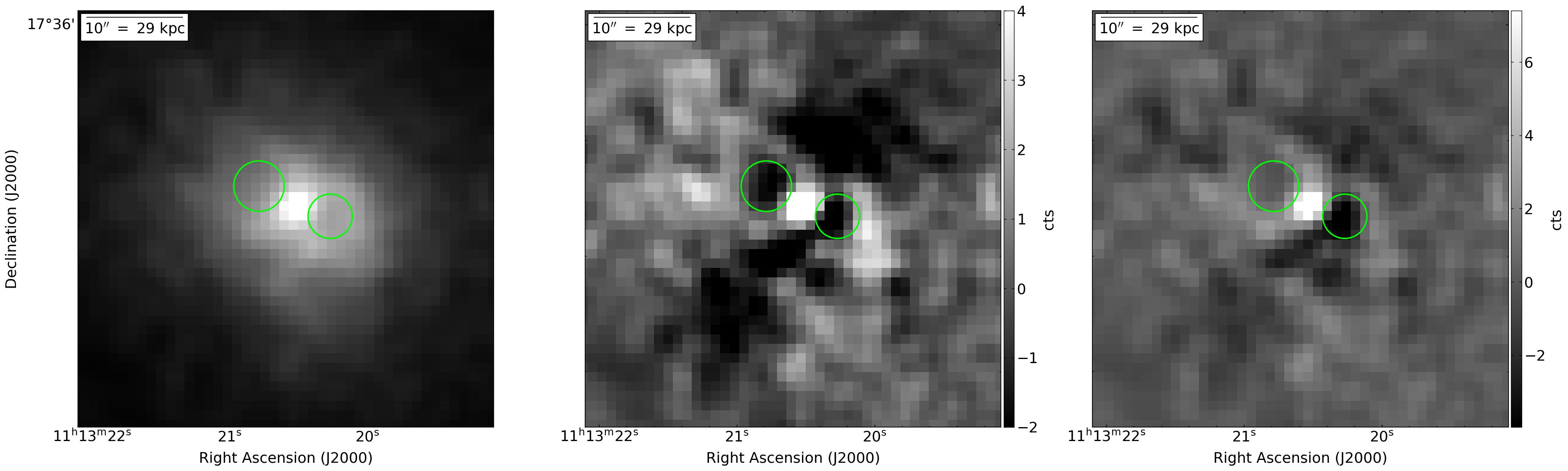}
\caption{A1204 cavity detection. \emph{Left panel:} \emph{Chandra} X-ray image smoothed on a scale of $1\arcsec$. \emph{Central panel:} Residual image after the subtraction of a circular $\beta$-model. (\emph{Right panel:}) Residual image after the subtraction of a circular double $\beta$-model. In all the panels, green circles indicate the position of the cavities.}
\label{fig:A1204_cav}
\end{figure*}

\begin{figure*}
\centering
\includegraphics[width=17cm]{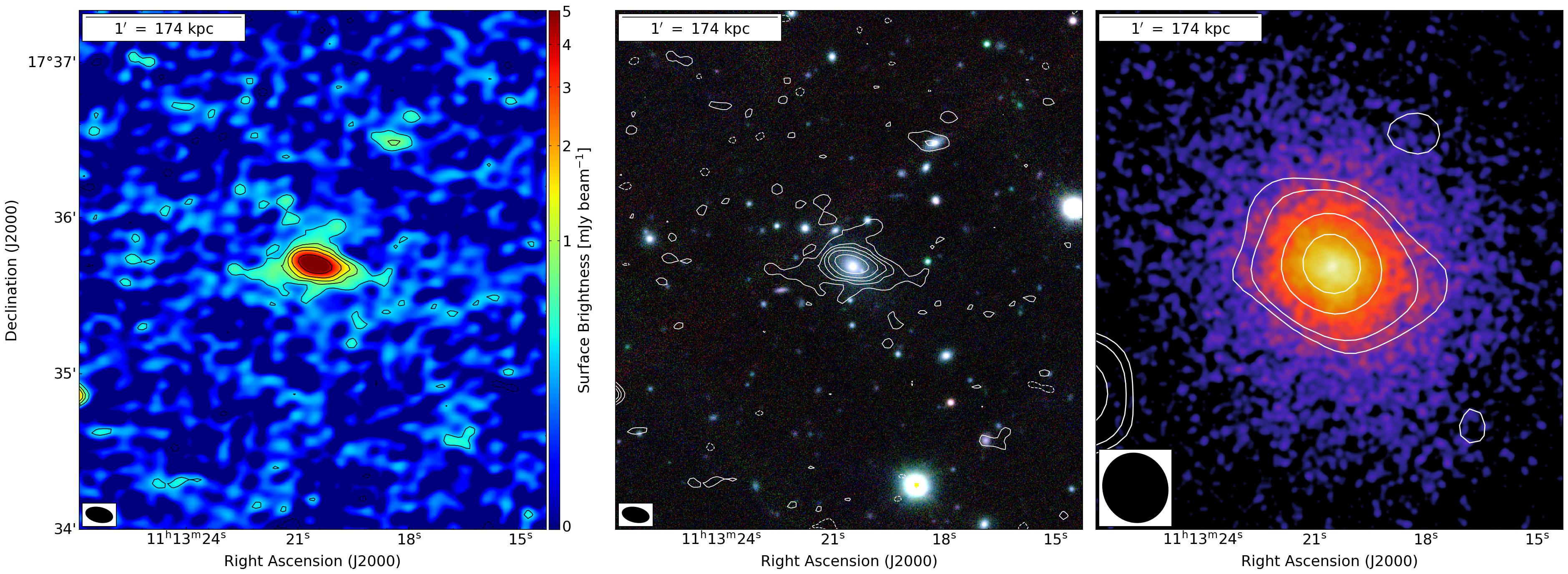}
\caption{Multi-wavelength images of A1204. \emph{Left panel}: High-resolution 144 MHz LOFAR image. Levels:[-3, 3, 6, 9, 15, 30]$\times\sigma$ (where $\sigma=0.13$ mJy $\rm{beam}^{-1}$). The beam is $10.6\arcsec\times5.7\arcsec$ and is shown in the bottom left corner of the image. 
\emph{Central panel}: Optical \emph{Pan-STARRS} RGB image with high-resolution LOFAR contours overlaid. \emph{Right panel}: \emph{Chandra} X-ray image smoothed on a scale of $1\arcsec$ with overlaid low-resolution source-subtracted LOFAR contours (Levels=[-3, 3, 5, 15, 30]$\times\sigma$, where $\sigma=0.29$ mJy $\rm{beam}^{-1}$ and the beam is $27.1\arcsec\times24.9\arcsec$). All the panels depict the same region of the sky.}
\label{fig:A1204}
\end{figure*}

\begin{figure*}
\centering
\includegraphics[width=17cm]{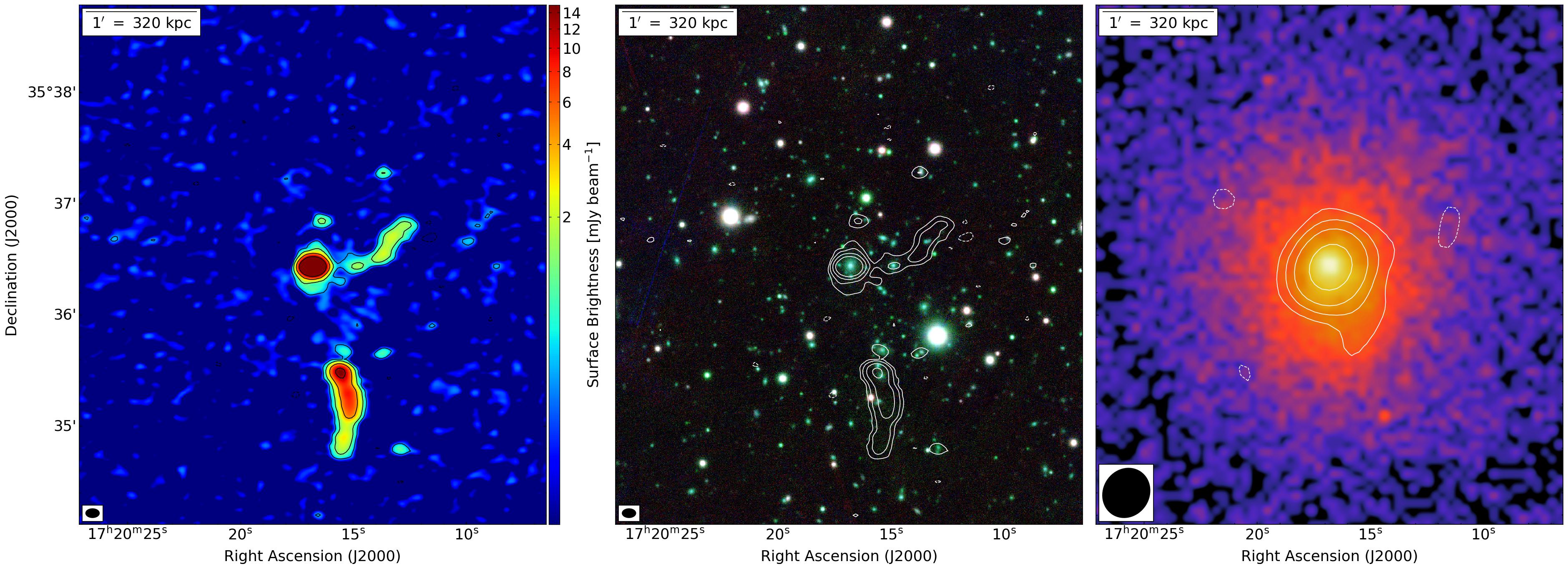}
\caption{Multi-wavelength images of MACS J1720.2+3536. \emph{Left panel}: High-resolution 144 MHz LOFAR image. Levels:[-3, 3, 9, 30, 90]$\times\sigma$ (where $\sigma=0.14$ mJy $\rm{beam}^{-1}$). The beam is $7.3\arcsec\times4.8\arcsec$ and is shown in the bottom left corner of the image. 
\emph{Central panel}: Optical \emph{Pan-STARRS} RGB image with high-resolution LOFAR contours overlaid. \emph{Right panel}: \emph{Chandra} X-ray image with overlaid low-resolution source-subtracted LOFAR contours (Levels=[-3, 3, 9, 20, 50]$\times\sigma$, where $\sigma=0.30$ mJy $\rm{beam}^{-1}$ and the beam is $27.4\arcsec\times24.7\arcsec$). All the panels depict the same region of the sky.}
\label{fig:RXJ1720_2}
\end{figure*}

\subsubsection{A1204}
The X-ray morphology of the cluster A1204 (z=0.171) is highly spherical with a bright central core. We noticed the presence of two X-ray depressions near the cluster centre (Fig. \ref{fig:A1204_cav}).
The residual images, obtained by subtracting both a single (central panel) and double $\beta$-model (right panel), emphasise the holes.
The projected size and position of cavities were determined by eye by approximating the depressions with circular regions (green circles in Fig. \ref{fig:A1204_cav}).
The eastern cavity is best represented by a circle with a radius $r\sim2.6\arcsec\cong7.6\rm{kpc}$, and projected distance from the X-ray peak, $R\sim4\arcsec\cong11.6\rm{kpc}$. The western cavity, instead, has a radius of $r\sim2.3\arcsec\cong6.7\rm{kpc}$ and is located at the same projected distance from the cluster centre.    
The cluster was observed in the radio band with VLA at 1.4 and 5 GHz. It presents an unresolved central source with a flat spectrum, while no central diffuse emission was detected \citep[][]{Giacintucci2017}. 
The central source was also detected by the FIRST survey at a higher resolution of $5.4\arcsec$, and it has a flux density of $S_{\rm 1400~MHz}=1.9$ mJy.
At LOFAR resolution ($10\arcsec\times6\arcsec$, Fig. \ref{fig:A1204}) we found that the bright central source is slightly elongated in the east-west direction (size $D_{\rm max}\sim140$ kpc), possibly due to the presence of radio lobes, but the elliptical shape of the beam may influence the morphology.
The flux density of the central compact source is $S_{\rm 144~MHz} = 9.9$ mJy.
Therefore, it has a spectral index of $\alpha\sim0.7$ between 144$-$1400 MHz, as is typically observed in the core of active galaxies. A slight increase in surface brightness is measured at lower resolution, which however is not significant. The low-resolution radio contours in Fig. \ref{fig:A1204} are obtained after the subtraction of the central source, modelled with uniform weighting, and re-imaged with an $uv$-taper of 20\arcsec.
Neither our high-resolution nor low-resolution maps show clear evidence of diffuse emission corresponding to a MH or RH-like structure around the central galaxy.

\subsubsection{MACS J1720.2+3536}
The thermal emission of MACS J1720.2+3536 (hereafter MACSJ1720.2, z=0.391) is slightly elongated towards the south. 
\cite{Hlavacek2012} reported the presence of two cavities in the central regions of this cluster, one to the north (at a distance of 19 kpc from the core) and a smaller and less significant one to the south-east (6.7 kpc from the core).
\cite{Giacintucci2017} catalogued this source as a candidate MH, without, however, reporting further indications on the detection.
With our LOFAR data (Fig. \ref{fig:RXJ1720_2}) we did not detect a MH at the centre of this cluster, but only a central radio source with a small extension in the north-south direction (total extension $D_{\rm max}\sim160$ kpc), that could be associated to the AGN activity. A head-tail radio galaxy is located south of the cluster centre, while on the west we detected a compact radio galaxy and even further away another radio galaxy with two radio lobes as bright as the core. 
The central compact source has a flux density of $S_{\rm 144~MHz}= 155\ \rm{mJy}$. It is detected also by the FIRST survey with a flux density of $S_{\rm 1400~MHz}= 16\ \rm{mJy}$. Therefore it has a spectral index of $\alpha\sim1.0$. \cite{Timmerman2022} has imaged the LOFAR data of this source including the international baselines, resolving the central source in three compact components of non-identified nature.
To obtain the low-resolution radio contours (see Fig. \ref{fig:RXJ1720_2}, right panel) we have subtracted first the surrounding radio sources and then the central unresolved source, modelled with an inner $uv$-cut of $8k\lambda\cong136$ kpc. There are still some residuals after the subtraction, but the flux density comparison has indicated that no further diffuse emission is detected at lower resolution.



\section{Analysis}\label{sec:analysis}
For the clusters presenting diffuse radio emission, we extracted radial profiles to investigate whether multiple components are present. 
We also estimated the radio power associated with putative RHs and derived upper limits in the case of a non-detection.

\subsection{Radio brightness radial profiles}
The surface brightness of radio halos decreases with increasing distance from the cluster centre and their radial profiles can normally be fitted by an exponential law \citep[e.g.][]{Murgia2009} in the form:
\begin{equation}
    I(r) = I_0 e^{-r/r_e},
    \label{exp}
\end{equation}
where $I_0$ is the central surface brightness and $r_e$ is the $e$-folding radius.

\begin{table*}
    \centering
    \caption{Radial profile fits.}
    \renewcommand\arraystretch{1.2}
    \begin{tabular}{cccccccccc} \hline
Cluster name  &FWHM &dr &Sector &Fit &$I_{01}$ &$r_{e1}$ &$I_{02}$ &$r_{e2}$ &$\chi^2_{red}$\\
 &[arcsec] &[arcsec] && &[$\mu Jy/arcsec^2$] &[kpc] &[$\mu Jy/arcsec^2$] &[kpc] &\\\hline
PSZ1G139.61+24  &35 &12  &NW &2 exp &$23\pm9$ &$51\pm20$ &$1\pm1$ &$881\pm2045$ &0.2    \\
&&&SE &1 exp &$11\pm2$ &$116\pm13$ &-&- &0.1 \\
A1068 &35 &5 &NW &2 exp &$300\pm565$ &$9\pm9$ &$4\pm3$ &$122\pm84$ &0.4   \\ 
&&&SE &2 exp &$248\pm270$ &$12\pm8$ &$7\pm7$ &$83\pm39$ &0.1 \\
MS 1455.0+2232   &15 &5 &N &1 exp &$246\pm2$ &$43.7\pm0.3$ &-&- &63.4      \\ 
&&&S &2 exp &$269\pm8$ &$24\pm1$ &$36\pm7$ &$85\pm7$ &1.6\\
RX J1720.1+2638   &15 &3 &SW &2 exp &$7877\pm431$ &$12.0\pm0.2$ &$23\pm2$ &$137\pm10$ &3.4        \\ 
RBS797 &28 &7 &whole &1 exp &$1456\pm70$ &$16.5\pm0.5$&-&- &0.7 \\
RXJ1532 &28 &7 &whole &1 exp &$817\pm21$ &$20.7\pm0.3$ &-&- &0.9\\\hline
    \end{tabular}
    \tablefoot{ The profiles are extracted on sectors with small angular aperture to highlight the presence of multiple components, so the best-fit parameter values are not representative of the cluster.}
    \label{tab:radial_profile_fit}
\end{table*}

\begin{figure*}[h!]
\centering
\subfloat{
\includegraphics[width=5.6cm]{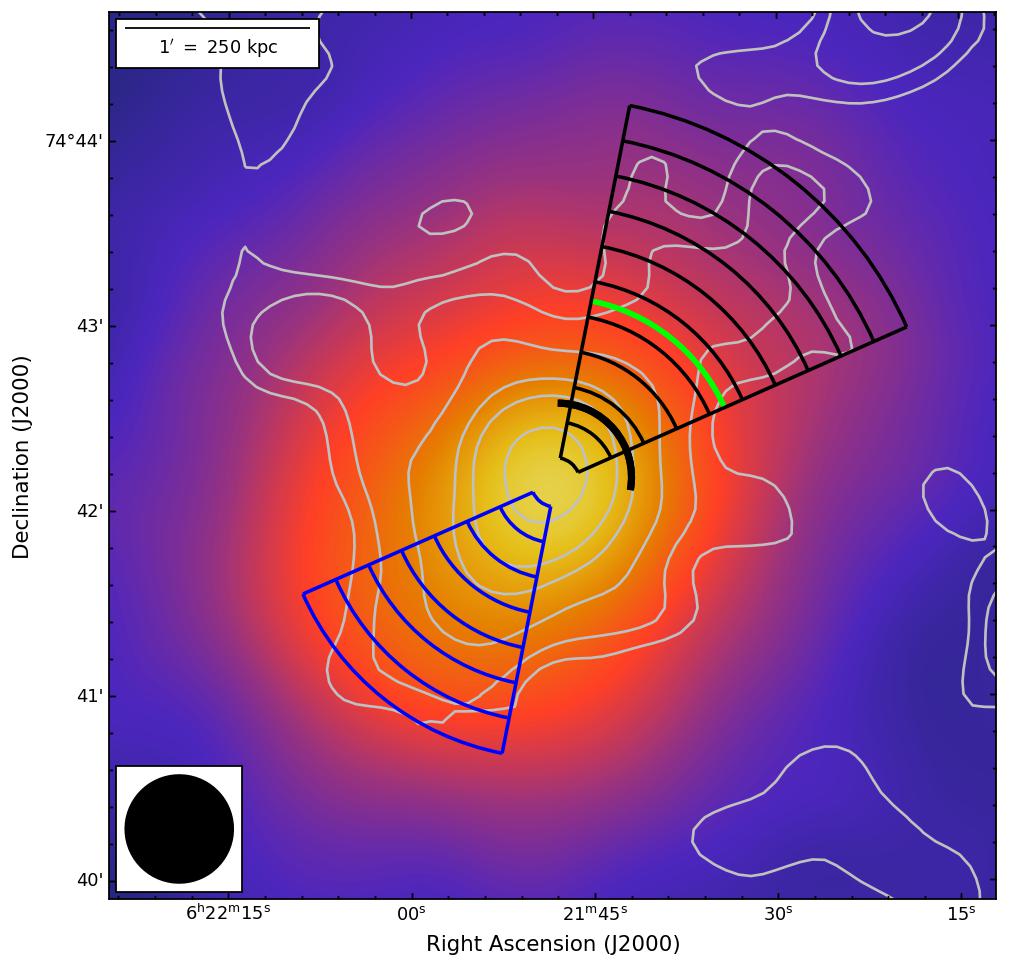}
\includegraphics[width=5.6cm]{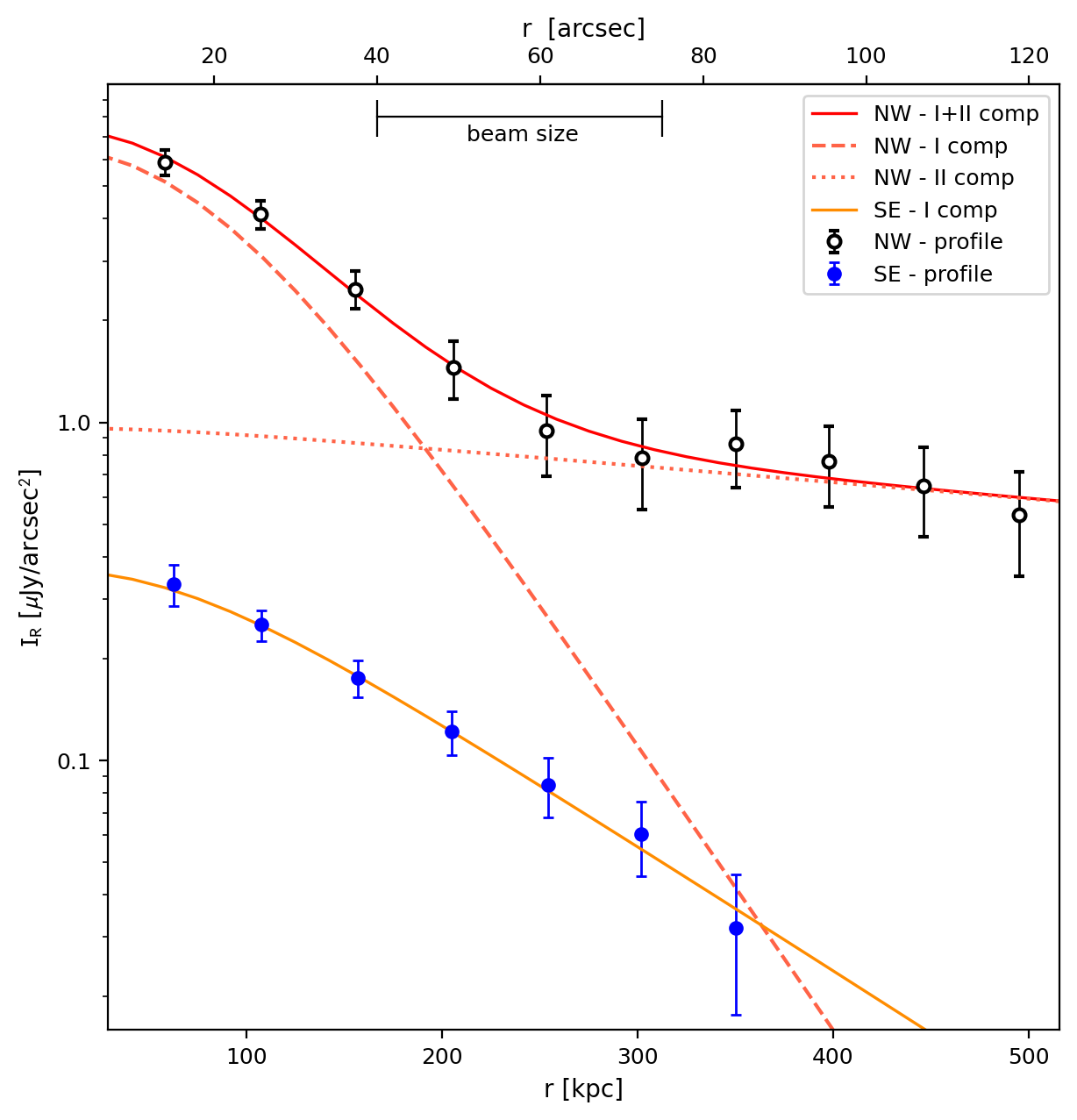}
\includegraphics[width=5.6cm]{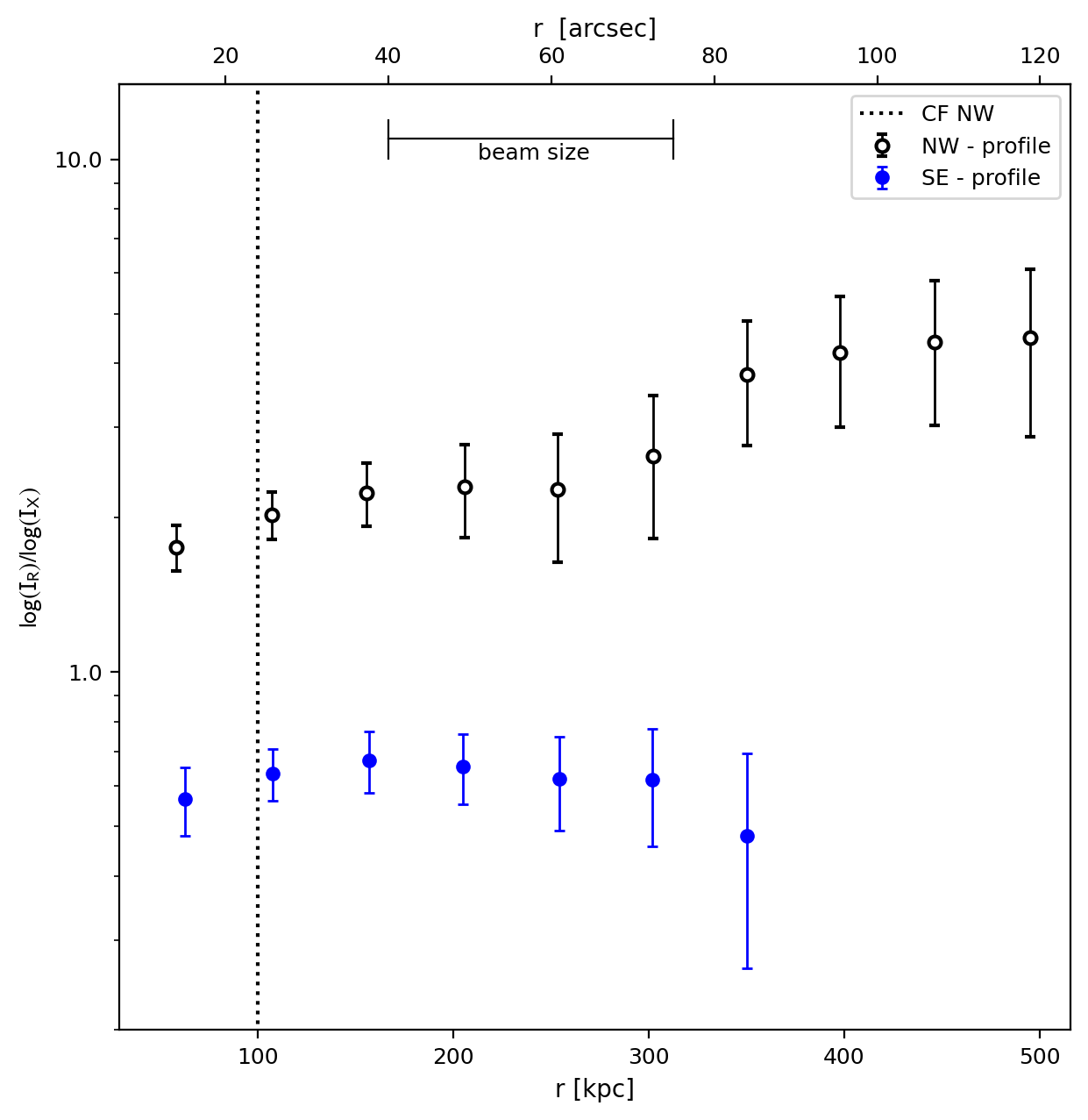}}\\
\subfloat{
\includegraphics[width=5.6cm]{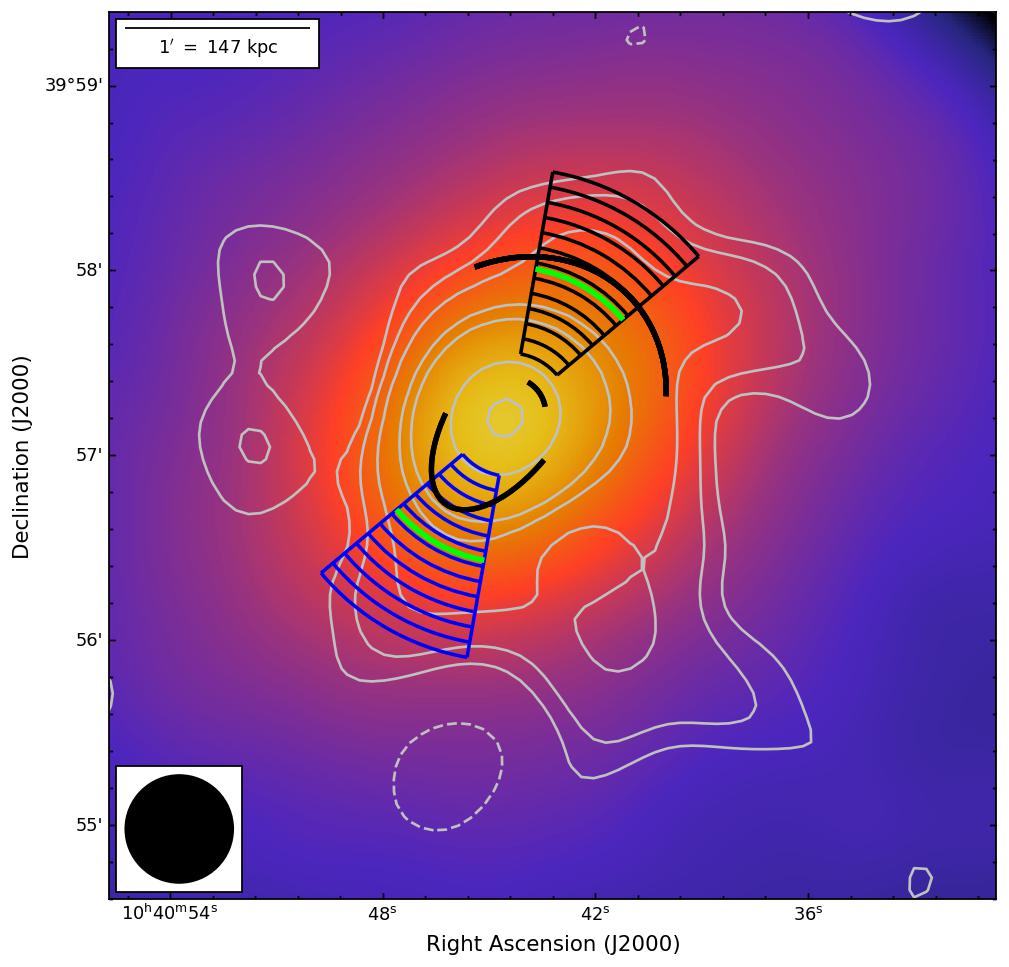}
\includegraphics[width=5.6cm]{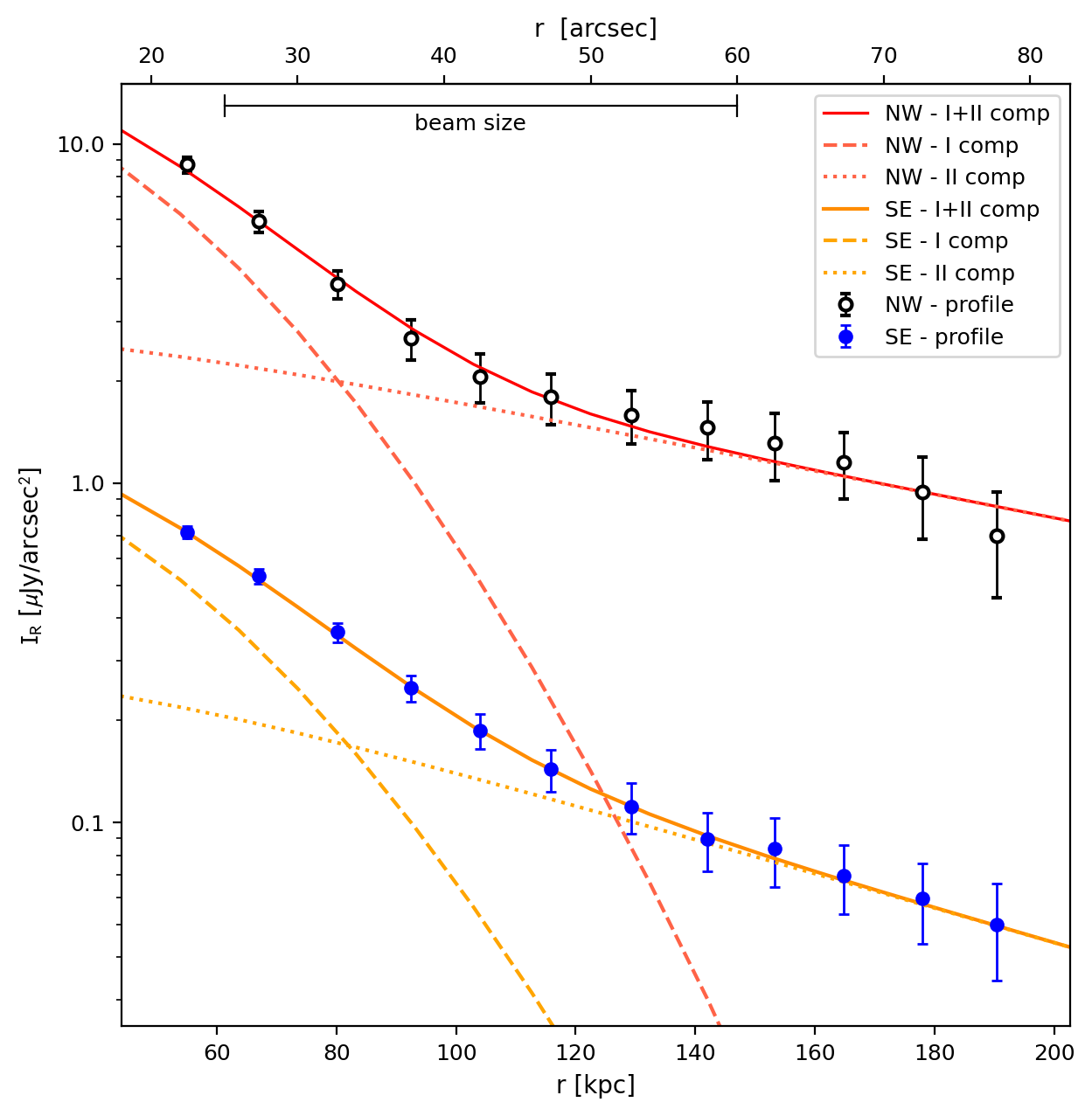}
\includegraphics[width=5.6cm]{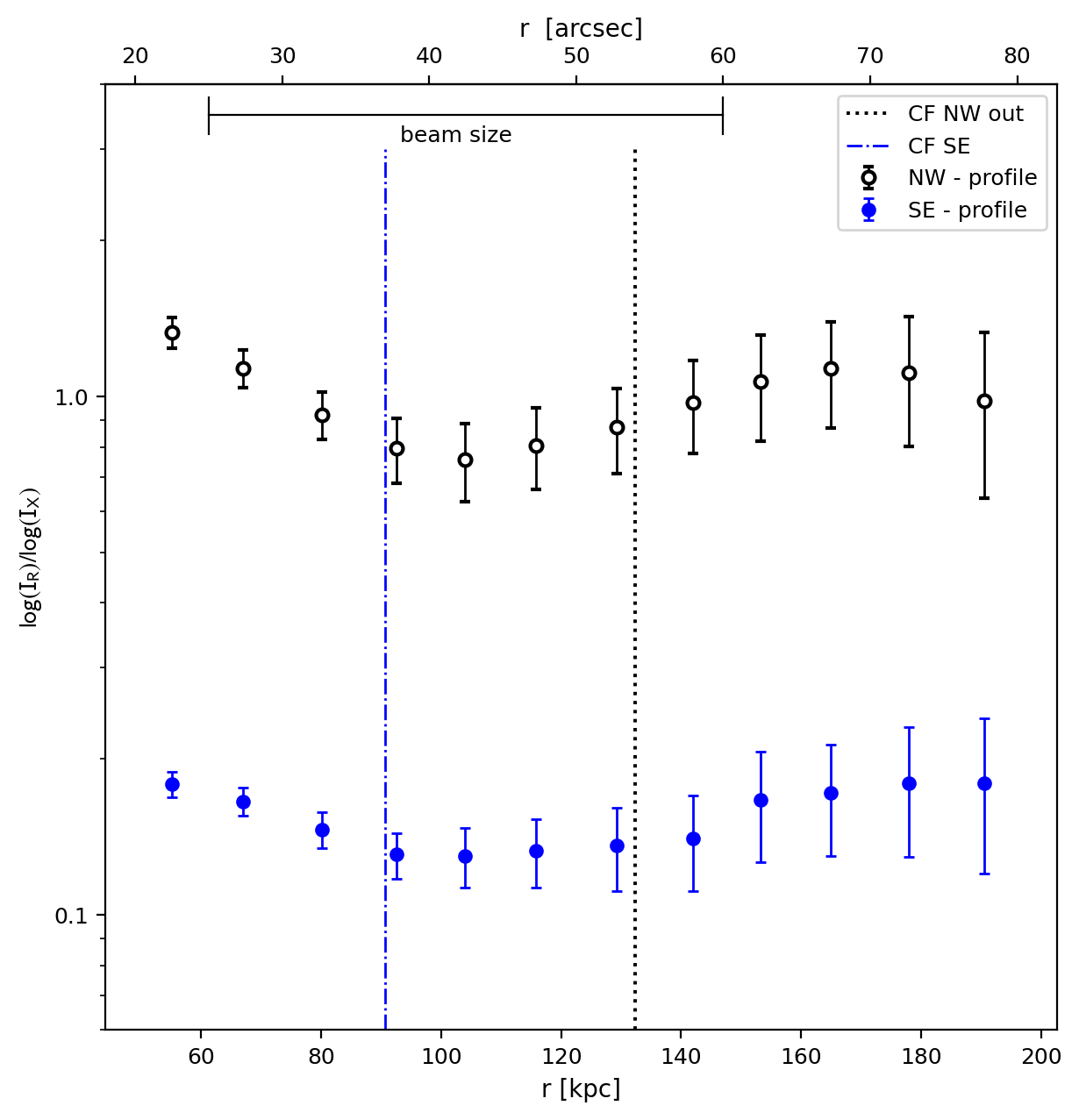}}\\
\subfloat{
\includegraphics[width=5.6cm]{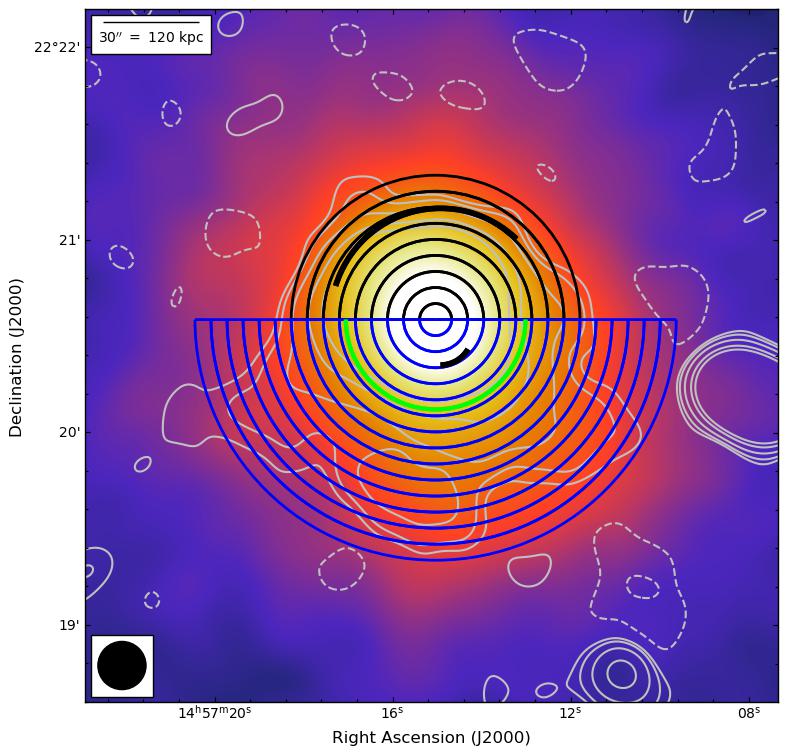}
\includegraphics[width=5.6cm]{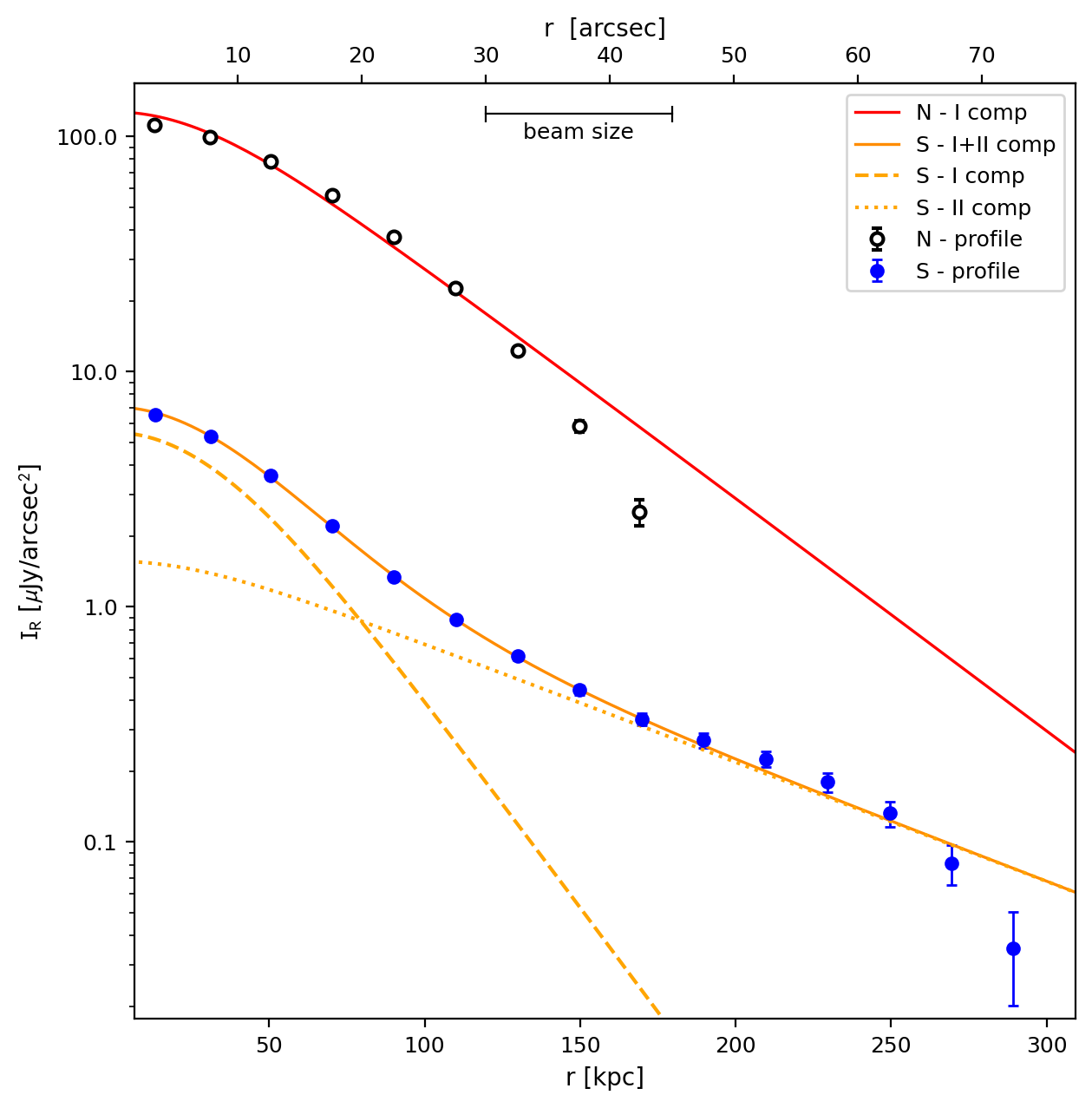}
\includegraphics[width=5.6cm]{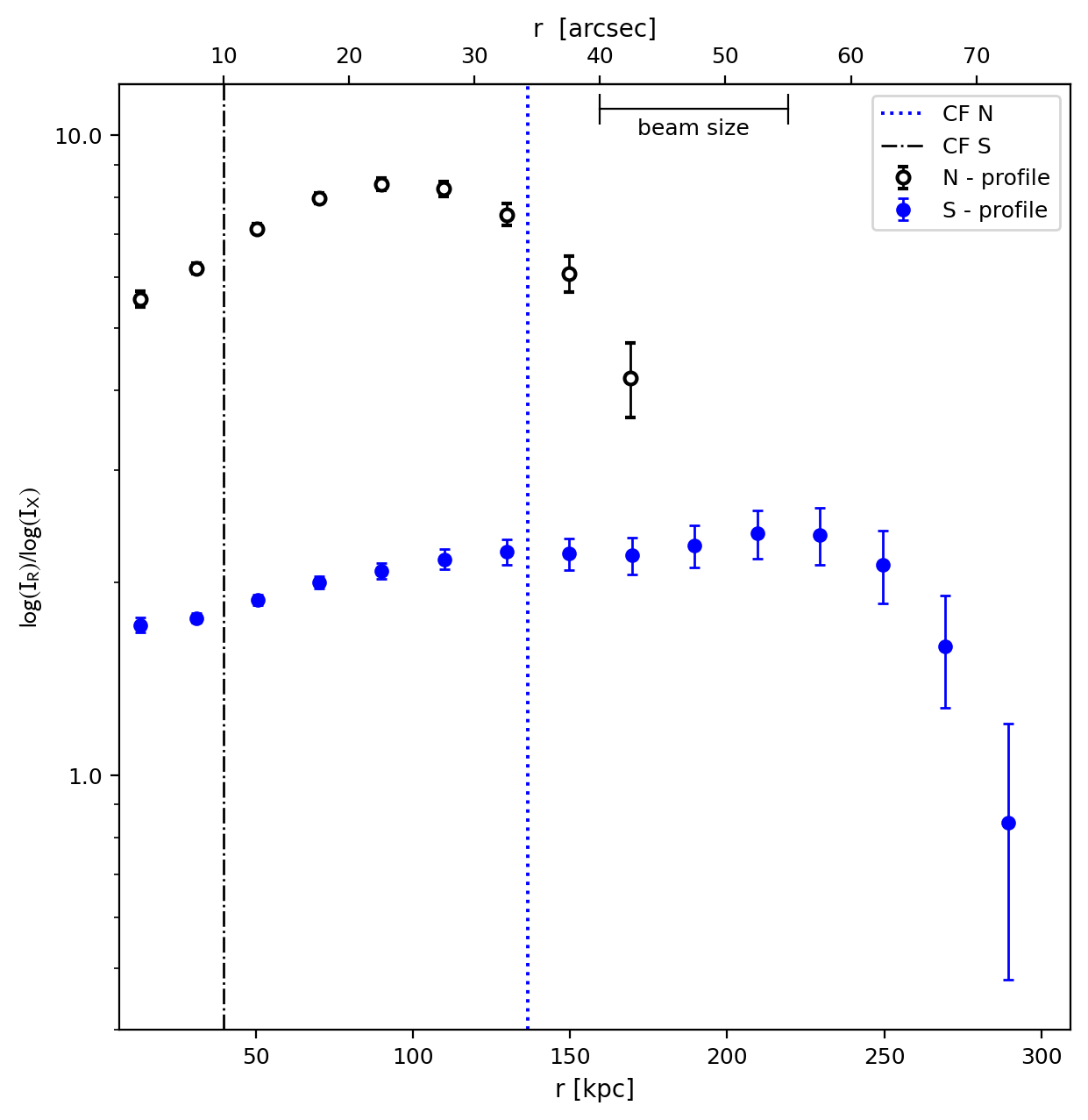}}
\caption{Radial profiles of PSZ1G139.61+24 (\emph{First row}), A1068 (\emph{Second row}), and MS 1455.0+2232 (\emph{Third row}). \emph{Left panels:} \emph{Chandra} X-ray maps smoothed to radio resolution (FWHM=35\arcsec for PSZ1G139 and A1068, FWHM=15\arcsec for MS1455), with overlaid LOFAR contours and sectors used to extract radial profiles. The black arcs indicate the position of the cold fronts, while the green arcs indicate the location of the radio discontinuities. \emph{Central panels:} Radio radial profiles extracted along the sectors indicated in the left panels. Profiles from different sectors have different normalisation for ease of viewing. Solid lines represent the best-fitting model, while dashed and dotted lines represent the best-fitting model of single components of the double exponential, convolved with the radio beam. \emph{Right panels:} Ratio of the radio and X-ray radial profiles extracted along the same regions reported in the left panels. The dotted vertical lines indicate the position of cold fronts.}
\label{fig:profili}
\end{figure*}

We want to investigate if the radio emission in clusters with cluster-scale diffuse emission is made of a single or two different components, as found in RXJ1720.1 by \cite{Biava2021b}. 
We also examined, for comparison, the radial profiles of the two clusters with a MH.
With this aim, we derived the averaged surface brightness radial profiles, using low-resolution point-source-subtracted images.
When the subtraction of surrounding sources has not produced sufficiently accurate results, we have masked their emission and we have excluded the masked pixels from the calculation of the surface brightness. Furthermore, we have excluded the central points from the fit when the subtraction of the central source has left some residuals.
We investigated the brightness profiles in various directions to check for spatial differences, as in most cases the radio emission is not symmetric with respect to the cluster centre.
We extracted the profiles in concentric sectors at increasing distance from the cluster centre. The sector width was chosen specifically for each cluster as a fraction of the beam FWHM to better trace the radial profile, as surface brightness variations are on small scales, and improve the fit statistics.
For each sector, we computed the mean radio surface brightness and associated uncertainty of $\delta I_R = rms/\sqrt{N_{beam}}$ (where $rms$ is the noise of the radio image and $N_{beam}$ is the number of beams in each sector).
Following the procedure used by \cite{Murgia2009}, and commonly used later on by other authors \citep[e.g.][]{Cuciti2022,Bruno2023b}, we have performed a fit of the extracted radial profiles taking into account the convolution by the beam. This procedure allows one to derive best-fit parameters that are not sensitive to the resolution of the observations. Furthermore, it allows to investigate the radial profiles on scales smaller than the beam size. Specifically, we have fitted the radio profiles with a single and double exponential function convolved by a Gaussian having major and minor axis as the restoring beam.
The regions of extraction and the associated surface brightness profiles are reported in Fig. \ref{fig:profili}, left and central panels, respectively, while the deconvolved best-fitting results are listed in Table \ref{tab:radial_profile_fit}.
We note that for some fits we obtained small chi-squared values, especially for those clusters with smaller surface brightness and then larger errors. This is due to the fact that the chi-squared value is weighted for the surface brightness errors. 
We also want to point out that the fitting parameters reported in Table \ref{tab:radial_profile_fit} are not representative of the clusters, which are not spherically symmetric. The fits are only used to identify the presence of a second component. The profiles were in fact extracted on very narrow sectors to enhance the detection of surface brightness discontinuities.

We found a discontinuity in the radial radio profile of all clusters which present cluster-scale diffuse radio emission (PSZ1G139, A1068, MS1455 and RXJ1720.1), but not in the clusters with a MH (RBS797 and RXJ1532), where a single exponential represent the best-fitting model. In the clusters with discontinuities, the radio emission decreases exponentially with increasing radius, from the cluster centre up to a certain distance beyond which there is a surface brightness excess, indicating the presence of a second component, so a double exponential represents the best-fitting model. This discontinuity, however, is not always present along all the directions considered, as we discuss in the following text on a per-cluster basis. 


\begin{figure*}[h!]
\centering
\subfloat{
\includegraphics[width=5.6cm]{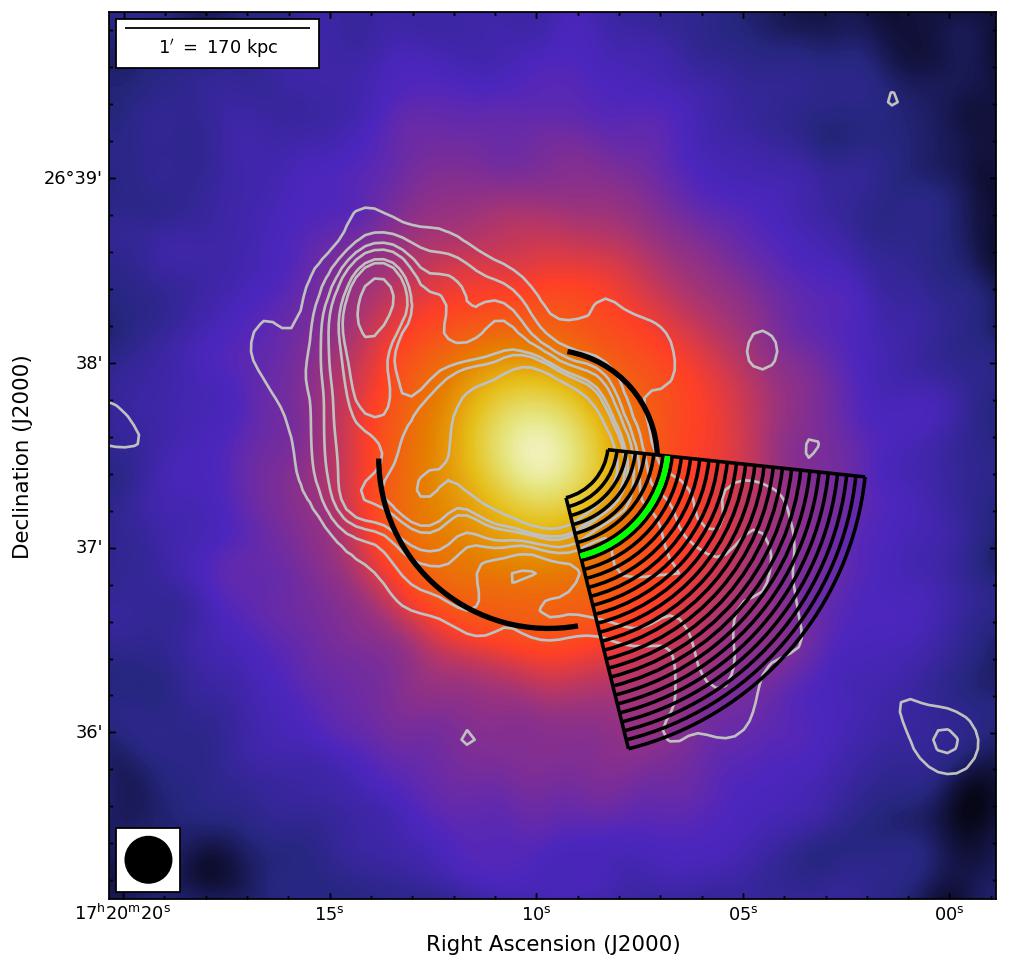}
\includegraphics[width=5.6cm]{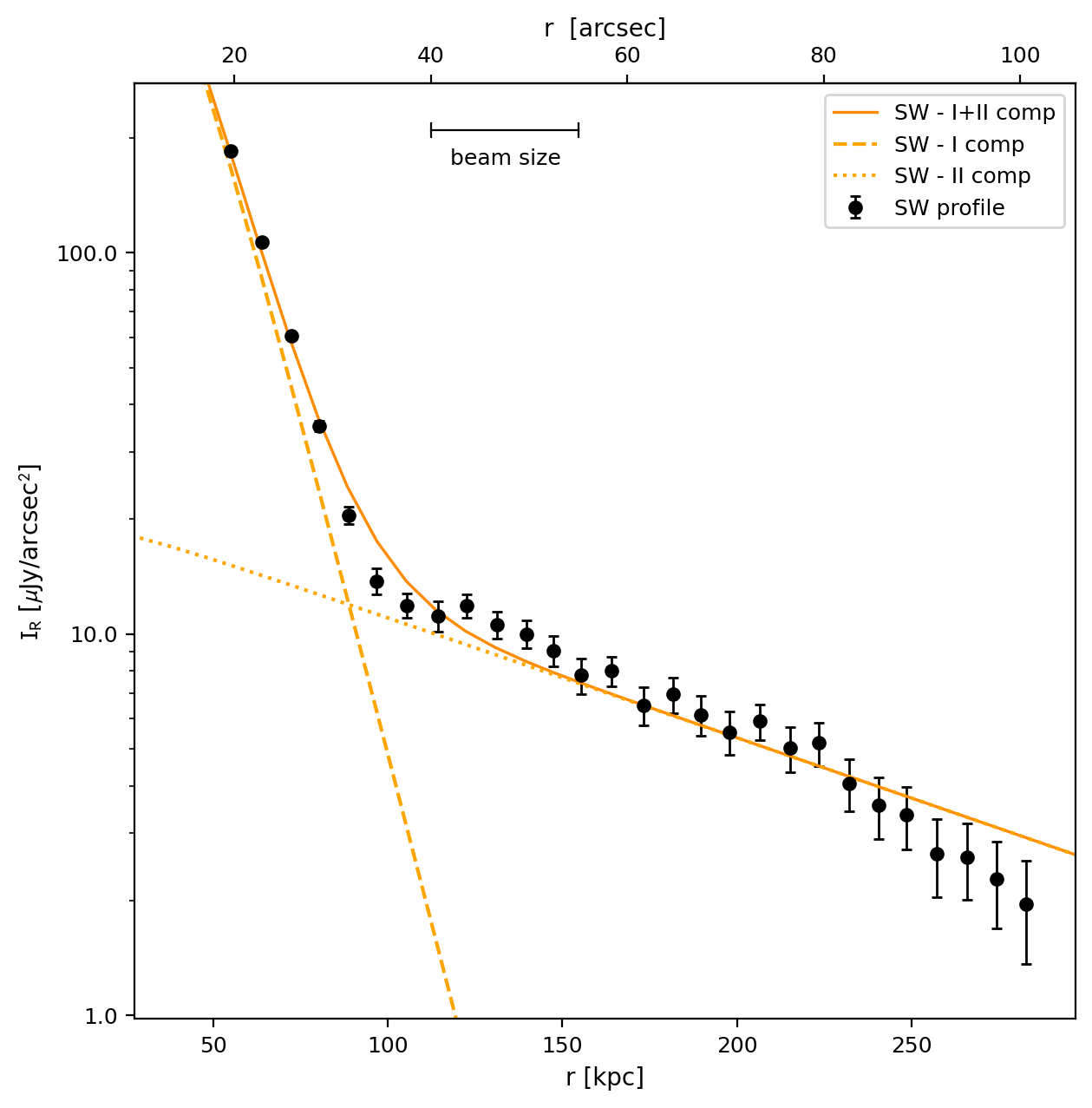}
\includegraphics[width=5.6cm]{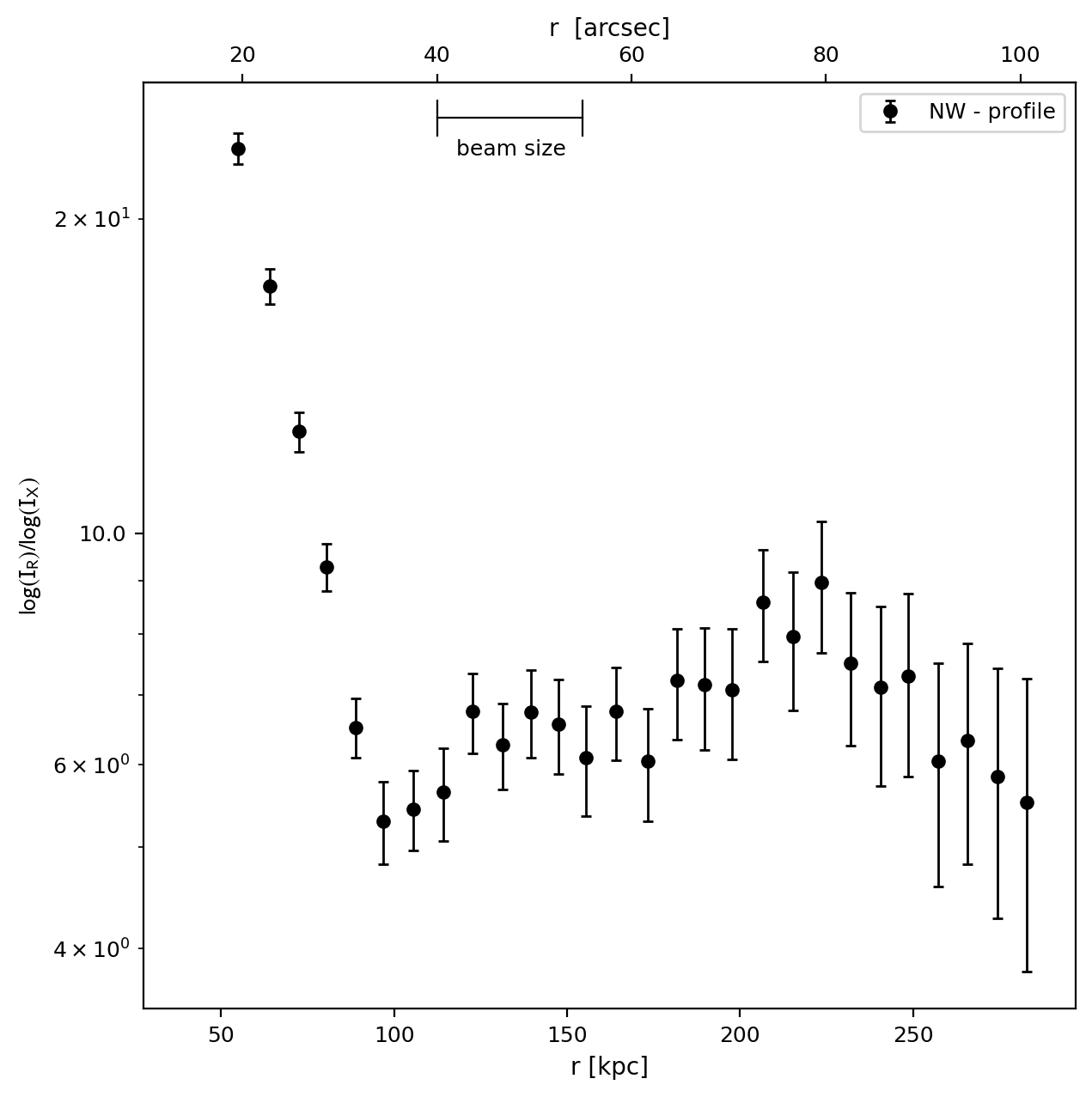}}\\
\subfloat{
\includegraphics[width=5.6cm]{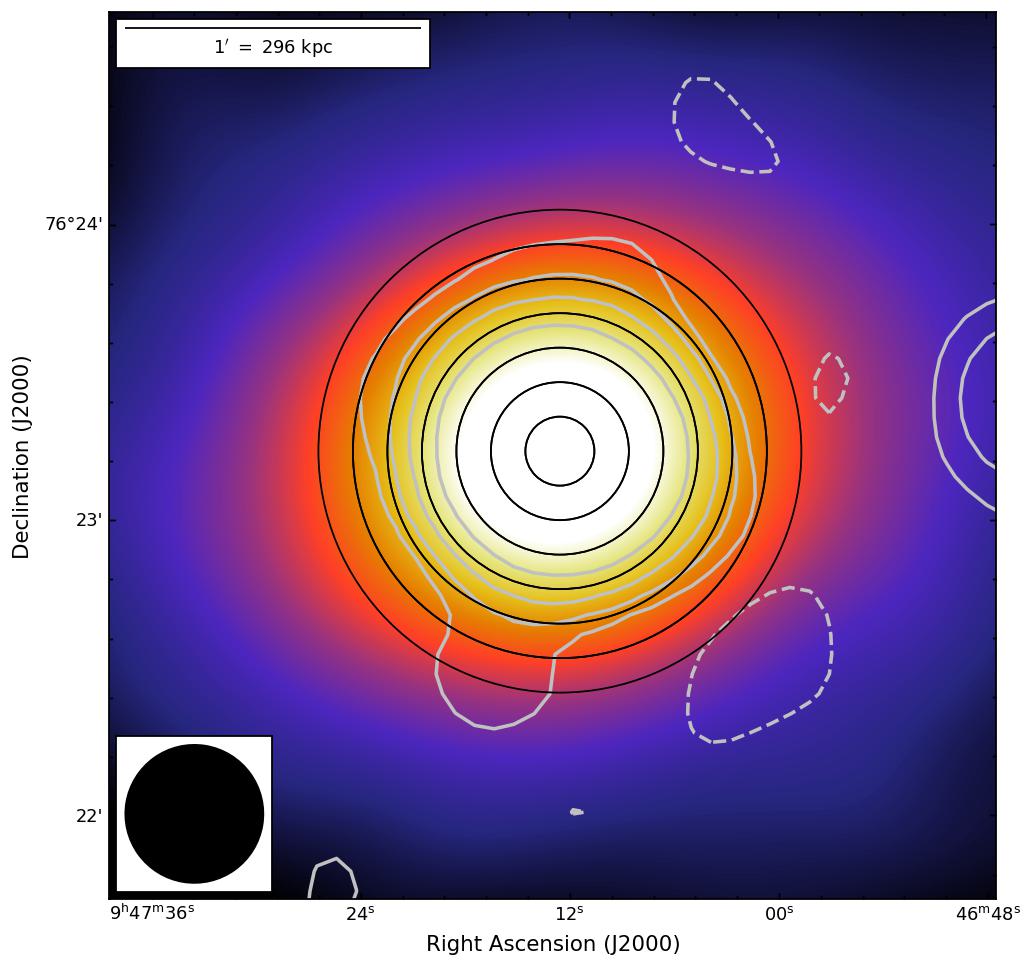}
\includegraphics[width=5.6cm]{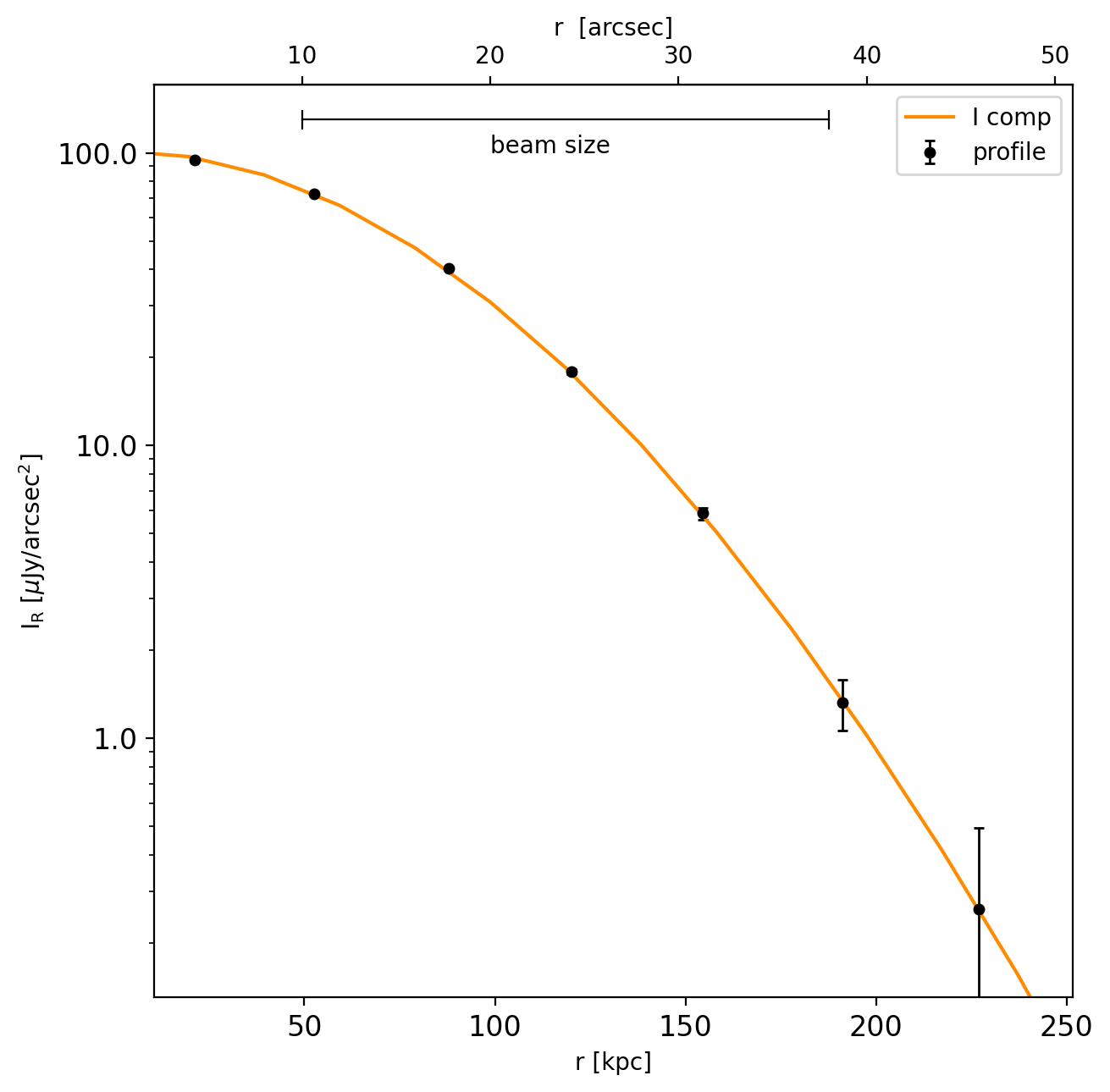}
\includegraphics[width=5.6cm]{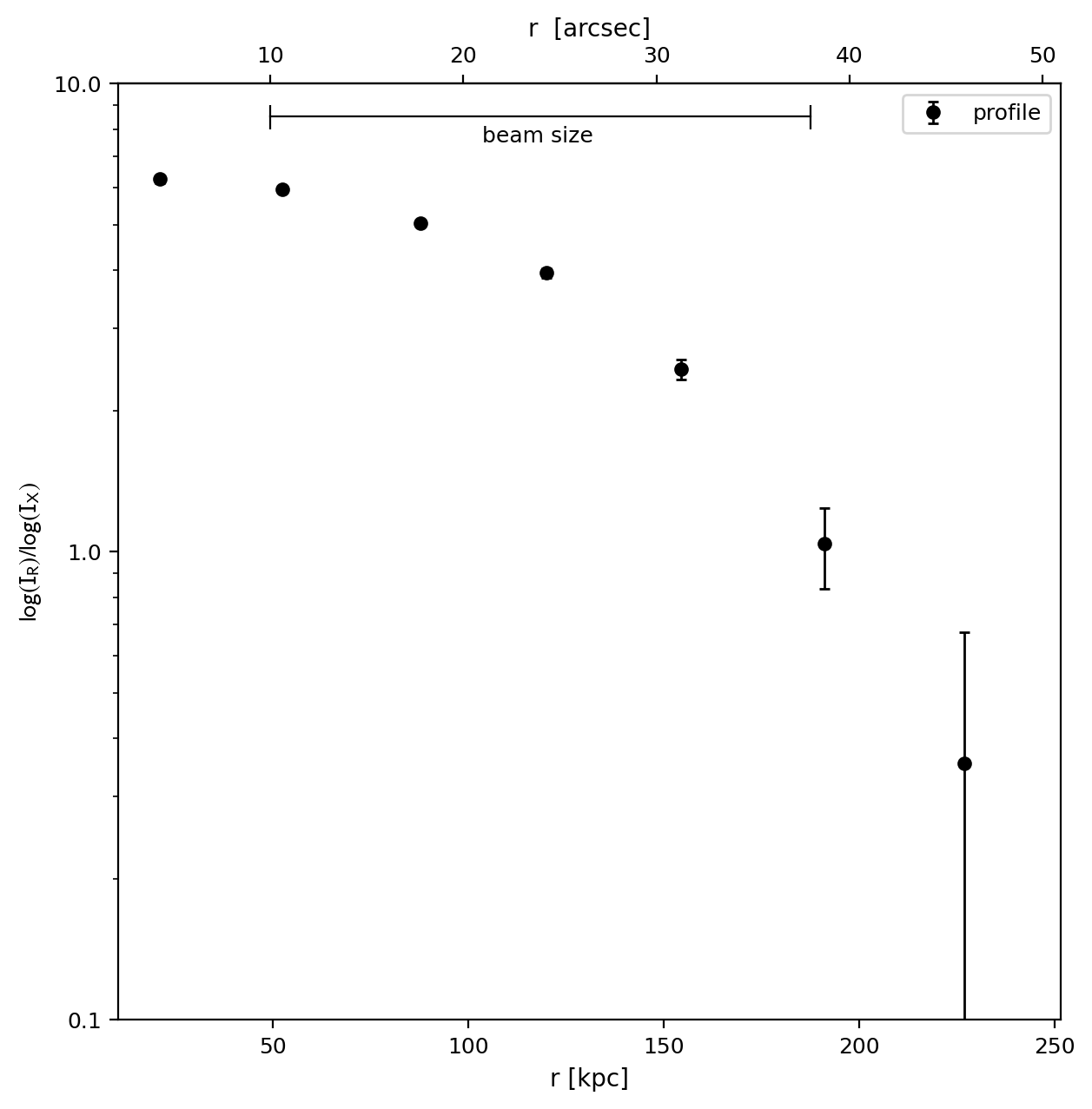}
}\\
\subfloat{
\includegraphics[width=5.6cm]{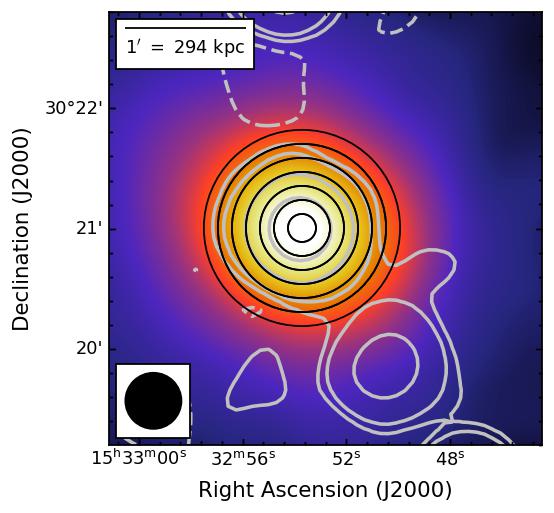}
\includegraphics[width=5.6cm]{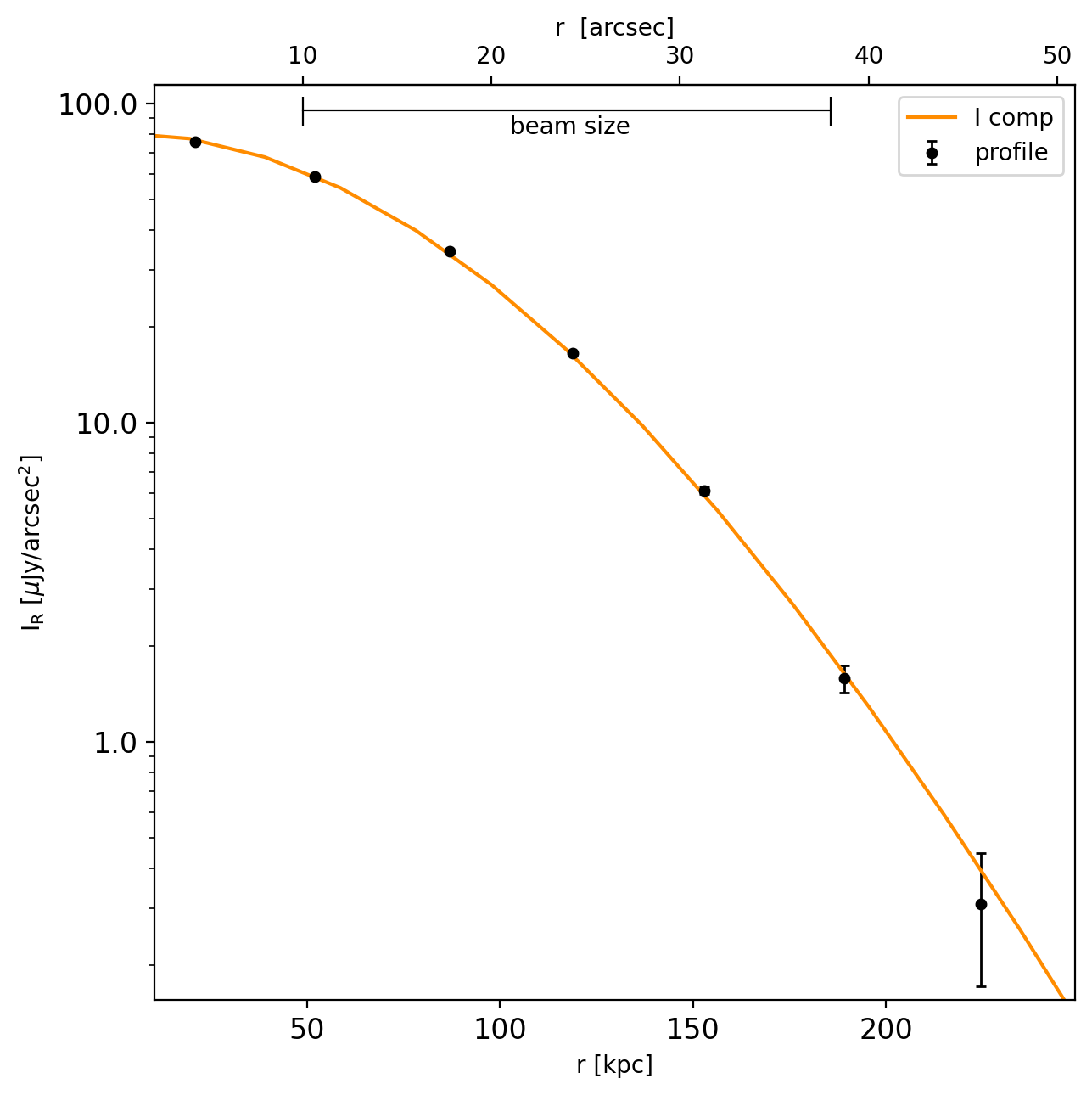}
\includegraphics[width=5.6cm]{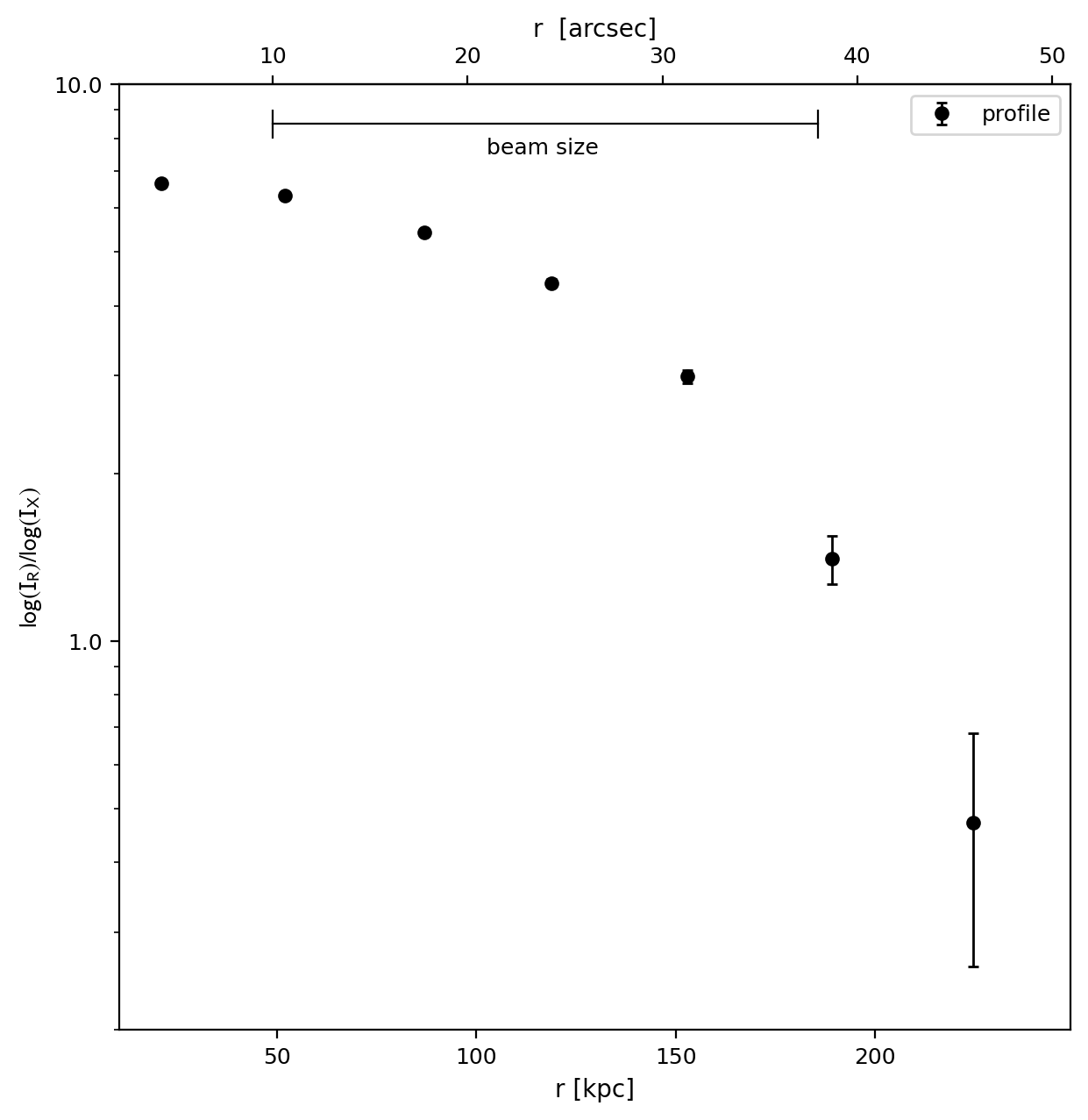}}
    \caption{Same as Fig. \ref{fig:profili} for RX J1720.1+2638 (\emph{First row}, FWHM=15\arcsec), RBS797 (\emph{Second row}, FWHM=28\arcsec), and RX J1532.9+3021 (\emph{Third row}, FWHM=28\arcsec).}
\end{figure*}

For PSZ1G139 (Fig. \ref{fig:profili}, first row), we detected a discontinuity only in the north-west sector, where there is a cold front. The change in the profile, however, does not occur at the cold front position ($\sim100$ kpc from cluster centre, black arc), but at a greater distance from cluster centre ($\sim295$ kpc, green arc). The south-east profile, instead, is well-fitted by a single exponential.

In A1068 (Fig. \ref{fig:profili}, second row) we detected a discontinuity both in the north-west and south-east sectors, roughly at the same distance from the cluster centre (105\,kpc).
The radio discontinuities (green arcs) are located in the proximity of the outer X-ray cold fronts (black arcs).


MS1455 (Fig. \ref{fig:profili}, third row) presents a discontinuity in the southern sector in coincidence with the sloshing boundary ($r=120$ kpc). In the northern sector, instead, there is no evident discontinuity, however, the profile does not follow an exponential law, as the radio surface brightness decreases more rapidly than an exponential profile.
This difference may be because the diffuse emission extends mainly towards the south.

For RXJ1720.1 (Fig. \ref{fig:profili}, second page first row), we extracted a radial profile only in the south-west direction, since at north-east there is a head-tail galaxy superimposed on the diffuse emission, as found by \cite{Biava2021b}, that we have not been able to subtract.
Also in this cluster, we found a clear departure from a single exponential, indicating an excess of surface brightness in the external regions. A double exponential, however, is not a good fit in that case, as the cluster-scale component follows a more complex trend. The discontinuity in the profile coincides with the transition zone between the MH and cluster-scale emission ($r=88$ kpc), indicative of a double radio component, as already shown by the net change in the spectral index inside and outside the cluster core \citep[][]{Biava2021b}.

For the clusters RBS797 and RXJ1532, we extracted the profiles considering the whole extension of diffuse emission, as it is circularly symmetric (Fig. \ref{fig:profili}, second page second and third row). No discontinuities are present in the profiles, which are well-fitted by a single exponential model, confirming a single component, i.e. a MH, is present. We also investigated the possible presence of dissimilarities in the profiles in different directions, extracting the profiles in sectors with 90-degree apertures, but all the profiles follow the same trend.


We further analysed the radial properties of these clusters, comparing their radio and X-ray surface brightness. 
According to literature studies, the thermal and non-thermal components of clusters presenting radio halos appear to be correlated, with a correlation between the radio ($I_R$) and X-ray ($I_X$) surface brightness of the form:
\begin{equation}
    I_R \propto I_X^{\beta}\ ,
\end{equation}
where RHs follow a linear or sub-linear correlation \citep[$\beta\le1$, e.g.][]{Govoni2001,Botteon2020,Rajpurohit2021a,Rajpurohit2021b,Riseley2022b}, while a super-linear correlation is found for MHs \citep[$\beta > 1$, e.g.][]{Ignesti2020,Riseley2022,Riseley2023}.
In RXJ1720.1 \cite{Biava2021b} found different trends in the two components: a super-linear correlation for the MH and a sub-linear correlation for the cluster-scale halo component. For MS1455, instead, \cite{Riseley2022} found the radio emission inside and outside the sloshing region could be quite well represented by a single correlation with a slope of $\beta = 1.15$.

Here, we investigated the presence of radial scaling and possibly different trends in the two components of the halo emission, considering how the ratio between the thermal and non-thermal components varies across the cluster extension.
If a single $I_R$-$I_X$ correlation is representative of the whole diffuse emission, we would expect a constant value of $I_R/I_X^{\beta}$ across the cluster size. If instead, given a fixed slope $\beta$, we observe an increasing (decreasing) behaviour, we are overestimating (underestimating) $\beta$, and a lower (higher) value is necessary to correctly reproduce the $I_R$-$I_X$ correlation \citep[][]{Bonafede2022, Bruno2023b, Rajpurohit2023,Balboni2024}.

We computed X-ray profiles using the same regions of radio profiles on X-ray maps smoothed to the same resolution of radio images.
The ratio of radio and X-ray surface brightness, considering a fixed linear slope ($\beta=1$), are reported in the right panels of Fig. \ref{fig:profili}.
The $I_R$/$I_X$ trend is not constant throughout the halos extension, but there are changes in the slope, indicating a unique linear correlation is not representative of the whole halos. Moreover, the radial pattern is different for each cluster.

The cluster PSZ1G139 presents the most constant behaviour of $I_R$/$I_X$ among the clusters with a double component. The profile of the south-east sector is roughly constant with distance, indicating that the radio and X-ray emission are linearly correlated. The north-west profile, instead, is slightly increasing with distance from the cluster centre, with a small jump in correspondence of the radio discontinuity; the variations, however, are not significant due to the large errors, especially at large radii.

The north-west and south-east profiles of A1068 follow the same trend: there is a decreasing profile in the central regions, then at 100 kpc there is a discontinuity as in the radio profile and an inversion of the trend, but still compatible with a constant profile due to large errors. Therefore, we are underestimating the correlation slope in the MH component, which is better described by a super-linear correlation, while the large diffuse component follows a more linear slope.

In MS1455 there is a different trend in the opposite directions. The northern profile presents a clear increasing trend, which abruptly decreases around 100 kpc. The change in slope is not coincident with the position of the cold front, which however could be affected by projection effects.
The southern profile, instead, follows a slightly increasing profile, and then a drastic decrease in the last points.
This result is at odds with the single super-linear correlation found by \cite{Riseley2022} performing point-to-point analysis, but they considered the whole halo area, so they are possibly averaging different trends.

In RXJ1720.1 the $I_R$/$I_X$ trend is decreasing in the central region (and so a linear slope is underestimated), then it becomes roughly constant in the cluster-scale halo component. This confirms the point-to-point correlation results found by \cite{Biava2021b}, where the MH follows a super-linear correlation, while the cluster-scale halo has a flatter slope. 

To summarise, in A1068 and RXJ1720.1 the MH and cluster-scale diffuse emission follows different $I_R/I_X$ trends, with a super-linear correlation for the central emission and a near linear correlation for the external component. PSZ1G139 and MS1455, instead, present a radio-to-X-ray ratio always increasing from the cluster centre to the outskirt. Further analysis is required to investigate the cause of this dissimilarity. 

Finally, for the two clusters presenting a classical MH, no slope changes are observed in their ratio profiles. These profiles slowly decline at increasing radius, indicating a super-linear slope as typically found in MHs \citep{Ignesti2020}.

\subsection{Radio power and upper limits}\label{sec:UL}
Radio observations and radial profile analysis have revealed the presence of a double radio component, which we interpret as MH + cluster-scale halo, in four of the 12 clusters in our sample. 

\begin{figure*}
\sidecaption
\includegraphics[width=12cm]{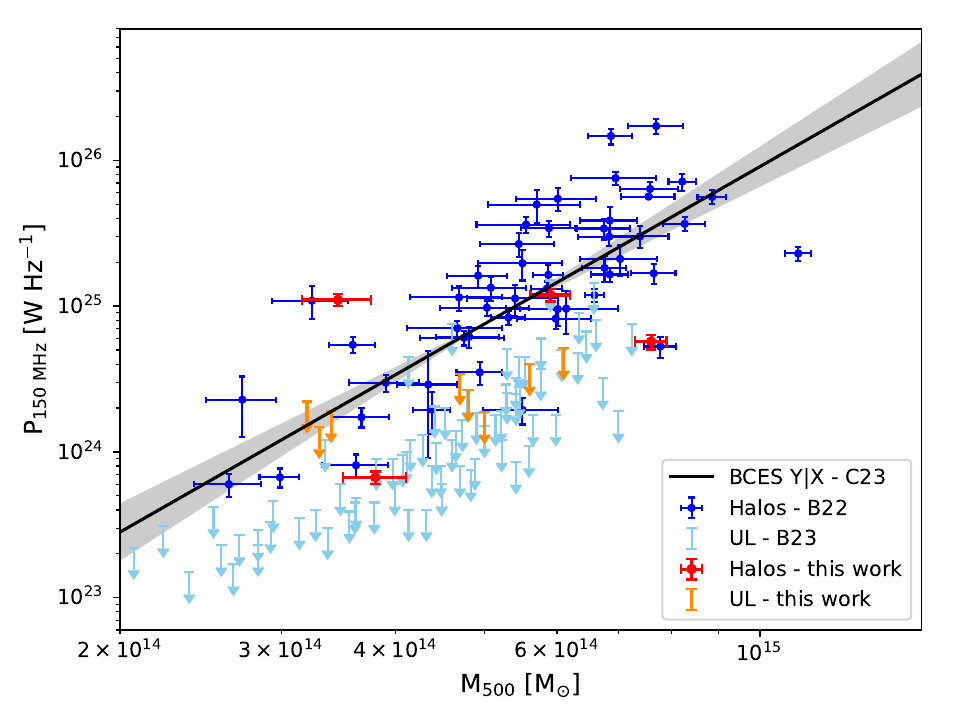}
\caption{Radio power at 150 MHz vs cluster mass ($M_{\rm 500}$). The clusters with a cluster-scale halo in our sample are indicated in red, while the upper limits we derived for the other clusters in the sample are indicated with orange arrows. RHs in \textit{Planck} clusters in LoTSS-DR2 \protect\citep{Botteon2022a} are indicated in blue and their upper limits \citep{Bruno2023a} in cyan. The black line indicates the correlation found by \cite{Cuciti2023}, using a BCES Y|X regression method, while the shaded region shows the associated $3\sigma$ region.}
\label{fig:M-P}
\end{figure*}

In this section, we want to compute the radio power of cluster-scale halos in our sample and make a comparison with the power of known RHs. Due to the irregular morphology of the emission and the presence of contaminating sources difficult to exclude, we have not been able to extract radial profiles on the whole extension of the radio emission, but only on small sectors, so the obtained best-fit parameters are not representative of the whole emission and cannot be used to compute the total power of the diffuse emission.
Therefore, we estimated the integrated flux densities of cluster-scale halos in these clusters from the low-resolution LOFAR images, considering all the emission enclosed by $3\sigma$ surface brightness contours and then we subtracted the contribution of MH and radio galaxies. We note that in that way we are providing a lower limit to the true integrated flux of the cluster-scale halo, as it is present also in the central regions of the cluster, superimposed on the MH.
So, we added the contribution of the cluster-scale halo in the central regions, estimating it has the same mean surface brightness as in the external regions. The final values are reported in Table \ref{tab:P_myAloni}, where we also indicated the expected radio power according to the $M_{500}-P_{150\ \rm{MHz}}$ correlation found by \cite{Cuciti2023} on radio halos in \textit{Planck} clusters in the LoTSS-DR2, for comparison.
We then investigated the distribution of the observed radio halos in the radio power–mass diagram (Fig. \ref{fig:M-P}). 
We represented our detection with a red dot, while in blue the radio halos in \textit{Planck} clusters in the LoTSS-DR2 \citep{Botteon2022a}.
We also reported in the plot the $M_{500}-P_{150\ \rm{MHz}}$ correlation found by \cite{Cuciti2023}, using a BCES Y|X regression method (black solid line), with corresponding $3\sigma$ region (grey shaded area).
We note that the halos detected in this work are far from the M-P correlation, except for RXJ1720.1 which lies on the correlation, but their powers are still consistent with the scatter in detected halos in \cite{Botteon2022a}.
The different spectra and sizes of radio halos and the different dynamical stages of clusters in the sample contribute to the scatter of the correlation, as investigated by \cite{Cuciti2023}.

\begin{table}
    \centering
    \caption{Radio halo power.}
    \renewcommand\arraystretch{1.2}
    \begin{tabular}{cccccccc} \hline
Cluster &$M_{500}$ &$P_{\rm 150,\ exp}$ &$P_{\rm 150,\ meas}$ \\ 
&[$10^{14}\ \rm{M_{\odot}}]$ &[$10^{24}$ W/Hz] &[$10^{24}$ W/Hz] \\\hline
PSZ1G139.61+24 &7.6 &34 &5.7\\
A1068 &3.8 &2.8 &0.67\\
MS 1455.0+2232 &3.5 &2.0 &11\\
RX J1720.1+2638 &5.9 &14 &12\\\hline
    \end{tabular}\\
    \tablefoot{ Col. 1: Cluster name; Col. 2: Cluster mass; Col. 3: Expected power at 150 MHz according to the correlation reported in \cite{Cuciti2023}; Col. 4: Measured power at 150 MHz. }
    \label{tab:P_myAloni}
\end{table}

\begin{table*}
    \centering
    \caption{Radio halo upper limits.}
    \renewcommand\arraystretch{1.2}
    \begin{tabular}{cccccc} \hline
Cluster &$M_{500}$ &$P_{\rm 150,\ exp}$ &$P_{\rm 150,\ UL,\ exp}$ &$P_{\rm 150,\ UL,\ meas}$ &$S_{\rm 150,\ UL,\ meas}$\\ 
&[$10^{14}\ \rm{M_{\odot}}]$ &[$10^{24}$ W/Hz] &[$10^{24}$ W/Hz] &[$10^{24}$ W/Hz] &[mJy] \\\hline
MS 0735.6+7421  &5.0 &7.6 &0.9 &1.9 &17.1\\ 
MS 0839.8+2938  &3.4 &1.9 &1.3 &1.9 &22.8\\ 
Z2089           &3.2 &1.5 &1.3 &2.2 &11.4\\ 
RBS 0797        &5.6 &11.3 &2.0 &4.0 &7.6\\
A1204           &3.3 &1.7 &1.0 &1.5 &14.6\\ 
RX J1532.9+3021 &4.7 &6.1 &2.0 &3.4 &9.5\\ 
MACS J1720.2+3536       &6.1 &15.4 &2.6 &5.1 &11.4\\ 
MACS J2245.0+2637    &4.8 &6.5 &1.4 &2.6 &9.5\\\hline
    \end{tabular}\\
    \tablefoot{ Col. 1: Cluster name; Col. 2: Cluster mass; Col. 3: Expected power at 150 MHz, according to \cite{Cuciti2023} correlation; Col. 4: Expected upper limit power at 150 MHz, according to  \cite{Bruno2023a} correlation; Col. 5: Measured upper limit power at 150 MHz; Col. 6: Measured upper limit flux density at 150 MHz.}
    \label{tab:UL}
\end{table*}

For the clusters in the sample where no cluster-scale diffuse emission is detected, we provided upper limits on the flux densities.  
Following the procedure described in \cite{Bruno2023a}, we derived upper limits injecting a mock radio halo in the $uv$-plane of each cluster using the \verb|MUVIT|\footnote{https://github.com/lucabruno2501/MUVIT} code.
We modelled the radio brightness profile of the radio halos as an exponential law described by Eq. \ref{exp}. 
As in \cite{Bruno2023a} we used a mass-independent value of $r_e=186$ kpc, equal to the mean value of the $e$-folding radius of the \textit{Planck} clusters in LoTSS-DR2 in our mass range. To derive a starting value of $I_0=S_{\rm UL}/2\pi r^2_e$, instead, we used the correlation found in \cite{Bruno2023a} between the flux density of the upper limits ($S_{\rm UL}$) and the noise of the image and the number of beams within the injected mock halo. 
We decided to not inject the mock halo at the cluster centre, but in a nearby region void of sources, for the presence of diffuse emission or radio galaxies that are difficult to subtract in most of the clusters.
The mock halo is Fourier transformed to add it to LOFAR $uv$-data that we imaged using a $uv$-taper of $20\arcsec$ to improve the sensitivity to diffuse emission.
We then increased or decreased the injected value of $I_0$ until the largest linear scale at $2\sigma$ contours ($D_{\rm 2\sigma}$) of the injected halo is $\sim R_{\rm H} \sim 2 \times r_e$, to consider it as extended emission.
We then used the obtained $I_0$ to derive the upper limit to the halo power scaled at 150 MHz, for comparison with other works, using a typical spectral index for radio halos of $\alpha=1.3$. The results are reported in Table \ref{tab:UL}, where we also reported the expected upper limits according to the correlation found by \cite{Bruno2023a} and the expected power according to the M-P correlation found by \cite{Cuciti2023}. We notice that the obtained values are barely dependent on the injection position for source-free regions \citep[see discussion in][]{Bruno2023a}. 
Discrepancies (typically being of factors 1.5-2) between expected and derived upper limits are likely mainly driven by the considered tapering \citep[lower-resolution images were typically used by][]{Bruno2023a}.
We represented the obtained upper limits in the M-P diagram (Fig. \ref{fig:M-P}) with an orange arrow and in cyan the upper limits found by \cite{Bruno2023a} for \textit{Planck} clusters in LoTSS-DR2, for comparison.
The upper limits of the low-mass clusters A1204, Z2089 and MS0839 lie on the correlation, therefore, we cannot exclude the presence of a radio halo in these clusters. Although the upper limits of the other clusters, with $M_{500}\ge4\times10^{14}M_{\odot}$, are consistent with the scatter of radio halos in \cite{Botteon2022a}, they lie below the correlation, in the upper limits region identified on the LoTSS-DR2 sample \citep{Bruno2023a}. \cite{Cuciti2023}, analysing the distribution of LoTSS-DR2 clusters with or without halos (i.e. upper limits) in the M-P diagram, concluded that it is not compatible with a single correlation or with a uniform distribution of clusters below the correlation. Thus deeper observations on a larger sample of cool-core clusters are necessary to confirm that the upper limits we derived are not compatible with the mass-power correlation.


\begin{figure}
\resizebox{\hsize}{!}{\includegraphics{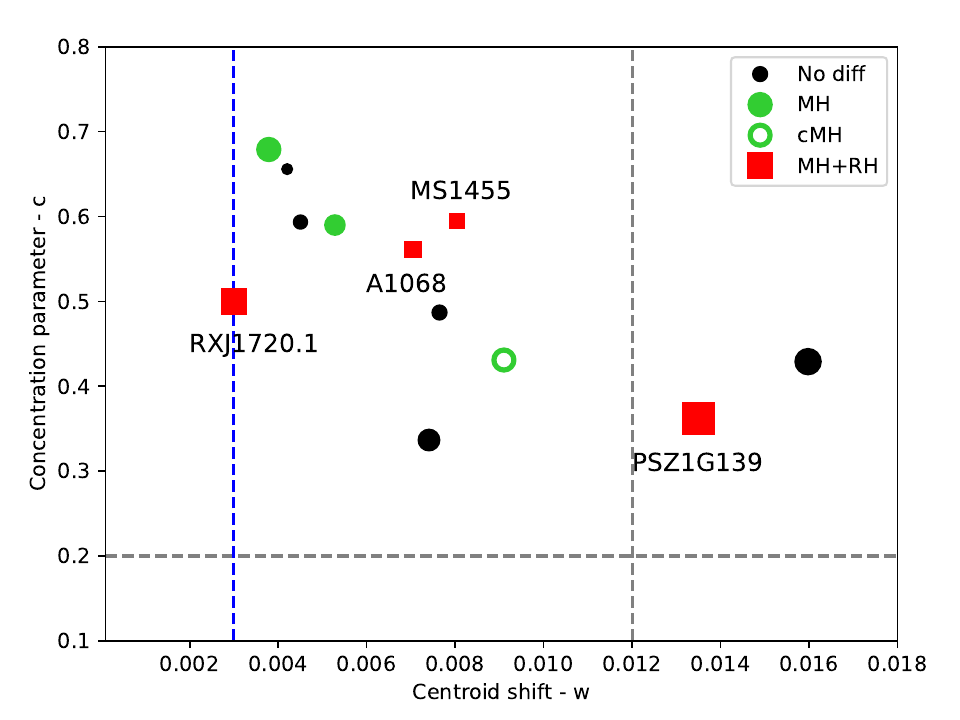}}
\caption{Morphological parameters' plot of the clusters in the sample. Colours are associated with the radio morphology detected with LOFAR: clusters without diffuse emission are represented in black, clusters with a MH in green (for the candidate MH we used a marker not filled), and clusters with a double component radio emission (MH+RH) in red.
Clusters with detected cold fronts are represented with a square, otherwise a circle is used. The marker size is proportional to the cluster mass. The dashed blue line indicates the $w$ value
used to select clusters in our sample. Dashed grey lines indicate the $c$ and $w$ values which separate clusters with a RH ($c\le0.2$ and $w\ge0.012$) from clusters without a RH, according to \citep{Cassano2010}.}
\label{fig:c-w_my}
\end{figure}

\section{Discussion}\label{sec:discussion}
In this article we presented LOFAR HBA observations of twelve cool-core clusters selected for a range of dynamical states, to search for diffuse radio emission on cluster scales.
In this section we discuss our results to investigate two aspects: whether diffuse emission beyond the core region is common in cool-core clusters with MHs and whether there is a connection between radio emission and the gas dynamic in these sources.

\subsection{Radio morphology}
Our study revealed that cluster-scale diffuse radio emission is not always present in cool-core clusters when observed at low frequencies.
The clusters in our sample exhibit a variety of radio morphologies, which could be grouped into three different categories:

\begin{itemize}
    \item \textbf{Clusters with a double radio component:} \\Four clusters present diffuse radio emission extending outside the cluster core: RXJ1720.1 and PSZ1G139 were already identified by \cite{Savini2019} and MS1455 by \cite{Riseley2022}; we found cluster-scale diffuse radio emission in a further cluster, A1068, which was previously classified as a candidate MH \citep{Govoni2009}. We note that the radio emission of all these clusters is not circularly symmetric with respect to the centre, but it is elongated roughly following the X-ray surface brightness distribution. This similarity suggests a strong connection with gas dynamics. As all these clusters present cold fronts and signs of dynamical interactions, the irregular morphology of the non-thermal emission presumably reflects the direction of the turbulent motion induced by a merger event. An exception is RXJ1720.1, where the radio emission is elongated in the north-east--south-west direction, while the X-ray emission has a more regular morphology. Considering the cluster-scale diffuse radio emission of that object has an ultra-steep spectrum \citep[$\alpha\sim3$;][]{Biava2021a}, the turbulence and the particles in this cluster may be dying out after having lost the connection with the dynamics of the gas.
    Our analysis evidences a discontinuity in the radio radial profile of all these four clusters, indicating the presence of a double radio component. We computed the power of the cluster-scale component and we found that is comparable to the power of RHs detected in merging clusters with the same cluster mass. Therefore, a central MH and a cluster-scale halo co-exist in these systems. 
    Recently, a second component has been observed also in merging systems, where the RH is embedded in a more extended diffuse radio emission \citep{Cuciti2022}. However, it is difficult to make an analogy with these systems, as they present a different dynamical state and their diffuse emission is on a much larger scale and fainter intensity, with respect to our cases.
    
    \item \textbf{clusters with a central mini halo:}
    \\We confirmed the presence of a central MH in the clusters RBS0797 \citep{Gitti2006} and RXJ1532 \citep{Hlavacek2013}. The radial profiles evidence the presence of a single radio component. In the cluster MACSJ2245 we detected a hint of diffuse emission around the central source at $2\sigma$ significance. We catalogued this source as a candidate MH. Deeper observations are necessary to confirm the presence of diffuse emission in this cluster.
    
    \item \textbf{clusters without diffuse emission:}
    \\In the remaining five clusters of the sample, we have not detected halo-like diffuse radio emission. We then derived upper limits to the radio power of possible halos. Although the limits are comparable with the scatter of the correlation and do not allow us to exclude the presence of a RH below the noise level, the values derived for the higher mass clusters ($M_{500}\ge4\times10^{14}\ \rm{M_{\odot}}$) are below the correlation, in the upper limit region of LoTSS-DR2 clusters \citep{Bruno2023a}. In this category we have grouped those clusters that present radio emission that appears to be associated with the central AGN. MS0735 presents two giant radio lobes (total extension $\sim550$ kpc) and other smaller-scale lobes associated with subsequent phases of the jet activity from the BCG. MS0839 also features radio lobes, only resolved at $3\arcsec$ resolution. Finally, the clusters Z2089, A1204, and MACSJ1720.2 present small-scale radio emission ($R_{max}\le80$ kpc) oriented in one direction, extending from the unresolved central source. This emission may be linked to an old phase of AGN activity or alternatively originated from hadronic interactions between particles emitted by the central AGN and those present in the ICM. 
    These sources should be further investigated at higher resolution using LOFAR international stations to classify this emission.
\end{itemize}



\subsection{Dynamical state of the gas}
We selected the clusters in the sample based on their morphological parameters, to investigate whether the radio morphology is connected to the dynamical state of the gas.
In Fig. \ref{fig:c-w_my} we re-examine the $c-w$ plot, indicating the clusters of the sample with different colours depending on the radio emission detected with LOFAR and with the marker size proportional to the cluster mass.
We detected cluster-scale diffuse radio emission both in the most massive and less massive clusters in the sample, so the radio morphology is not related to the cluster mass.
The radio morphology also appears uncorrelated to the dynamical state of the cluster, as traced by the morphological parameters.
Clusters with a double component are spread in the $c$-$w$ plot. PSZ1G139 presents a large value of the centroid shift, while A1068, MS1455, and RXJ1720.1 are more relaxed and are located in the region where \cite{Cassano2010} observed no RH. In the last few years, however, this region is becoming more populated with radio halos thanks to low-frequency radio observations \citep[][]{Savini2019,Cassano2023}. The most disturbed cluster in the sample, MACSJ1720.2, instead, does not present halo-like diffuse emission. 
We want to underline that the $c$ and $w$ definitions depend on the scale of extraction. Here we have used a largest scale of 500\,kpc, as typically used in literature \citep[e.g.][]{Cassano2010,Cuciti2021,Cassano2023,Zhang2023}, which however could not be optimal to trace the dynamical state of these objects. The radio halos we detected in cool-core clusters have, indeed, a smaller linear size (between 400-760 kpc) with respect to RHs detected in merging systems, which have typical sizes of about 1–2 Mpc \citep{vanWeeren2019}, suggesting the turbulence has been distributed on a smaller volume.
Larger samples are required to infer more appropriate scales and derive statistically solid considerations. 

There is, however, another signature of gas dynamics that allows us to distinguish clusters with cluster-scale diffuse emission from other clusters in the sample: the presence of cold fronts.
Our analysis shows a tight connection between cluster-scale diffuse emission and sloshing features, as we detected a double component diffuse emission only in those clusters with cold fronts. Conversely, in the other clusters, hosting a MH or without halo-like diffuse emission, there is no clear evidence of sloshing. Even if not all the clusters are observed by \emph{Chandra} with the same exposure time, we considered that the data is good enough to look for surface brightness discontinuities. Cold fronts have also been observed in clusters with relatively shallow observations. Only for the cluster MACSJ2245, deeper observations would help to better classify its dynamical state. Radio observations instead are more homogeneous, as all the clusters have been observed for the same amount of time, and radio images have a comparable noise. However, the upper limits derived do not allow us to exclude the presence of very steep spectrum radio halos in the clusters without cluster-scale diffuse emission.

So far, there has been evidence that diffuse emission in relaxed clusters has been confined by cold fronts \citep[e.g.][]{Mazzotta2008}.
With LOFAR, instead, we detect emission on a larger scale possibly associated with weaker turbulence that is created outside the sloshing region.
According to simulations by \cite{ZuHone2013}, the sloshing motions generate turbulence in the core regions strong enough to re-accelerate seed electrons and produce diffuse synchrotron emission on scales coincident with the sloshing region. In these simulations, however, the turbulence is set to zero outside the sloshing region. 

The co-existence of cluster-scale diffuse radio emission and cold fronts in these cool-core systems, implies the presence of sloshing motions induced by a specific scenario, that allows to generate enough turbulence to re-accelerate particles on large scales while preserving the central cool-core. We speculate these sources experienced an off-axis merger with a large impact parameter. This configuration avoids the destruction of the central cool core, favours the creation of cold fronts, supplying angular momentum to the gas and therefore triggering sloshing motions, and can generate more efficient turbulence in the peripheral zones.
To constrain the turbulence and reproduce the thermal and non-thermal properties of these clusters ad hoc simulations are needed. 

There are a few other cases reported in the literature of relaxed clusters with detected diffuse emission on large scale \citep{Sommer2017,Venturi2017, Kale2019, Raja2020, Bruno2023b, Riseley2023,Lusetti2023, Riseley2024}, but those are transition systems between cool-core and non-cool-core clusters, as they have large values for the core entropy. The clusters studied in this work, instead, have a cool-core, so the sloshing perturbation has not heated significantly the central regions of these clusters, keeping the core entropy at low levels. Based on the dynamical state of the clusters and the level of perturbation of the gas we can then identify a sequence that goes from completely relaxed clusters without radio halos, cool-core clusters with cluster-scale emission, clusters with partially destroyed cores and cluster-scale emission, and finally the non-cool-core clusters exhibiting RHs, where the impact of the merger completely destroyed the central core.


Another point that deserves to be discussed is the absence of signatures of sloshing in the two clusters which host a central MH, single component small-scale diffuse emission. One of the proposed scenarios for the origin of MHs is the re-acceleration of particles by turbulence \citep{ZuHone2013}. The absence of cold fronts in these clusters seems to indicate that they have not recently experienced a merger event, undermining this interpretation. However, these may be objects at a later evolutionary stage than cool-core clusters with cold fronts and cluster-scale diffuse emission, where the turbulence has now dissipated, the sloshing is over and the spectra of the cluster-scale diffuse emission have become too steep to be detected.
Alternatively, it has been speculated that the turbulence that gives rise to the MHs does not come from an external source, but is generated by the central AGN \citep{Bravi2016,RL2020} or that shocks generated by AGN activity may contribute to particle re-acceleration \citep{Bonafede2023_RBS797}. A further hypothesis is the hadronic model, where the MHs are produced by continuous injection of electrons from collisions between thermal and cosmic ray protons \citep{Pfrommer2004}. This possibility is supported by the estimate that expected levels of gamma rays produced by hadronic collisions are lower than the detection limits of \emph{Fermi}-LAT \citep{Ignesti2020,Biava2021a}.
With our results, we could only speculate on the origin of diffuse emission in these systems. Further observations over a wide frequency range is necessary to discern between different scenarios.


\section{Conclusions}\label{sec:conclusions}
In this article, we have presented LOFAR HBA observations and \emph{Chandra} X-ray analysis of a sample of 12 cool-core clusters, with redshifts in the range z=0.17$-$0.39 and masses in the range $M_{500}$=3.2$-$7.6$ \times10^{14}\ \rm{M_{\odot}}$. These clusters were selected for a range of dynamical states to verify the occurrence of cluster-scale diffuse radio emission in cool-core clusters and to understand if this emission is related to the thermal gas condition.
The main results are summarised below:
\begin{itemize}

    \item Cluster-scale diffuse radio emission, extending beyond the cluster core, is detected in 30\% of the cool-core clusters analysed. Those are A1068, a new detection, MS 1455.0+2232, RX J1720.1+2638, and PSZ1G139.61+24. 
    Two clusters (RBS 0797 and RX J1532.9+3021) host a classical MH, diffuse radio emission confined to the cluster core. We detected a new candidate MH (MACS J2245.0+2637). The remaining five clusters (MS 0735.6+7421, MS0839.8+2938, Z2089, A1204, and MACS J1720.2+3536) do not have halo-like diffuse radio emission.

    \item Radio radial profiles show an excess of radio emission at large radii in the four clusters with cluster-scale diffuse emission, indicating the presence of two different components: a central MH and diffuse emission on a cluster scale that extends beyond the sloshing region. A single component is instead found in clusters with a classical MH.
    
    \item The power at 144 MHz of the radio halo component in clusters with cluster-scale diffuse emission is similar to the power of RH detected in merging clusters. For the other clusters, we derived upper limits to the halo power. The limits for the low-mass clusters ($M_{500}\le4\times10^{14}\ \rm{M_{\odot}}$) are consistent with the correlation found by \cite{Cuciti2023}, while those of high-mass clusters lie below the correlation and have the same power of the upper limits derived by \cite{Bruno2023a}. However, given the large scatter of the correlation, we are not able to exclude the presence of a RH in these systems.

    \item An X-ray analysis indicates the presence of cold fronts in the four clusters with double component radio emission, while no sloshing features are found in the other clusters. This connection between cluster-scale diffuse radio emission and cold fronts in cool-core clusters suggests these clusters might have experienced an off-axis minor merger that is able to generate enough turbulence to re-accelerate particles on large scales without disrupting the central cool core. Whereas, the classical MHs hosted in clusters without sloshing features could represent a later phase of evolution, where the turbulence and sloshing have already switched off, or they could originate from other processes such as re-acceleration by turbulence or shocks from the central AGN or hadronic processes.
    
\end{itemize}


The overall picture that is emerging from our study and literature studies \citep{Bonafede2014,Savini2019,Cassano2023} is that there is evidence of cluster-scale diffuse emission in relatively relaxed systems (c>0.2 and w>0.003). Our study found that this emission is located in systems showing signs of sloshing from X-ray observations, possibly entailing a connection between low-energetic (off-axis) mergers and cluster-scale diffuse emission. Additional studies involving larger samples are necessary to further support this hypothesis. Most importantly, spectral studies are necessary to prove the ultra-steep spectra nature of such emission, which is expected in the framework of the merger-induced re-acceleration scenario \citep[i.e.][]{Cassano2006,Cassano2023}.

\section*{Acknowledgements}
NB, A Bonafede, CJR acknowledge support from the ERC through the grant ERC-Stg DRANOEL number 714245. 
FG, GB, RC, MR, acknowledge support from INAF mainstream project “Galaxy Clusters Science with LOFAR” 1.05.01.86.05.
A Botteon acknowledges financial support from the European Union - Next Generation EU.
MB acknowledges financial support from the agreement ASI-INAF n. 2017-14-H.O and from the PRIN MIUR 2017PH3WAT “Blackout”.
FdG acknowledges support from the ERC Consolidator Grant ULU 101086378.
AD acknowledges support by the BMBF Verbundforschung under the grant 05A20STA.
ACE acknowledges support from STFC grant ST/P00541/1.
RJvW acknowledges support from the ERC Starting Grant ClusterWeb 804208. 
LOFAR, the Low Frequency Array designed and constructed by ASTRON (Netherlands Institute for Radio Astronomy), has facilities in several countries, that are owned by various parties (each with their own funding sources), and that are collectively operated by the International LOFAR Telescope (ILT) foundation under a joint scientific policy. The Jülich LOFAR Long Term Archive and the German LOFAR network are both coordinated and operated by the Jülich Supercomputing Centre (JSC), and computing resources on the supercomputer JUWELS at JSC were provided by the Gauss Centre for Supercomputing e.V. (grant CHTB00) through the John von Neumann Institute for Computing (NIC).
This research made use of the Italian LOFAR IT computing infrastructure supported and operated by INAF, Italy.
The scientific results reported in this article are based on observations made by the \emph{Chandra} X-ray Observatory data obtained from the \emph{Chandra} Data Archive. This research has made
use of data and/or software provided by the High Energy Astrophysics Science Archive Research Center (HEASARC), which is a service of the Astrophysics Science Division at NASA/GSFC.
This research made use of different packages for Python: APLpy \citep{Robitaille2012}, Astropy \citep{Astropy2013}, NumPy \citep{vanderwalt2011} and SciPy \citep{Scipy2020}.

\bibliographystyle{aa}
\bibliography{biblio}

\begin{thebibliography}{118}
\expandafter\ifx\csname natexlab\endcsname\relax\def\natexlab#1{#1}\fi

\bibitem[{{Anders} \& {Grevesse}(1989)}]{Anders1989}
{Anders}, E. \& {Grevesse}, N. 1989, \gca, 53, 197

\bibitem[{{Astropy Collaboration} {et~al.}(2013){Astropy Collaboration}, {Robitaille}, {Tollerud}, {Greenfield}, {Droettboom}, {Bray}, {Aldcroft}, {Davis}, {Ginsburg}, {Price-Whelan}, {Kerzendorf}, {Conley}, {Crighton}, {Barbary}, {Muna}, {Ferguson}, {Grollier}, {Parikh}, {Nair}, {Unther}, {Deil}, {Woillez}, {Conseil}, {Kramer}, {Turner}, {Singer}, {Fox}, {Weaver}, {Zabalza}, {Edwards}, {Azalee Bostroem}, {Burke}, {Casey}, {Crawford}, {Dencheva}, {Ely}, {Jenness}, {Labrie}, {Lim}, {Pierfederici}, {Pontzen}, {Ptak}, {Refsdal}, {Servillat}, \& {Streicher}}]{Astropy2013}
{Astropy Collaboration}, {Robitaille}, T.~P., {Tollerud}, E.~J., {et~al.} 2013, \aap, 558, A33

\bibitem[{{Balboni} {et~al.}(2024){Balboni}, {Gastaldello}, {Bonafede}, {Botteon}, {Bartalucci}, {Bourdin}, {Brunetti}, {Cassano}, {De Grandi}, {De Luca}, {Gitti}, {Johnston-Hollitt}, {Mazzotta}, {Rossetti}, {Ettori}, {Ghizzardi}, {Iqbal}, {Lovisari}, {Molendi}, {Pointecoteau}, {Pratt}, {Riva}, {Rottgering}, {Sereno}, {van Weeren}, {Venturi}, \& {Veronesi}}]{Balboni2024}
{Balboni}, M., {Gastaldello}, F., {Bonafede}, A., {et~al.} 2024, arXiv e-prints, arXiv:2402.18654

\bibitem[{{Becker} {et~al.}(1995){Becker}, {White}, \& {Helfand}}]{Becker1995}
{Becker}, R.~H., {White}, R.~L., \& {Helfand}, D.~J. 1995, \apj, 450, 559

\bibitem[{{B{\'e}gin} {et~al.}(2023){B{\'e}gin}, {Hlavacek-Larrondo}, {Rhea}, {Gendron-Marsolais}, {McNamara}, {van Weeren}, {Richard-Laferri{\`e}re}, {Guit{\'e}}, {Prasow-{\'E}mond}, \& {Haggard}}]{Begin2023}
{B{\'e}gin}, T., {Hlavacek-Larrondo}, J., {Rhea}, C.~L., {et~al.} 2023, \mnras, 519, 767

\bibitem[{{Biava} {et~al.}(2021{\natexlab{a}}){Biava}, {Brienza}, {Bonafede}, {Gitti}, {Bonnassieux}, {Harwood}, {Edge}, {Riseley}, \& {Vantyghem}}]{Biava2021a}
{Biava}, N., {Brienza}, M., {Bonafede}, A., {et~al.} 2021{\natexlab{a}}, \aap, 650, A170

\bibitem[{{Biava} {et~al.}(2021{\natexlab{b}}){Biava}, {de Gasperin}, {Bonafede}, {Edler}, {Giacintucci}, {Mazzotta}, {Brunetti}, {Botteon}, {Br{\"u}ggen}, {Cassano}, {Drabent}, {Edge}, {En{\ss}lin}, {Gastaldello}, {Riseley}, {Rossetti}, {Rottgering}, {Shimwell}, {Tasse}, \& {van Weeren}}]{Biava2021b}
{Biava}, N., {de Gasperin}, F., {Bonafede}, A., {et~al.} 2021{\natexlab{b}}, \mnras, 508, 3995

\bibitem[{{B{\^\i}rzan} {et~al.}(2020){B{\^\i}rzan}, {Rafferty}, {Br{\"u}ggen}, {Botteon}, {Brunetti}, {Cuciti}, {Edge}, {Morganti}, {R{\"o}ttgering}, \& {Shimwell}}]{Birzan2020}
{B{\^\i}rzan}, L., {Rafferty}, D.~A., {Br{\"u}ggen}, M., {et~al.} 2020, \mnras, 496, 2613

\bibitem[{{Bonafede} {et~al.}(2022){Bonafede}, {Brunetti}, {Rudnick}, {Vazza}, {Bourdin}, {Giovannini}, {Shimwell}, {Zhang}, {Mazzotta}, {Simionescu}, {Biava}, {Bonnassieux}, {Brienza}, {Br{\"u}ggen}, {Rajpurohit}, {Riseley}, {Stuardi}, {Feretti}, {Tasse}, {Botteon}, {Carretti}, {Cassano}, {Cuciti}, {de Gasperin}, {Gastaldello}, {Rossetti}, {Rottgering}, {Venturi}, \& {van Weeren}}]{Bonafede2022}
{Bonafede}, A., {Brunetti}, G., {Rudnick}, L., {et~al.} 2022, \apj, 933, 218

\bibitem[{{Bonafede} {et~al.}(2023){Bonafede}, {Gitti}, {La Bella}, {Biava}, {Ubertosi}, {Brunetti}, {Lusetti}, {Brienza}, {Riseley}, {Stuardi}, {Botteon}, {Ignesti}, {R{\"o}ttgering}, \& {van Weeren}}]{Bonafede2023_RBS797}
{Bonafede}, A., {Gitti}, M., {La Bella}, N., {et~al.} 2023, \aap, 680, A5

\bibitem[{{Bonafede} {et~al.}(2014){Bonafede}, {Intema}, {Bruggen}, {Russell}, {Ogrean}, {Basu}, {Sommer}, {van Weeren}, {Cassano}, {Fabian}, \& {Rottgering}}]{Bonafede2014}
{Bonafede}, A., {Intema}, H.~T., {Bruggen}, M., {et~al.} 2014, \mnras, 444, L44

\bibitem[{{Botteon} {et~al.}(2020){Botteon}, {Brunetti}, {van Weeren}, {Shimwell}, {Pizzo}, {Cassano}, {Iacobelli}, {Gastaldello}, {B{\^\i}rzan}, {Bonafede}, {Br{\"u}ggen}, {Cuciti}, {Dallacasa}, {de Gasperin}, {Di Gennaro}, {Drabent}, {Hardcastle}, {Hoeft}, {Mandal}, {R{\"o}ttgering}, \& {Simionescu}}]{Botteon2020}
{Botteon}, A., {Brunetti}, G., {van Weeren}, R.~J., {et~al.} 2020, \apj, 897, 93

\bibitem[{{Botteon} {et~al.}(2022{\natexlab{a}}){Botteon}, {Shimwell}, {Cassano}, {Cuciti}, {Zhang}, {Bruno}, {Camillini}, {Natale}, {Jones}, {Gastaldello}, {Simionescu}, {Rossetti}, {Akamatsu}, {van Weeren}, {Brunetti}, {Br{\"u}ggen}, {Groeneveld}, {Hoang}, {Hardcastle}, {Ignesti}, {Di Gennaro}, {Bonafede}, {Drabent}, {R{\"o}ttgering}, {Hoeft}, \& {de Gasperin}}]{Botteon2022a}
{Botteon}, A., {Shimwell}, T.~W., {Cassano}, R., {et~al.} 2022{\natexlab{a}}, \aap, 660, A78

\bibitem[{{Botteon} {et~al.}(2022{\natexlab{b}}){Botteon}, {van Weeren}, {Brunetti}, {Vazza}, {Shimwell}, {Br{\"u}ggen}, {R{\"o}ttgering}, {de Gasperin}, {Akamatsu}, {Bonafede}, {Cassano}, {Cuciti}, {Dallacasa}, {Di Gennaro}, \& {Gastaldello}}]{Botteon2022b}
{Botteon}, A., {van Weeren}, R.~J., {Brunetti}, G., {et~al.} 2022{\natexlab{b}}, Science Advances, 8, eabq7623

\bibitem[{{Bravi} {et~al.}(2016){Bravi}, {Gitti}, \& {Brunetti}}]{Bravi2016}
{Bravi}, L., {Gitti}, M., \& {Brunetti}, G. 2016, \mnras, 455, L41

\bibitem[{{Brunetti} {et~al.}(2008){Brunetti}, {Giacintucci}, {Cassano}, {Lane}, {Dallacasa}, {Venturi}, {Kassim}, {Setti}, {Cotton}, \& {Markevitch}}]{Brunetti2008}
{Brunetti}, G., {Giacintucci}, S., {Cassano}, R., {et~al.} 2008, \nat, 455, 944

\bibitem[{{Brunetti} \& {Lazarian}(2016)}]{Brunetti2016}
{Brunetti}, G. \& {Lazarian}, A. 2016, \mnras, 458, 2584

\bibitem[{{Brunetti} {et~al.}(2001){Brunetti}, {Setti}, {Feretti}, \& {Giovannini}}]{Brunetti2001}
{Brunetti}, G., {Setti}, G., {Feretti}, L., \& {Giovannini}, G. 2001, \mnras, 320, 365

\bibitem[{{Bruno} {et~al.}(2023{\natexlab{a}}){Bruno}, {Botteon}, {Shimwell}, {Cuciti}, {de Gasperin}, {Brunetti}, {Dallacasa}, {Gastaldello}, {Rossetti}, {van Weeren}, {Venturi}, {Russo}, {Taffoni}, {Cassano}, {Biava}, {Lusetti}, {Bonafede}, {Ghizzardi}, \& {De Grandi}}]{Bruno2023b}
{Bruno}, L., {Botteon}, A., {Shimwell}, T., {et~al.} 2023{\natexlab{a}}, \aap, 678, A133

\bibitem[{{Bruno} {et~al.}(2023{\natexlab{b}}){Bruno}, {Brunetti}, {Botteon}, {Cuciti}, {Dallacasa}, {Cassano}, {van Weeren}, {Shimwell}, {Taffoni}, {Russo}, {Bonafede}, {Br{\"u}ggen}, {Hoang}, {Rottgering}, \& {Tasse}}]{Bruno2023a}
{Bruno}, L., {Brunetti}, G., {Botteon}, A., {et~al.} 2023{\natexlab{b}}, \aap, 672, A41

\bibitem[{{Buote}(2001)}]{Buote2001}
{Buote}, D.~A. 2001, \apjl, 553, L15

\bibitem[{{Cash}(1979)}]{Cash1979}
{Cash}, W. 1979, \apj, 228, 939

\bibitem[{{Cassano} {et~al.}(2016){Cassano}, {Brunetti}, {Giocoli}, \& {Ettori}}]{Cassano2016}
{Cassano}, R., {Brunetti}, G., {Giocoli}, C., \& {Ettori}, S. 2016, \aap, 593, A81

\bibitem[{{Cassano} {et~al.}(2006){Cassano}, {Brunetti}, \& {Setti}}]{Cassano2006}
{Cassano}, R., {Brunetti}, G., \& {Setti}, G. 2006, \mnras, 369, 1577

\bibitem[{{Cassano} {et~al.}(2023){Cassano}, {Cuciti}, {Brunetti}, {Botteon}, {Rossetti}, {Bruno}, {Simionescu}, {Gastaldello}, {van Weeren}, {Br{\"u}ggen}, {Dallacasa}, {Zhang}, {Akamatsu}, {Bonafede}, {Di Gennaro}, {Shimwell}, {de Gasperin}, {R{\"o}ttgering}, \& {Jones}}]{Cassano2023}
{Cassano}, R., {Cuciti}, V., {Brunetti}, G., {et~al.} 2023, \aap, 672, A43

\bibitem[{{Cassano} {et~al.}(2010){Cassano}, {Ettori}, {Giacintucci}, {Brunetti}, {Markevitch}, {Venturi}, \& {Gitti}}]{Cassano2010}
{Cassano}, R., {Ettori}, S., {Giacintucci}, S., {et~al.} 2010, \apjl, 721, L82

\bibitem[{{Cavagnolo} {et~al.}(2009){Cavagnolo}, {Donahue}, {Voit}, \& {Sun}}]{Cavagnolo2009}
{Cavagnolo}, K.~W., {Donahue}, M., {Voit}, G.~M., \& {Sun}, M. 2009, \apjs, 182, 12

\bibitem[{{Cohen} {et~al.}(2005){Cohen}, {Clarke}, {Feretti}, \& {Kassim}}]{Cohen2005}
{Cohen}, A.~S., {Clarke}, T.~E., {Feretti}, L., \& {Kassim}, N.~E. 2005, \apjl, 620, L5

\bibitem[{{Cuciti} {et~al.}(2021){Cuciti}, {Cassano}, {Brunetti}, {Dallacasa}, {de Gasperin}, {Ettori}, {Giacintucci}, {Kale}, {Pratt}, {van Weeren}, \& {Venturi}}]{Cuciti2021}
{Cuciti}, V., {Cassano}, R., {Brunetti}, G., {et~al.} 2021, \aap, 647, A51

\bibitem[{{Cuciti} {et~al.}(2023){Cuciti}, {Cassano}, {Sereno}, {Brunetti}, {Botteon}, {Shimwell}, {Bruno}, {Gastaldello}, {Rossetti}, {Zhang}, {Simionescu}, {Br{\"u}ggen}, {van Weeren}, {Jones}, {Akamatsu}, {Bonafede}, {De Gasperin}, {Di Gennaro}, {Pasini}, \& {R{\"o}ttgering}}]{Cuciti2023}
{Cuciti}, V., {Cassano}, R., {Sereno}, M., {et~al.} 2023, \aap, 680, A30

\bibitem[{{Cuciti} {et~al.}(2022){Cuciti}, {de Gasperin}, {Br{\"u}ggen}, {Vazza}, {Brunetti}, {Shimwell}, {Edler}, {van Weeren}, {Botteon}, {Cassano}, {Di Gennaro}, {Gastaldello}, {Drabent}, {R{\"o}ttgering}, \& {Tasse}}]{Cuciti2022}
{Cuciti}, V., {de Gasperin}, F., {Br{\"u}ggen}, M., {et~al.} 2022, \nat, 609, 911

\bibitem[{{de Gasperin} {et~al.}(2021){de Gasperin}, {Williams}, {Best}, {Br{\"u}ggen}, {Brunetti}, {Cuciti}, {Dijkema}, {Hardcastle}, {Norden}, {Offringa}, {Shimwell}, {van Weeren}, {Bomans}, {Bonafede}, {Botteon}, {Callingham}, {Cassano}, {Chy{\.z}y}, {Emig}, {Edler}, {Haverkorn}, {Heald}, {Heesen}, {Iacobelli}, {Intema}, {Kadler}, {Ma{\l}ek}, {Mevius}, {Miley}, {Mingo}, {Morabito}, {Sabater}, {Morganti}, {Orr{\'u}}, {Pizzo}, {Prandoni}, {Shulevski}, {Tasse}, {Vaccari}, {Zarka}, \& {R{\"o}ttgering}}]{deGasperin2021}
{de Gasperin}, F., {Williams}, W.~L., {Best}, P., {et~al.} 2021, \aap, 648, A104

\bibitem[{{Doria} {et~al.}(2012){Doria}, {Gitti}, {Ettori}, {Brighenti}, {Nulsen}, \& {McNamara}}]{Doria2012}
{Doria}, A., {Gitti}, M., {Ettori}, S., {et~al.} 2012, \apj, 753, 47

\bibitem[{{Ebeling} {et~al.}(2010){Ebeling}, {Edge}, {Mantz}, {Barrett}, {Henry}, {Ma}, \& {van Speybroeck}}]{Ebeling2010}
{Ebeling}, H., {Edge}, A.~C., {Mantz}, A., {et~al.} 2010, \mnras, 407, 83

\bibitem[{{Eckert} {et~al.}(2020){Eckert}, {Finoguenov}, {Ghirardini}, {Grandis}, {Kaefer}, {Sanders}, \& {Ramos-Ceja}}]{Eckert2020}
{Eckert}, D., {Finoguenov}, A., {Ghirardini}, V., {et~al.} 2020, The Open Journal of Astrophysics, 3, 12

\bibitem[{{Edler} {et~al.}(2022){Edler}, {de Gasperin}, {Brunetti}, {Botteon}, {Cuciti}, {van Weeren}, {Cassano}, {Shimwell}, {Br{\"u}ggen}, \& {Drabent}}]{Edler2022}
{Edler}, H.~W., {de Gasperin}, F., {Brunetti}, G., {et~al.} 2022, \aap, 666, A3

\bibitem[{{Gastaldello} {et~al.}(2013){Gastaldello}, {Di Gesu}, {Ghizzardi}, {Giacintucci}, {Girardi}, {Roediger}, {Rossetti}, {Brighenti}, {Buote}, {Eckert}, {Ettori}, {Humphrey}, \& {Mathews}}]{Gastaldello2013}
{Gastaldello}, F., {Di Gesu}, L., {Ghizzardi}, S., {et~al.} 2013, \apj, 770, 56

\bibitem[{{Giacintucci} {et~al.}(2014{\natexlab{a}}){Giacintucci}, {Markevitch}, {Brunetti}, {ZuHone}, {Venturi}, {Mazzotta}, \& {Bourdin}}]{Giacintucci2014a}
{Giacintucci}, S., {Markevitch}, M., {Brunetti}, G., {et~al.} 2014{\natexlab{a}}, \apj, 795, 73

\bibitem[{{Giacintucci} {et~al.}(2017){Giacintucci}, {Markevitch}, {Cassano}, {Venturi}, {Clarke}, \& {Brunetti}}]{Giacintucci2017}
{Giacintucci}, S., {Markevitch}, M., {Cassano}, R., {et~al.} 2017, \apj, 841, 71

\bibitem[{{Giacintucci} {et~al.}(2019){Giacintucci}, {Markevitch}, {Cassano}, {Venturi}, {Clarke}, {Kale}, \& {Cuciti}}]{Giacintucci2019}
{Giacintucci}, S., {Markevitch}, M., {Cassano}, R., {et~al.} 2019, \apj, 880, 70

\bibitem[{{Giacintucci} {et~al.}(2014{\natexlab{b}}){Giacintucci}, {Markevitch}, {Venturi}, {Clarke}, {Cassano}, \& {Mazzotta}}]{Giacintucci2014b}
{Giacintucci}, S., {Markevitch}, M., {Venturi}, T., {et~al.} 2014{\natexlab{b}}, \apj, 781, 9

\bibitem[{{Giacintucci} {et~al.}(2024){Giacintucci}, {Venturi}, {Markevitch}, {Brunetti}, {Clarke}, \& {Kale}}]{Giacintucci2024}
{Giacintucci}, S., {Venturi}, T., {Markevitch}, M., {et~al.} 2024, \apj, 961, 133

\bibitem[{{Gitti} {et~al.}(2006){Gitti}, {Feretti}, \& {Schindler}}]{Gitti2006}
{Gitti}, M., {Feretti}, L., \& {Schindler}, S. 2006, \aap, 448, 853

\bibitem[{{Govoni} {et~al.}(2001){Govoni}, {Feretti}, {Giovannini}, {B{\"o}hringer}, {Reiprich}, \& {Murgia}}]{Govoni2001}
{Govoni}, F., {Feretti}, L., {Giovannini}, G., {et~al.} 2001, \aap, 376, 803

\bibitem[{{Govoni} {et~al.}(2009){Govoni}, {Murgia}, {Markevitch}, {Feretti}, {Giovannini}, {Taylor}, \& {Carretti}}]{Govoni2009}
{Govoni}, F., {Murgia}, M., {Markevitch}, M., {et~al.} 2009, \aap, 499, 371

\bibitem[{{Hao} {et~al.}(2010){Hao}, {McKay}, {Koester}, {Rykoff}, {Rozo}, {Annis}, {Wechsler}, {Evrard}, {Siegel}, {Becker}, {Busha}, {Gerdes}, {Johnston}, \& {Sheldon}}]{Hao2010}
{Hao}, J., {McKay}, T.~A., {Koester}, B.~P., {et~al.} 2010, \apjs, 191, 254

\bibitem[{{Hardcastle} {et~al.}(2021){Hardcastle}, {Shimwell}, {Tasse}, {Best}, {Drabent}, {Jarvis}, {Prandoni}, {R{\"o}ttgering}, {Sabater}, \& {Schwarz}}]{Hardcastle2021}
{Hardcastle}, M.~J., {Shimwell}, T.~W., {Tasse}, C., {et~al.} 2021, \aap, 648, A10

\bibitem[{{HI4PI Collaboration} {et~al.}(2016){HI4PI Collaboration}, {Ben Bekhti}, {Fl{\"o}er}, {Keller}, {Kerp}, {Lenz}, {Winkel}, {Bailin}, {Calabretta}, {Dedes}, {Ford}, {Gibson}, {Haud}, {Janowiecki}, {Kalberla}, {Lockman}, {McClure-Griffiths}, {Murphy}, {Nakanishi}, {Pisano}, \& {Staveley-Smith}}]{HI4PI2016}
{HI4PI Collaboration}, {Ben Bekhti}, N., {Fl{\"o}er}, L., {et~al.} 2016, \aap, 594, A116

\bibitem[{{Hlavacek-Larrondo} {et~al.}(2013){Hlavacek-Larrondo}, {Allen}, {Taylor}, {Fabian}, {Canning}, {Werner}, {Sanders}, {Grimes}, {Ehlert}, \& {von der Linden}}]{Hlavacek2013}
{Hlavacek-Larrondo}, J., {Allen}, S.~W., {Taylor}, G.~B., {et~al.} 2013, \apj, 777, 163

\bibitem[{{Hlavacek-Larrondo} {et~al.}(2012){Hlavacek-Larrondo}, {Fabian}, {Edge}, {Ebeling}, {Sanders}, {Hogan}, \& {Taylor}}]{Hlavacek2012}
{Hlavacek-Larrondo}, J., {Fabian}, A.~C., {Edge}, A.~C., {et~al.} 2012, \mnras, 421, 1360

\bibitem[{{Ignesti} {et~al.}(2020){Ignesti}, {Brunetti}, {Gitti}, \& {Giacintucci}}]{Ignesti2020}
{Ignesti}, A., {Brunetti}, G., {Gitti}, M., \& {Giacintucci}, S. 2020, \aap, 640, A37

\bibitem[{{Kale} {et~al.}(2019){Kale}, {Shende}, \& {Parekh}}]{Kale2019}
{Kale}, R., {Shende}, K.~M., \& {Parekh}, V. 2019, \mnras, 486, L80

\bibitem[{{Kravtsov} \& {Borgani}(2012)}]{Borgani2012}
{Kravtsov}, A.~V. \& {Borgani}, S. 2012, \araa, 50, 353

\bibitem[{{Lovisari} {et~al.}(2017){Lovisari}, {Forman}, {Jones}, {Ettori}, {Andrade-Santos}, {Arnaud}, {D{\'e}mocl{\`e}s}, {Pratt}, {Randall}, \& {Kraft}}]{Lovisari2017}
{Lovisari}, L., {Forman}, W.~R., {Jones}, C., {et~al.} 2017, \apj, 846, 51

\bibitem[{{Lusetti} {et~al.}(2023){Lusetti}, {Bonafede}, {Lovisari}, {Gitti}, {Ettori}, {Cassano}, {Riseley}, {Govoni}, {Br{\"u}ggen}, {Bruno}, {van Weeren}, {Botteon}, {Hoang}, {Gastaldello}, {Ignesti}, {Rossetti}, \& {Shimwell}}]{Lusetti2023}
{Lusetti}, G., {Bonafede}, A., {Lovisari}, L., {et~al.} 2023, arXiv e-prints, arXiv:2308.01884

\bibitem[{{Mann} \& {Ebeling}(2012)}]{Mann2012}
{Mann}, A.~W. \& {Ebeling}, H. 2012, \mnras, 420, 2120

\bibitem[{{Markevitch} \& {Vikhlinin}(2007)}]{Markevitch2007}
{Markevitch}, M. \& {Vikhlinin}, A. 2007, \physrep, 443, 1

\bibitem[{{Maughan} {et~al.}(2008){Maughan}, {Jones}, {Forman}, \& {Van Speybroeck}}]{Maughan2008}
{Maughan}, B.~J., {Jones}, C., {Forman}, W., \& {Van Speybroeck}, L. 2008, \apjs, 174, 117

\bibitem[{{Mazzotta} \& {Giacintucci}(2008)}]{Mazzotta2008}
{Mazzotta}, P. \& {Giacintucci}, S. 2008, \apjl, 675, L9

\bibitem[{{Mazzotta} {et~al.}(2001){Mazzotta}, {Markevitch}, {Vikhlinin}, {Forman}, {David}, \& {van Speybroeck}}]{Mazzotta2001}
{Mazzotta}, P., {Markevitch}, M., {Vikhlinin}, A., {et~al.} 2001, \apj, 555, 205

\bibitem[{{McNamara} {et~al.}(2005){McNamara}, {Nulsen}, {Wise}, {Rafferty}, {Carilli}, {Sarazin}, \& {Blanton}}]{McNamara2005}
{McNamara}, B.~R., {Nulsen}, P.~E.~J., {Wise}, M.~W., {et~al.} 2005, \nat, 433, 45

\bibitem[{{McNamara} {et~al.}(2004){McNamara}, {Wise}, \& {Murray}}]{McNamara2004}
{McNamara}, B.~R., {Wise}, M.~W., \& {Murray}, S.~S. 2004, \apj, 601, 173

\bibitem[{{Murgia} {et~al.}(2009){Murgia}, {Govoni}, {Markevitch}, {Feretti}, {Giovannini}, {Taylor}, \& {Carretti}}]{Murgia2009}
{Murgia}, M., {Govoni}, F., {Markevitch}, M., {et~al.} 2009, \aap, 499, 679

\bibitem[{{Offringa} {et~al.}(2014){Offringa}, {McKinley}, {Hurley-Walker}, {Briggs}, {Wayth}, {Kaplan}, {Bell}, {Feng}, {Neben}, {Hughes}, {Rhee}, {Murphy}, {Bhat}, {Bernardi}, {Bowman}, {Cappallo}, {Corey}, {Deshpand e}, {Emrich}, {Ewall-Wice}, {Gaensler}, {Goeke}, {Greenhill}, {Hazelton}, {Hindson}, {Johnston-Hollitt}, {Jacobs}, {Kasper}, {Kratzenberg}, {Lenc}, {Lonsdale}, {Lynch}, {McWhirter}, {Mitchell}, {Morales}, {Morgan}, {Kudryavtseva}, {Oberoi}, {Ord}, {Pindor}, {Procopio}, {Prabu}, {Riding}, {Roshi}, {Shankar}, {Srivani}, {Subrahmanyan}, {Tingay}, {Waterson}, {Webster}, {Whitney}, {Williams}, \& {Williams}}]{Offringa2014}
{Offringa}, A.~R., {McKinley}, B., {Hurley-Walker}, N., {et~al.} 2014, \mnras, 444, 606

\bibitem[{{Owers} {et~al.}(2011){Owers}, {Nulsen}, \& {Couch}}]{Owers2011}
{Owers}, M.~S., {Nulsen}, P. E.~J., \& {Couch}, W.~J. 2011, \apj, 741, 122

\bibitem[{{Petrosian}(2001)}]{Petrosian2001}
{Petrosian}, V. 2001, \apj, 557, 560

\bibitem[{{Pfrommer} \& {En{\ss}lin}(2004)}]{Pfrommer2004}
{Pfrommer}, C. \& {En{\ss}lin}, T.~A. 2004, Journal of Korean Astronomical Society, 37, 455

\bibitem[{{Planck Collaboration} {et~al.}(2016){Planck Collaboration}, {Ade}, {Aghanim}, {Arnaud}, {Ashdown}, {Aumont}, {Baccigalupi}, {Banday}, {Barreiro}, {Barrena}, {Bartlett}, {Bartolo}, {Battaner}, {Battye}, {Benabed}, {Beno{\^\i}t}, {Benoit-L{\'e}vy}, {Bernard}, {Bersanelli}, {Bielewicz}, {Bikmaev}, {B{\"o}hringer}, {Bonaldi}, {Bonavera}, {Bond}, {Borrill}, {Bouchet}, {Bucher}, {Burenin}, {Burigana}, {Butler}, {Calabrese}, {Cardoso}, {Carvalho}, {Catalano}, {Challinor}, {Chamballu}, {Chary}, {Chiang}, {Chon}, {Christensen}, {Clements}, {Colombi}, {Colombo}, {Combet}, {Comis}, {Couchot}, {Coulais}, {Crill}, {Curto}, {Cuttaia}, {Dahle}, {Danese}, {Davies}, {Davis}, {de Bernardis}, {de Rosa}, {de Zotti}, {Delabrouille}, {D{\'e}sert}, {Dickinson}, {Diego}, {Dolag}, {Dole}, {Donzelli}, {Dor{\'e}}, {Douspis}, {Ducout}, {Dupac}, {Efstathiou}, {Eisenhardt}, {Elsner}, {En{\ss}lin}, {Eriksen}, {Falgarone}, {Fergusson}, {Feroz}, {Ferragamo}, {Finelli}, {Forni}, {Frailis}, {Fraisse}, {Franceschi}, {Frejsel},
  {Galeotta}, {Galli}, {Ganga}, {G{\'e}nova-Santos}, {Giard}, {Giraud-H{\'e}raud}, {Gjerl{\o}w}, {Gonz{\'a}lez-Nuevo}, {G{\'o}rski}, {Grainge}, {Gratton}, {Gregorio}, {Gruppuso}, {Gudmundsson}, {Hansen}, {Hanson}, {Harrison}, {Hempel}, {Henrot-Versill{\'e}}, {Hern{\'a}ndez-Monteagudo}, {Herranz}, {Hildebrandt}, {Hivon}, {Hobson}, {Holmes}, {Hornstrup}, {Hovest}, {Huffenberger}, {Hurier}, {Jaffe}, {Jaffe}, {Jin}, {Jones}, {Juvela}, {Keih{\"a}nen}, {Keskitalo}, {Khamitov}, {Kisner}, {Kneissl}, {Knoche}, {Kunz}, {Kurki-Suonio}, {Lagache}, {Lamarre}, {Lasenby}, {Lattanzi}, {Lawrence}, {Leonardi}, {Lesgourgues}, {Levrier}, {Liguori}, {Lilje}, {Linden-V{\o}rnle}, {L{\'o}pez-Caniego}, {Lubin}, {Mac{\'\i}as-P{\'e}rez}, {Maggio}, {Maino}, {Mak}, {Mandolesi}, {Mangilli}, {Martin}, {Mart{\'\i}nez-Gonz{\'a}lez}, {Masi}, {Matarrese}, {Mazzotta}, {McGehee}, {Mei}, {Melchiorri}, {Melin}, {Mendes}, {Mennella}, {Migliaccio}, {Mitra}, {Miville-Desch{\^e}nes}, {Moneti}, {Montier}, {Morgante}, {Mortlock}, {Moss}, {Munshi},
  {Murphy}, {Naselsky}, {Nastasi}, {Nati}, {Natoli}, {Netterfield}, {N{\o}rgaard-Nielsen}, {Noviello}, {Novikov}, {Novikov}, {Olamaie}, {Oxborrow}, {Paci}, {Pagano}, {Pajot}, {Paoletti}, {Pasian}, {Patanchon}, {Pearson}, {Perdereau}, {Perotto}, {Perrott}, {Perrotta}, {Pettorino}, {Piacentini}, {Piat}, {Pierpaoli}, {Pietrobon}, {Plaszczynski}, {Pointecouteau}, {Polenta}, {Pratt}, {Pr{\'e}zeau}, {Prunet}, {Puget}, {Rachen}, {Reach}, {Rebolo}, {Reinecke}, {Remazeilles}, {Renault}, {Renzi}, {Ristorcelli}, {Rocha}, {Rosset}, {Rossetti}, {Roudier}, {Rozo}, {Rubi{\~n}o-Mart{\'\i}n}, {Rumsey}, {Rusholme}, {Rykoff}, {Sandri}, {Santos}, {Saunders}, {Savelainen}, {Savini}, {Schammel}, {Scott}, {Seiffert}, {Shellard}, {Shimwell}, {Spencer}, {Stanford}, {Stern}, {Stolyarov}, {Stompor}, {Streblyanska}, {Sudiwala}, {Sunyaev}, {Sutton}, {Suur-Uski}, {Sygnet}, {Tauber}, {Terenzi}, {Toffolatti}, {Tomasi}, {Tramonte}, {Tristram}, {Tucci}, {Tuovinen}, {Umana}, {Valenziano}, {Valiviita}, {Van Tent}, {Vielva}, {Villa}, {Wade},
  {Wandelt}, {Wehus}, {White}, {Wright}, {Yvon}, {Zacchei}, \& {Zonca}}]{Planck2016}
{Planck Collaboration}, {Ade}, P.~A.~R., {Aghanim}, N., {et~al.} 2016, \aap, 594, A27

\bibitem[{{Poole} {et~al.}(2006){Poole}, {Fardal}, {Babul}, {McCarthy}, {Quinn}, \& {Wadsley}}]{Poole2006}
{Poole}, G.~B., {Fardal}, M.~A., {Babul}, A., {et~al.} 2006, \mnras, 373, 881

\bibitem[{{Raja} {et~al.}(2020){Raja}, {Rahaman}, {Datta}, {Burns}, {Intema}, {van Weeren}, {Hallman}, {Rapetti}, \& {Paul}}]{Raja2020}
{Raja}, R., {Rahaman}, M., {Datta}, A., {et~al.} 2020, \mnras, 493, L28

\bibitem[{{Rajpurohit} {et~al.}(2021{\natexlab{a}}){Rajpurohit}, {Brunetti}, {Bonafede}, {van Weeren}, {Botteon}, {Vazza}, {Hoeft}, {Riseley}, {Bonnassieux}, {Brienza}, {Forman}, {R{\"o}ttgering}, {Rajpurohit}, {Locatelli}, {Shimwell}, {Cassano}, {Di Gennaro}, {Br{\"u}ggen}, {Wittor}, {Drabent}, \& {Ignesti}}]{Rajpurohit2021b}
{Rajpurohit}, K., {Brunetti}, G., {Bonafede}, A., {et~al.} 2021{\natexlab{a}}, \aap, 646, A135

\bibitem[{{Rajpurohit} {et~al.}(2023){Rajpurohit}, {Osinga}, {Brienza}, {Botteon}, {Brunetti}, {Forman}, {Riseley}, {Vazza}, {Bonafede}, {van Weeren}, {Br{\"u}ggen}, {Rajpurohit}, {Drabent}, {Dallacasa}, {Rossetti}, {Rajpurohit}, {Hoeft}, {Bonnassieux}, {Cassano}, \& {Miley}}]{Rajpurohit2023}
{Rajpurohit}, K., {Osinga}, E., {Brienza}, M., {et~al.} 2023, \aap, 669, A1

\bibitem[{{Rajpurohit} {et~al.}(2021{\natexlab{b}}){Rajpurohit}, {Vazza}, {van Weeren}, {Hoeft}, {Brienza}, {Bonnassieux}, {Riseley}, {Brunetti}, {Bonafede}, {Br{\"u}ggen}, {Formann}, {Rajpurohit}, {R{\"o}ttgering}, {Drabent}, {Dom{\'\i}nguez-Fern{\'a}ndez}, {Wittor}, \& {Andrade-Santos}}]{Rajpurohit2021a}
{Rajpurohit}, K., {Vazza}, F., {van Weeren}, R.~J., {et~al.} 2021{\natexlab{b}}, \aap, 654, A41

\bibitem[{{Richard-Laferri{\`e}re} {et~al.}(2020){Richard-Laferri{\`e}re}, {Hlavacek-Larrondo}, {Nemmen}, {Rhea}, {Taylor}, {Prasow-{\'E}mond}, {Gendron-Marsolais}, {Latulippe}, {Edge}, {Fabian}, {Sanders}, {Hogan}, \& {Demontigny}}]{RL2020}
{Richard-Laferri{\`e}re}, A., {Hlavacek-Larrondo}, J., {Nemmen}, R.~S., {et~al.} 2020, \mnras, 499, 2934

\bibitem[{{Riseley} {et~al.}(2023){Riseley}, {Biava}, {Lusetti}, {Bonafede}, {Bonnassieux}, {Botteon}, {Loi}, {Brunetti}, {Cassano}, {Osinga}, {Rajpurohit}, {R{\"o}ttgering}, {Shimwell}, {Timmerman}, \& {van Weeren}}]{Riseley2023}
{Riseley}, C.~J., {Biava}, N., {Lusetti}, G., {et~al.} 2023, \mnras, 524, 6052

\bibitem[{{Riseley} {et~al.}(2024){Riseley}, {Bonafede}, {Bruno}, {Botteon}, {Rossetti}, {Biava}, {Bonnassieux}, {Loi}, {Vernstrom}, \& {Balboni}}]{Riseley2024}
{Riseley}, C.~J., {Bonafede}, A., {Bruno}, L., {et~al.} 2024, arXiv e-prints, arXiv:2403.00414

\bibitem[{{Riseley} {et~al.}(2022{\natexlab{a}}){Riseley}, {Bonnassieux}, {Vernstrom}, {Galvin}, {Chokshi}, {Botteon}, {Rajpurohit}, {Duchesne}, {Bonafede}, {Rudnick}, {Hoeft}, {Quici}, {Eckert}, {Brienza}, {Tasse}, {Carretti}, {Collier}, {Diego}, {Di Mascolo}, {Hopkins}, {Johnston-Hollitt}, {Keel}, {Koribalski}, \& {Reiprich}}]{Riseley2022b}
{Riseley}, C.~J., {Bonnassieux}, E., {Vernstrom}, T., {et~al.} 2022{\natexlab{a}}, \mnras, 515, 1871

\bibitem[{{Riseley} {et~al.}(2022{\natexlab{b}}){Riseley}, {Rajpurohit}, {Loi}, {Botteon}, {Timmerman}, {Biava}, {Bonafede}, {Bonnassieux}, {Brunetti}, {En{\ss}lin}, {Di Gennaro}, {Ignesti}, {Shimwell}, {Stuardi}, {Vernstrom}, \& {van Weeren}}]{Riseley2022}
{Riseley}, C.~J., {Rajpurohit}, K., {Loi}, F., {et~al.} 2022{\natexlab{b}}, \mnras, 512, 4210

\bibitem[{{Robitaille} \& {Bressert}(2012)}]{Robitaille2012}
{Robitaille}, T. \& {Bressert}, E. 2012, {APLpy: Astronomical Plotting Library in Python}

\bibitem[{{Roediger} {et~al.}(2011){Roediger}, {Br{\"u}ggen}, {Simionescu}, {B{\"o}hringer}, {Churazov}, \& {Forman}}]{Roediger2011}
{Roediger}, E., {Br{\"u}ggen}, M., {Simionescu}, A., {et~al.} 2011, \mnras, 413, 2057

\bibitem[{{Rossetti} {et~al.}(2017){Rossetti}, {Gastaldello}, {Eckert}, {Della Torre}, {Pantiri}, {Cazzoletti}, \& {Molendi}}]{Rossetti2017}
{Rossetti}, M., {Gastaldello}, F., {Eckert}, D., {et~al.} 2017, \mnras, 468, 1917

\bibitem[{{Sanders} {et~al.}(2016){Sanders}, {Fabian}, {Russell}, {Walker}, \& {Blundell}}]{Sanders2016}
{Sanders}, J.~S., {Fabian}, A.~C., {Russell}, H.~R., {Walker}, S.~A., \& {Blundell}, K.~M. 2016, \mnras, 460, 1898

\bibitem[{{Santos} {et~al.}(2008){Santos}, {Rosati}, {Tozzi}, {B{\"o}hringer}, {Ettori}, \& {Bignamini}}]{Santos2008}
{Santos}, J.~S., {Rosati}, P., {Tozzi}, P., {et~al.} 2008, \aap, 483, 35

\bibitem[{{Savini} {et~al.}(2019){Savini}, {Bonafede}, {Br{\"u}ggen}, {Rafferty}, {Shimwell}, {Botteon}, {Brunetti}, {Intema}, {Wilber}, {Cassano}, {Vazza}, {van Weeren}, {Cuciti}, {De Gasperin}, {R{\"o}ttgering}, {Sommer}, {B{\^\i}rzan}, \& {Drabent}}]{Savini2019}
{Savini}, F., {Bonafede}, A., {Br{\"u}ggen}, M., {et~al.} 2019, \aap, 622, A24

\bibitem[{{Savini} {et~al.}(2018){Savini}, {Bonafede}, {Br{\"u}ggen}, {van Weeren}, {Brunetti}, {Intema}, {Botteon}, {Shimwell}, {Wilber}, {Rafferty}, {Giacintucci}, {Cassano}, {Cuciti}, {de Gasperin}, {R{\"o}ttgering}, {Hoeft}, \& {White}}]{Savini2018}
{Savini}, F., {Bonafede}, A., {Br{\"u}ggen}, M., {et~al.} 2018, \mnras, 478, 2234

\bibitem[{{Scaife} \& {Heald}(2012)}]{ScaifeHeald2012}
{Scaife}, A. M.~M. \& {Heald}, G.~H. 2012, \mnras, 423, L30

\bibitem[{{Schindler} {et~al.}(2001){Schindler}, {Castillo-Morales}, {De Filippis}, {Schwope}, \& {Wambsganss}}]{Schindler2001}
{Schindler}, S., {Castillo-Morales}, A., {De Filippis}, E., {Schwope}, A., \& {Wambsganss}, J. 2001, \aap, 376, L27

\bibitem[{{Shimwell} {et~al.}(2022){Shimwell}, {Hardcastle}, {Tasse}, {Best}, {R{\"o}ttgering}, {Williams}, {Botteon}, {Drabent}, {Mechev}, {Shulevski}, {van Weeren}, {Bester}, {Br{\"u}ggen}, {Brunetti}, {Callingham}, {Chy{\.z}y}, {Conway}, {Dijkema}, {Duncan}, {de Gasperin}, {Hale}, {Haverkorn}, {Hugo}, {Jackson}, {Mevius}, {Miley}, {Morabito}, {Morganti}, {Offringa}, {Oonk}, {Rafferty}, {Sabater}, {Smith}, {Schwarz}, {Smirnov}, {O'Sullivan}, {Vedantham}, {White}, {Albert}, {Alegre}, {Asabere}, {Bacon}, {Bonafede}, {Bonnassieux}, {Brienza}, {Bilicki}, {Bonato}, {Calistro Rivera}, {Cassano}, {Cochrane}, {Croston}, {Cuciti}, {Dallacasa}, {Danezi}, {Dettmar}, {Di Gennaro}, {Edler}, {En{\ss}lin}, {Emig}, {Franzen}, {Garc{\'\i}a-Vergara}, {Grange}, {G{\"u}rkan}, {Hajduk}, {Heald}, {Heesen}, {Hoang}, {Hoeft}, {Horellou}, {Iacobelli}, {Jamrozy}, {Jeli{\'c}}, {Kondapally}, {Kukreti}, {Kunert-Bajraszewska}, {Magliocchetti}, {Mahatma}, {Ma{\l}ek}, {Mandal}, {Massaro}, {Meyer-Zhao}, {Mingo}, {Mostert}, {Nair},
  {Nakoneczny}, {Nikiel-Wroczy{\'n}ski}, {Orr{\'u}}, {Pajdosz-{\'S}mierciak}, {Pasini}, {Prandoni}, {van Piggelen}, {Rajpurohit}, {Retana-Montenegro}, {Riseley}, {Rowlinson}, {Saxena}, {Schrijvers}, {Sweijen}, {Siewert}, {Timmerman}, {Vaccari}, {Vink}, {West}, {Wo{\l}owska}, {Zhang}, \& {Zheng}}]{Shimwell2022}
{Shimwell}, T.~W., {Hardcastle}, M.~J., {Tasse}, C., {et~al.} 2022, \aap, 659, A1

\bibitem[{{Shimwell} {et~al.}(2017){Shimwell}, {R{\"o}ttgering}, {Best}, {Williams}, {Dijkema}, {de Gasperin}, {Hardcastle}, {Heald}, {Hoang}, {Horneffer}, {Intema}, {Mahony}, {Mandal}, {Mechev}, {Morabito}, {Oonk}, {Rafferty}, {Retana-Montenegro}, {Sabater}, {Tasse}, {van Weeren}, {Br{\"u}ggen}, {Brunetti}, {Chy{\.z}y}, {Conway}, {Haverkorn}, {Jackson}, {Jarvis}, {McKean}, {Miley}, {Morganti}, {White}, {Wise}, {van Bemmel}, {Beck}, {Brienza}, {Bonafede}, {Calistro Rivera}, {Cassano}, {Clarke}, {Cseh}, {Deller}, {Drabent}, {van Driel}, {Engels}, {Falcke}, {Ferrari}, {Fr{\"o}hlich}, {Garrett}, {Harwood}, {Heesen}, {Hoeft}, {Horellou}, {Israel}, {Kapi{\'n}ska}, {Kunert-Bajraszewska}, {McKay}, {Mohan}, {Orr{\'u}}, {Pizzo}, {Prandoni}, {Schwarz}, {Shulevski}, {Sipior}, {Smith}, {Sridhar}, {Steinmetz}, {Stroe}, {Varenius}, {van der Werf}, {Zensus}, \& {Zwart}}]{Shimwell2017}
{Shimwell}, T.~W., {R{\"o}ttgering}, H.~J.~A., {Best}, P.~N., {et~al.} 2017, \aap, 598, A104

\bibitem[{{Shimwell} {et~al.}(2019){Shimwell}, {Tasse}, {Hardcastle}, {Mechev}, {Williams}, {Best}, {R{\"o}ttgering}, {Callingham}, {Dijkema}, {de Gasperin}, {Hoang}, {Hugo}, {Mirmont}, {Oonk}, {Prandoni}, {Rafferty}, {Sabater}, {Smirnov}, {van Weeren}, {White}, {Atemkeng}, {Bester}, {Bonnassieux}, {Br{\"u}ggen}, {Brunetti}, {Chy{\.z}y}, {Cochrane}, {Conway}, {Croston}, {Danezi}, {Duncan}, {Haverkorn}, {Heald}, {Iacobelli}, {Intema}, {Jackson}, {Jamrozy}, {Jarvis}, {Lakhoo}, {Mevius}, {Miley}, {Morabito}, {Morganti}, {Nisbet}, {Orr{\'u}}, {Perkins}, {Pizzo}, {Schrijvers}, {Smith}, {Vermeulen}, {Wise}, {Alegre}, {Bacon}, {van Bemmel}, {Beswick}, {Bonafede}, {Botteon}, {Bourke}, {Brienza}, {Calistro Rivera}, {Cassano}, {Clarke}, {Conselice}, {Dettmar}, {Drabent}, {Dumba}, {Emig}, {En{\ss}lin}, {Ferrari}, {Garrett}, {G{\'e}nova-Santos}, {Goyal}, {G{\"u}rkan}, {Hale}, {Harwood}, {Heesen}, {Hoeft}, {Horellou}, {Jackson}, {Kokotanekov}, {Kondapally}, {Kunert-Bajraszewska}, {Mahatma}, {Mahony}, {Mandal}, {McKean},
  {Merloni}, {Mingo}, {Miskolczi}, {Mooney}, {Nikiel-Wroczy{\'n}ski}, {O'Sullivan}, {Quinn}, {Reich}, {Roskowi{\'n}ski}, {Rowlinson}, {Savini}, {Saxena}, {Schwarz}, {Shulevski}, {Sridhar}, {Stacey}, {Urquhart}, {van der Wiel}, {Varenius}, {Webster}, \& {Wilber}}]{Shimwell2019}
{Shimwell}, T.~W., {Tasse}, C., {Hardcastle}, M.~J., {et~al.} 2019, \aap, 622, A1

\bibitem[{{Smirnov} \& {Tasse}(2015)}]{SmirnovTasse2015}
{Smirnov}, O.~M. \& {Tasse}, C. 2015, \mnras, 449, 2668

\bibitem[{{Sommer} {et~al.}(2017){Sommer}, {Basu}, {Intema}, {Pacaud}, {Bonafede}, {Babul}, \& {Bertoldi}}]{Sommer2017}
{Sommer}, M.~W., {Basu}, K., {Intema}, H., {et~al.} 2017, \mnras, 466, 996

\bibitem[{{Tasse}(2014{\natexlab{a}})}]{Tasse2014a}
{Tasse}, C. 2014{\natexlab{a}}, arXiv e-prints, arXiv:1410.8706

\bibitem[{{Tasse}(2014{\natexlab{b}})}]{Tasse2014b}
{Tasse}, C. 2014{\natexlab{b}}, \aap, 566, A127

\bibitem[{{Tasse} {et~al.}(2018){Tasse}, {Hugo}, {Mirmont}, {Smirnov}, {Atemkeng}, {Bester}, {Hardcastle}, {Lakhoo}, {Perkins}, \& {Shimwell}}]{Tasse2018}
{Tasse}, C., {Hugo}, B., {Mirmont}, M., {et~al.} 2018, \aap, 611, A87

\bibitem[{{Tasse} {et~al.}(2021){Tasse}, {Shimwell}, {Hardcastle}, {O'Sullivan}, {van Weeren}, {Best}, {Bester}, {Hugo}, {Smirnov}, {Sabater}, {Calistro-Rivera}, {de Gasperin}, {Morabito}, {R{\"o}ttgering}, {Williams}, {Bonato}, {Bondi}, {Botteon}, {Br{\"u}ggen}, {Brunetti}, {Chy{\.z}y}, {Garrett}, {G{\"u}rkan}, {Jarvis}, {Kondapally}, {Mandal}, {Prandoni}, {Repetti}, {Retana-Montenegro}, {Schwarz}, {Shulevski}, \& {Wiaux}}]{Tasse2021}
{Tasse}, C., {Shimwell}, T., {Hardcastle}, M.~J., {et~al.} 2021, \aap, 648, A1

\bibitem[{{Timmerman} {et~al.}(2022){Timmerman}, {van Weeren}, {Botteon}, {R{\"o}ttgering}, {McNamara}, {Sweijen}, {B{\^\i}rzan}, \& {Morabito}}]{Timmerman2022}
{Timmerman}, R., {van Weeren}, R.~J., {Botteon}, A., {et~al.} 2022, \aap, 668, A65

\bibitem[{{Ubertosi} {et~al.}(2021){Ubertosi}, {Gitti}, {Brighenti}, {Brunetti}, {McDonald}, {Nulsen}, {McNamara}, {Randall}, {Forman}, {Donahue}, {Ignesti}, {Gaspari}, {Ettori}, {Feretti}, {Blanton}, {Jones}, \& {Calzadilla}}]{Ubertosi2021}
{Ubertosi}, F., {Gitti}, M., {Brighenti}, F., {et~al.} 2021, \apjl, 923, L25

\bibitem[{{Ubertosi} {et~al.}(2023){Ubertosi}, {Gitti}, {Brighenti}, {McDonald}, {Nulsen}, {Donahue}, {Brunetti}, {Randall}, {Gaspari}, {Ettori}, {Calzadilla}, {Ignesti}, {Feretti}, \& {Blanton}}]{Ubertosi2023}
{Ubertosi}, F., {Gitti}, M., {Brighenti}, F., {et~al.} 2023, \apj, 944, 216

\bibitem[{{van der Walt} {et~al.}(2011){van der Walt}, {Colbert}, \& {Varoquaux}}]{vanderwalt2011}
{van der Walt}, S., {Colbert}, S.~C., \& {Varoquaux}, G. 2011, Computing in Science and Engineering, 13, 22

\bibitem[{{van Diepen} {et~al.}(2018){van Diepen}, {Dijkema}, \& {Offringa}}]{DPPP2018}
{van Diepen}, G., {Dijkema}, T.~J., \& {Offringa}, A. 2018, {DPPP: Default Pre-Processing Pipeline}

\bibitem[{{van Haarlem} {et~al.}(2013){van Haarlem}, {Wise}, {Gunst}, {Heald}, {McKean}, {Hessels}, {de Bruyn}, {Nijboer}, {Swinbank}, {Fallows}, {Brentjens}, {Nelles}, {Beck}, {Falcke}, {Fender}, {H{\"o}randel}, {Koopmans}, {Mann}, {Miley}, {R{\"o}ttgering}, {Stappers}, {Wijers}, {Zaroubi}, {van den Akker}, {Alexov}, {Anderson}, {Anderson}, {van Ardenne}, {Arts}, {Asgekar}, {Avruch}, {Batejat}, {B{\"a}hren}, {Bell}, {Bell}, {van Bemmel}, {Bennema}, {Bentum}, {Bernardi}, {Best}, {B{\^\i}rzan}, {Bonafede}, {Boonstra}, {Braun}, {Bregman}, {Breitling}, {van de Brink}, {Broderick}, {Broekema}, {Brouw}, {Br{\"u}ggen}, {Butcher}, {van Cappellen}, {Ciardi}, {Coenen}, {Conway}, {Coolen}, {Corstanje}, {Damstra}, {Davies}, {Deller}, {Dettmar}, {van Diepen}, {Dijkstra}, {Donker}, {Doorduin}, {Dromer}, {Drost}, {van Duin}, {Eisl{\"o}ffel}, {van Enst}, {Ferrari}, {Frieswijk}, {Gankema}, {Garrett}, {de Gasperin}, {Gerbers}, {de Geus}, {Grie{\ss}meier}, {Grit}, {Gruppen}, {Hamaker}, {Hassall}, {Hoeft}, {Holties},
  {Horneffer}, {van der Horst}, {van Houwelingen}, {Huijgen}, {Iacobelli}, {Intema}, {Jackson}, {Jelic}, {de Jong}, {Juette}, {Kant}, {Karastergiou}, {Koers}, {Kollen}, {Kondratiev}, {Kooistra}, {Koopman}, {Koster}, {Kuniyoshi}, {Kramer}, {Kuper}, {Lambropoulos}, {Law}, {van Leeuwen}, {Lemaitre}, {Loose}, {Maat}, {Macario}, {Markoff}, {Masters}, {McFadden}, {McKay-Bukowski}, {Meijering}, {Meulman}, {Mevius}, {Middelberg}, {Millenaar}, {Miller-Jones}, {Mohan}, {Mol}, {Morawietz}, {Morganti}, {Mulcahy}, {Mulder}, {Munk}, {Nieuwenhuis}, {van Nieuwpoort}, {Noordam}, {Norden}, {Noutsos}, {Offringa}, {Olofsson}, {Omar}, {Orr{\'u}}, {Overeem}, {Paas}, {Pand ey-Pommier}, {Pandey}, {Pizzo}, {Polatidis}, {Rafferty}, {Rawlings}, {Reich}, {de Reijer}, {Reitsma}, {Renting}, {Riemers}, {Rol}, {Romein}, {Roosjen}, {Ruiter}, {Scaife}, {van der Schaaf}, {Scheers}, {Schellart}, {Schoenmakers}, {Schoonderbeek}, {Serylak}, {Shulevski}, {Sluman}, {Smirnov}, {Sobey}, {Spreeuw}, {Steinmetz}, {Sterks}, {Stiepel}, {Stuurwold},
  {Tagger}, {Tang}, {Tasse}, {Thomas}, {Thoudam}, {Toribio}, {van der Tol}, {Usov}, {van Veelen}, {van der Veen}, {ter Veen}, {Verbiest}, {Vermeulen}, {Vermaas}, {Vocks}, {Vogt}, {de Vos}, {van der Wal}, {van Weeren}, {Weggemans}, {Weltevrede}, {White}, {Wijnholds}, {Wilhelmsson}, {Wucknitz}, {Yatawatta}, {Zarka}, {Zensus}, \& {van Zwieten}}]{vanHaarlem2013}
{van Haarlem}, M.~P., {Wise}, M.~W., {Gunst}, A.~W., {et~al.} 2013, \aap, 556, A2

\bibitem[{{van Weeren} {et~al.}(2019){van Weeren}, {de Gasperin}, {Akamatsu}, {Br{\"u}ggen}, {Feretti}, {Kang}, {Stroe}, \& {Zandanel}}]{vanWeeren2019}
{van Weeren}, R.~J., {de Gasperin}, F., {Akamatsu}, H., {et~al.} 2019, \ssr, 215, 16

\bibitem[{{van Weeren} {et~al.}(2021){van Weeren}, {Shimwell}, {Botteon}, {Brunetti}, {Br{\"u}ggen}, {Boxelaar}, {Cassano}, {Di Gennaro}, {Andrade-Santos}, {Bonnassieux}, {Bonafede}, {Cuciti}, {Dallacasa}, {de Gasperin}, {Gastaldello}, {Hardcastle}, {Hoeft}, {Kraft}, {Mandal}, {Rossetti}, {R{\"o}ttgering}, {Tasse}, \& {Wilber}}]{vanWeeren2021}
{van Weeren}, R.~J., {Shimwell}, T.~W., {Botteon}, A., {et~al.} 2021, \aap, 651, A115

\bibitem[{{van Weeren} {et~al.}(2016){van Weeren}, {Williams}, {Hardcastle}, {Shimwell}, {Rafferty}, {Sabater}, {Heald}, {Sridhar}, {Dijkema}, {Brunetti}, {Br{\"u}ggen}, {Andrade-Santos}, {Ogrean}, {R{\"o}ttgering}, {Dawson}, {Forman}, {de Gasperin}, {Jones}, {Miley}, {Rudnick}, {Sarazin}, {Bonafede}, {Best}, {B{\^\i}rzan}, {Cassano}, {Chy{\.z}y}, {Croston}, {Ensslin}, {Ferrari}, {Hoeft}, {Horellou}, {Jarvis}, {Kraft}, {Mevius}, {Intema}, {Murray}, {Orr{\'u}}, {Pizzo}, {Simionescu}, {Stroe}, {van der Tol}, \& {White}}]{vanWeeren2016}
{van Weeren}, R.~J., {Williams}, W.~L., {Hardcastle}, M.~J., {et~al.} 2016, \apjs, 223, 2

\bibitem[{{Vantyghem} {et~al.}(2014){Vantyghem}, {McNamara}, {Russell}, {Main}, {Nulsen}, {Wise}, {Hoekstra}, \& {Gitti}}]{Vantyghem2014}
{Vantyghem}, A.~N., {McNamara}, B.~R., {Russell}, H.~R., {et~al.} 2014, \mnras, 442, 3192

\bibitem[{{Vazza} {et~al.}(2012){Vazza}, {Roediger}, \& {Br{\"u}ggen}}]{Vazza2012}
{Vazza}, F., {Roediger}, E., \& {Br{\"u}ggen}, M. 2012, \aap, 544, A103

\bibitem[{{Venturi} {et~al.}(2008){Venturi}, {Giacintucci}, {Dallacasa}, {Cassano}, {Brunetti}, {Bardelli}, \& {Setti}}]{Venturi2008}
{Venturi}, T., {Giacintucci}, S., {Dallacasa}, D., {et~al.} 2008, \aap, 484, 327

\bibitem[{{Venturi} {et~al.}(2017){Venturi}, {Rossetti}, {Brunetti}, {Farnsworth}, {Gastaldello}, {Giacintucci}, {Lal}, {Rudnick}, {Shimwell}, {Eckert}, {Molendi}, \& {Owers}}]{Venturi2017}
{Venturi}, T., {Rossetti}, M., {Brunetti}, G., {et~al.} 2017, \aap, 603, A125

\bibitem[{{Vikhlinin} {et~al.}(2009){Vikhlinin}, {Burenin}, {Ebeling}, {Forman}, {Hornstrup}, {Jones}, {Kravtsov}, {Murray}, {Nagai}, {Quintana}, \& {Voevodkin}}]{Vikhlinin2009}
{Vikhlinin}, A., {Burenin}, R.~A., {Ebeling}, H., {et~al.} 2009, \apj, 692, 1033

\bibitem[{{Virtanen} {et~al.}(2020){Virtanen}, {Gommers}, {Oliphant}, {Haberland}, {Reddy}, {Cournapeau}, {Burovski}, {Peterson}, {Weckesser}, {Bright}, {van der Walt}, {Brett}, {Wilson}, {Millman}, {Mayorov}, {Nelson}, {Jones}, {Kern}, {Larson}, {Carey}, {Polat}, {Feng}, {Moore}, {VanderPlas}, {Laxalde}, {Perktold}, {Cimrman}, {Henriksen}, {Quintero}, {Harris}, {Archibald}, {Ribeiro}, {Pedregosa}, {van Mulbregt}, \& {SciPy 1. 0 Contributors}}]{Scipy2020}
{Virtanen}, P., {Gommers}, R., {Oliphant}, T.~E., {et~al.} 2020, Nature Methods, 17, 261

\bibitem[{{Wang} \& {Markevitch}(2018)}]{Markevitch2018}
{Wang}, Q. H.~S. \& {Markevitch}, M. 2018, \apj, 868, 45

\bibitem[{{Wen} \& {Han}(2013)}]{Wen2013}
{Wen}, Z.~L. \& {Han}, J.~L. 2013, \mnras, 436, 275

\bibitem[{{Williams} {et~al.}(2016){Williams}, {van Weeren}, {R{\"o}ttgering}, {Best}, {Dijkema}, {de Gasperin}, {Hardcastle}, {Heald}, {Prandoni}, {Sabater}, {Shimwell}, {Tasse}, {van Bemmel}, {Br{\"u}ggen}, {Brunetti}, {Conway}, {En{\ss}lin}, {Engels}, {Falcke}, {Ferrari}, {Haverkorn}, {Jackson}, {Jarvis}, {Kapi{\'n}ska}, {Mahony}, {Miley}, {Morabito}, {Morganti}, {Orr{\'u}}, {Retana-Montenegro}, {Sridhar}, {Toribio}, {White}, {Wise}, \& {Zwart}}]{Williams2016}
{Williams}, W.~L., {van Weeren}, R.~J., {R{\"o}ttgering}, H.~J.~A., {et~al.} 2016, \mnras, 460, 2385

\bibitem[{{Wise} {et~al.}(2004){Wise}, {McNamara}, \& {Murray}}]{Wise2004}
{Wise}, M.~W., {McNamara}, B.~R., \& {Murray}, S.~S. 2004, \apj, 601, 184

\bibitem[{{Zhang} {et~al.}(2023){Zhang}, {Simionescu}, {Gastaldello}, {Eckert}, {Camillini}, {Natale}, {Rossetti}, {Brunetti}, {Akamatsu}, {Botteon}, {Cassano}, {Cuciti}, {Bruno}, {Shimwell}, {Jones}, {Kaastra}, {Ettori}, {Br{\"u}ggen}, {de Gasperin}, {Drabent}, {van Weeren}, \& {R{\"o}ttgering}}]{Zhang2023}
{Zhang}, X., {Simionescu}, A., {Gastaldello}, F., {et~al.} 2023, \aap, 672, A42

\bibitem[{{ZuHone} {et~al.}(2013){ZuHone}, {Markevitch}, {Brunetti}, \& {Giacintucci}}]{ZuHone2013}
{ZuHone}, J.~A., {Markevitch}, M., {Brunetti}, G., \& {Giacintucci}, S. 2013, \apj, 762, 78

\bibitem[{{ZuHone} {et~al.}(2011){ZuHone}, {Markevitch}, \& {Lee}}]{ZuHone2011}
{ZuHone}, J.~A., {Markevitch}, M., \& {Lee}, D. 2011, \apj, 743, 16

\end{thebibliography}

\appendix
\section{Detailed X-ray analysis}\label{sec:appendix}
\begin{figure*}
\centering
\subfloat{
\includegraphics[width=8.5cm]{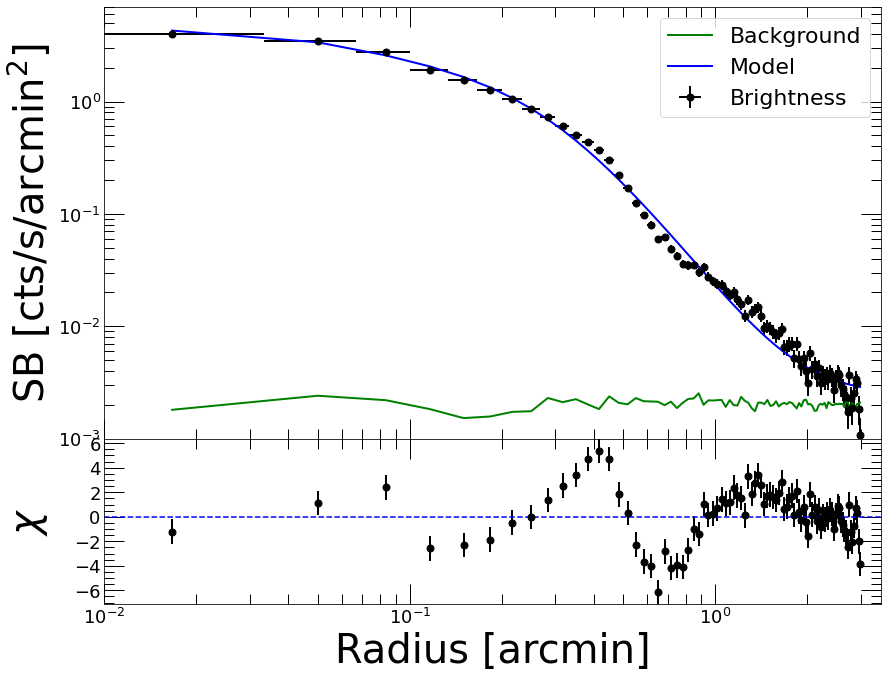}} \quad
\subfloat{
\includegraphics[width=8.5cm]{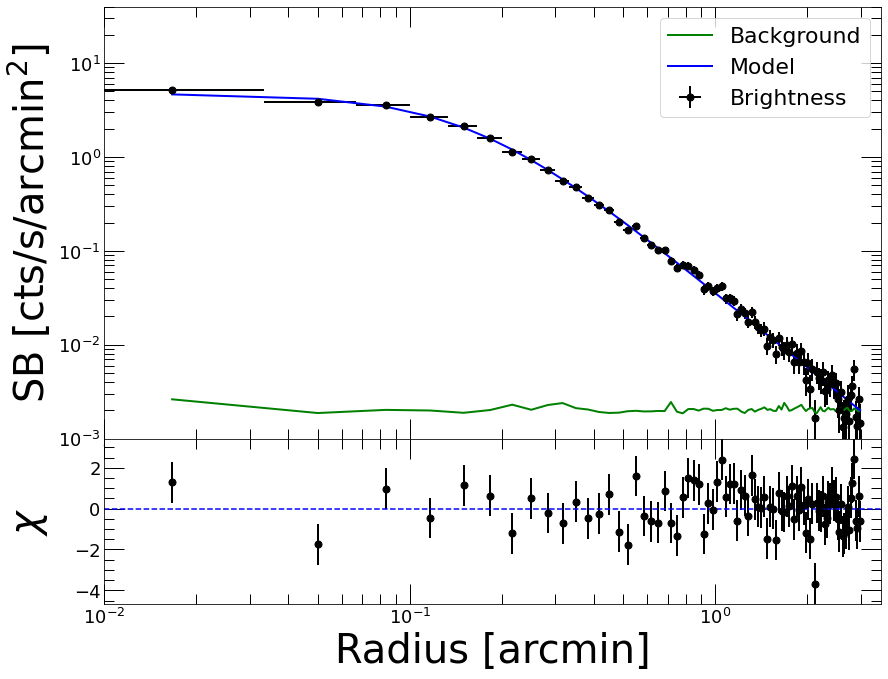}}
\caption{Comparison of a double $\beta$-model fits of a radial profile with (\emph{Left panel}) and without (\emph{Right panel}) a SB jump.}
\label{fig:double-beta}
\end{figure*}

We performed a systematic search of cold fronts in the selected clusters, re-analysing archival \emph{Chandra} data. This analysis aims to verify the presence of sloshing in the central regions of clusters. We did not determine the angular extent of cold fronts as this is beyond the scope of this work. Furthermore, we reserve for a future study the search of sloshing in the outer regions of clusters to check if also cluster-scale diffuse radio emission is confined by cold fronts. 
We proceed in the following way:
\begin{itemize}
    \item We visually inspected X-ray \emph{Chandra} images, residual images obtained subtracting a double $\beta$-model and GGM images, to identify possible surface brightness discontinuities.
    \item We extracted radial profiles along directions of features identified in X-ray images and in adjacent sectors to investigate the thermal emission at 360 degrees. We note that the presence of a cold front, if enough counts are present, is not diminished by the fact of considering 90-degree sectors with quite arbitrary orientation. We then fitted every profile with a double $\beta$-model to quantify the entity of SB jumps, evaluating the goodness of fit. We considered the double $\beta$-model as a good representation of a profile if $\chi^2_{red}\le1.5$ and if the residuals present a scattered trend. While we investigated further those features with a $\chi^2_{red}\ge1.5$ and with an evident jump in the residuals of the fit, performing a fit with a broken power-law restricted to the region of SB jump. See Fig. \ref{fig:double-beta} for a comparison of a profile well fitted with a double $\beta$-model vs a profile presenting an evident SB jump. 
    
    \item Finally, we performed spectral analysis of the newly detected SB discontinuities, i.e. SB profiles presenting a jump well fitted with a broken power-law, to identify the nature of the front. 
    A cold front is characterised by a downstream temperature cooler than the upstream temperature, the opposite of a shock front. 
\end{itemize}
For clusters with already known cold fronts, we performed the first two steps of the analysis, to verify if new sloshing features appear thanks to modern techniques of edge detection, like the GGM filter, but we decided to not repeat the spectral analysis, as there were no improvements in the data used.
In the following subsections, we briefly summarise the results of our analysis for each cluster.
In Table \ref{tab:double-beta}, we report the results of double $\beta$-model fits on the extracted profiles and the nature of detected fronts.

\begin{table}
    \caption{Dobule $\beta$-model fit of X-ray radial profiles.}
    \renewcommand\arraystretch{1.2}
    \centering
    \begin{tabular}{ccccc} \hline
Cluster name  &sector &$\chi^2_{red}$ &SB jump? &Front\\\hline
PSZ1G139.61+24   &NW &2.0 &Yes &CF\\
                 &NE &1.1 &No \\
                 &SE &1.7 &No \\
                 &SW &1.4 &No \\
MS 0735.6+7421   &N &107.0 &Yes   &Shock\\
                 &E &3.8 &Yes &Shock \\
                 &S &46.0 &Yes &Shock \\
                 &W &4.0 &Yes &Shock \\
MS 0839.8+2938   &N &1.0 &No    \\ 
                 &E &1.0 &No \\
                 &S &1.4 &No \\
                 &W &1.2 &No \\
Z2089            &NW &1.2 &No    \\
                 &NE &1.3 &No \\
                 &SE &1.2 &No \\
                 &SW &1.3 &No \\
RBS 0797         &N &1.5 &No    \\
                 &E &1.2 &No \\
                 &S &1.2 &No \\
                 &W &2.0 &No (Cavity) & \\
A1068            &SE in &2.8  &Yes &CF    \\ 
                 &SW in &2.0 &Yes\\
                 &NW out &4.6 &Yes & 2 CFs \\
                 &NE in &1.3 &No\\
                 &NE out &1.1 &No \\
A1204            &NW &1.2 &No   \\
                 &NE &1.2 &No \\
                 &SE &0.9 &No \\
                 &SW &1.1 &No \\
MS 1455.0+2232   &N &5.0 &Yes &CF  \\
                 &E &1.4 &No \\
                 &S &2.0 &Yes &CF\\
                 &W &1.2 &No \\
RX J1532.9+3021     &W &1.2 &No\\
                    &N &1.4 &No \\
                    &E &1.7 &No (Cavity)  &\\
                    &S &1.1 &No \\
RX J1720.1+2638         &NW &3.4 &Yes &CF\\
                    &NE &1.4 &No  \\
                    &SE &4.2 &Yes &CF \\
                    &SW &2.1 &Yes \\
MACS J1720.2+3536       &N &1.1 &No   \\
                    &E &1.3 &No \\
                    &S &1.2 &No \\
                    &W &1.0 &No \\
MACS J2245.0+2637       &NW &1.0 &No \\
                    &NE &1.1 &No \\
                    &SE &1.4 &No \\
                    &SW &1.2 &No \\\hline
    \end{tabular} \\
    \tablefoot{CF = cold front.} 
    \label{tab:double-beta}
\end{table}

\subsection{Detected cold fronts}
We detected cold fronts in four clusters: three of them were already known, while those in A1068 are new detection.

\begin{figure*}
\centering
\subfloat{
\includegraphics[width=17cm]{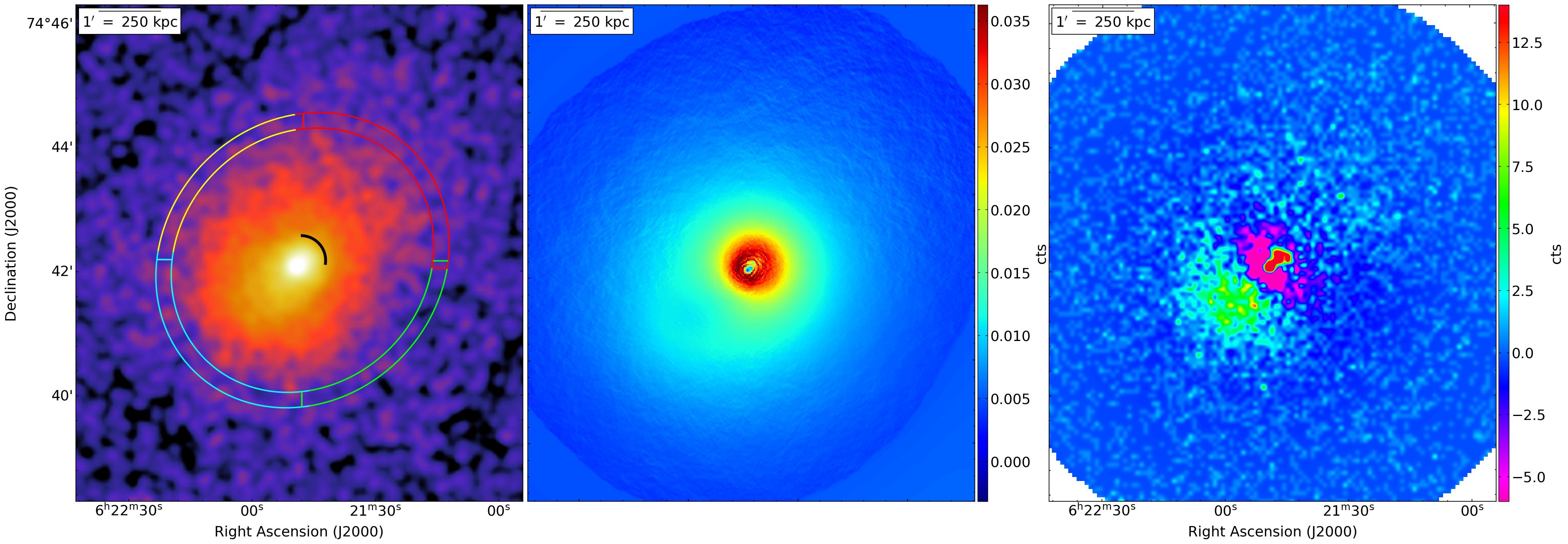}}\\
\subfloat{
\includegraphics[width=5.6cm]{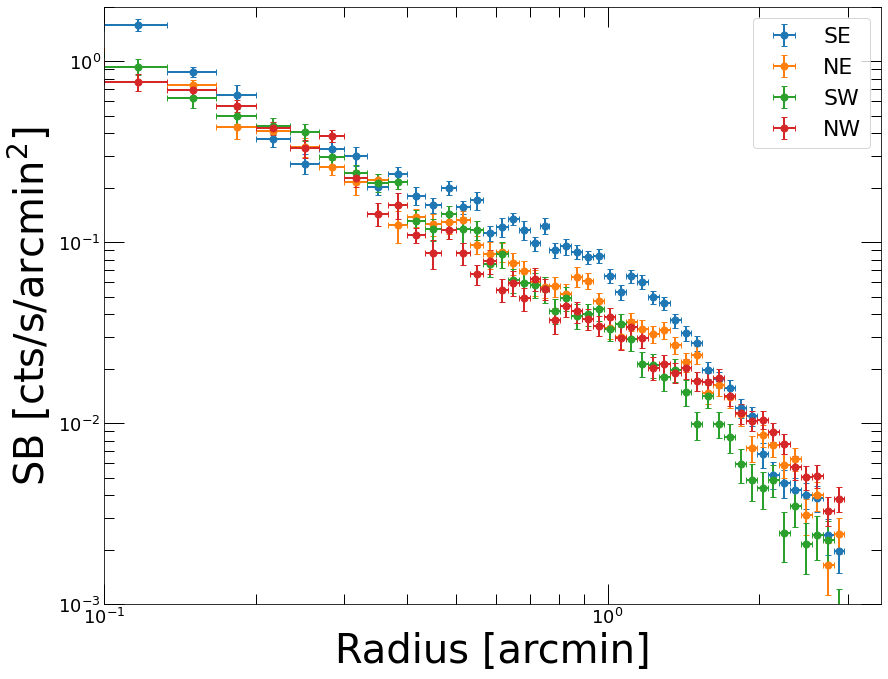} 
\includegraphics[width=5.6cm]{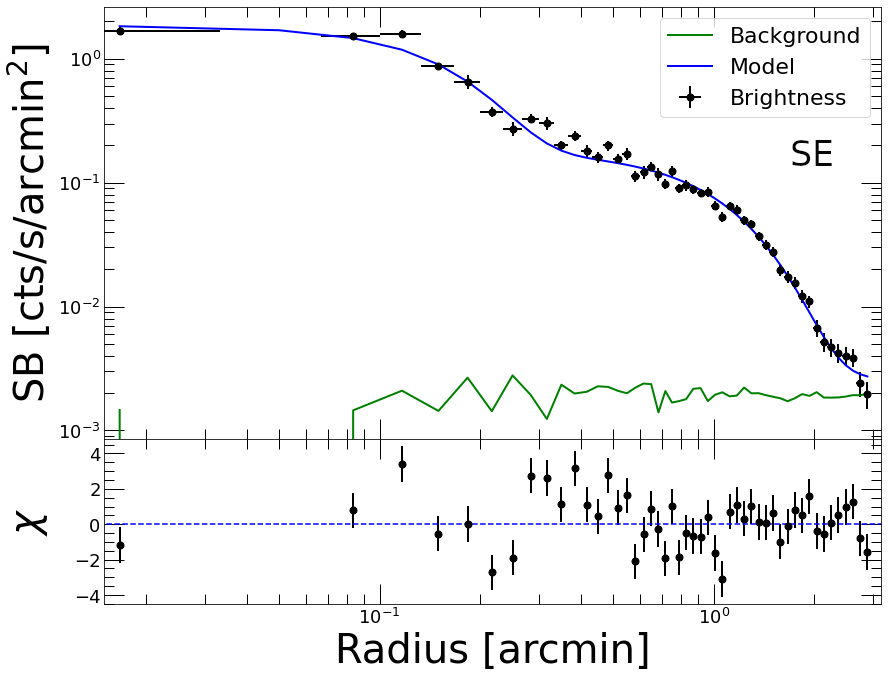}
\includegraphics[width=5.6cm]{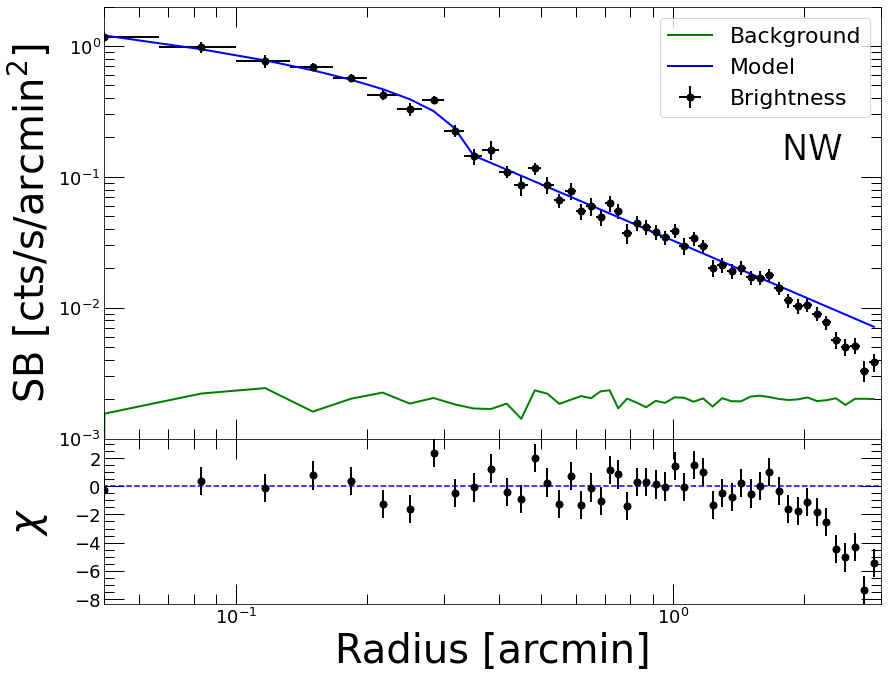}}
\caption{X-ray analysis of PSZ1G139.61+24. \emph{Upper left panel:} \emph{Chandra} 0.5–7.0 keV image, with represented a portion of the extraction regions. The black arc indicates the cold front position. \emph{Upper central panel:} GGM filtered image. \emph{Upper right panel:} Residual image after the subtraction of a double $\beta$-model. \emph{Lower left panel:} SB profiles extracted along the directions depicted in the upper left panel. \emph{Lower central panel:} best-fitting double $\beta$-model (blue line) with associated residuals of the SE profile. \emph{Lower right panel:} best-fitting broken power-law model (blue line) with associated residuals on the SB discontinuity in the NW profile.}
\label{fig:PSZ_Xmaps}
\end{figure*}

\subsubsection*{PSZ1G139.61+24}
The cluster was observed two times with \emph{Chandra} (ObsID 15139, 15297), for a total exposure time of 27.5 ks.
The thermal emission of this cluster is elongated towards SE (Fig. \ref{fig:PSZ_Xmaps}, upper left panel), as highlighted also by the excess of SB in the residual image (upper right panel).
We extracted SB profiles along elliptical annuli, to better trace the cluster morphology, in the SE direction and adjacent sectors with angular sizes of 90 degrees to cover the whole extension of thermal emission.
The SE profile clearly shows an excess of SB at large radii, with respect to the other sectors (lower left panel), but the SB jump cannot be fitted by a broken power-law, so do not present the characteristic shape of cold fronts. A surface brightness discontinuity is instead found in the NW direction, where there is a jump in SB of 1.68 at a distance of $r\sim20\arcsec$ (lower right panel).
This feature was already identified in \cite{Savini2018}, with an associated temperature jump of $\Delta T = 3.4$ keV, with higher temperature in the downstream region, indicative of a cold front.

\begin{figure*}
\centering
\subfloat{
\includegraphics[width=17cm]{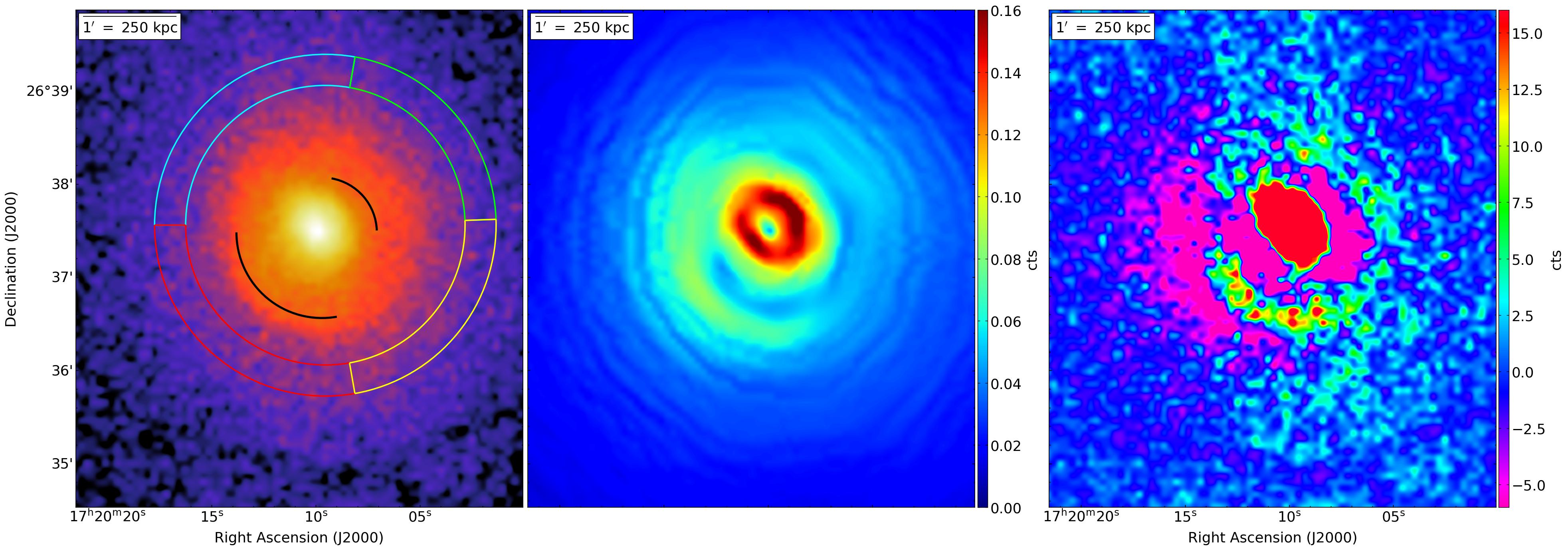}}\\
\subfloat{
\includegraphics[width=5.6cm]{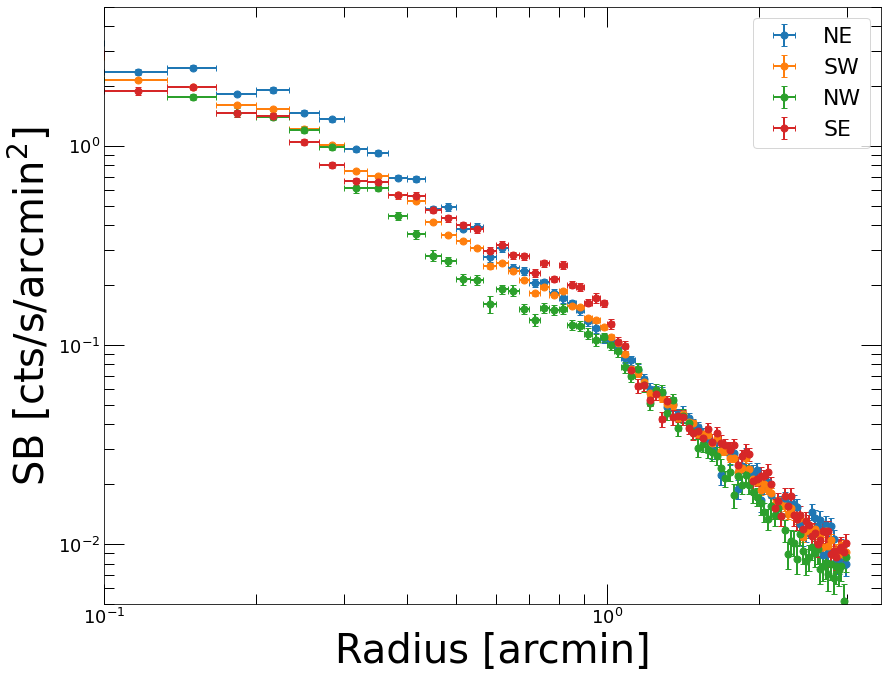} \
\includegraphics[width=5.6cm]{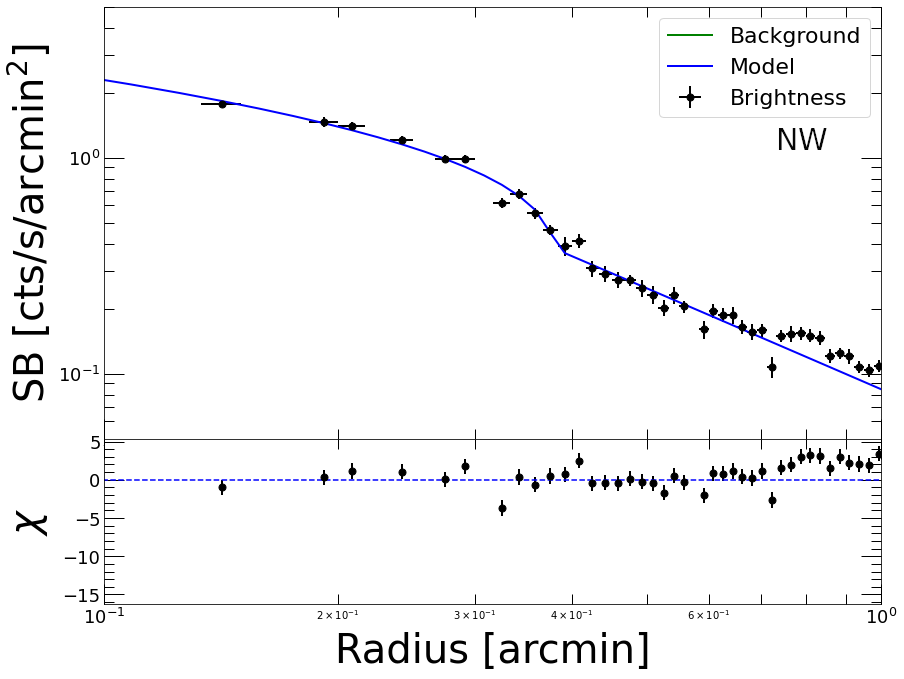} \ \includegraphics[width=5.6cm]{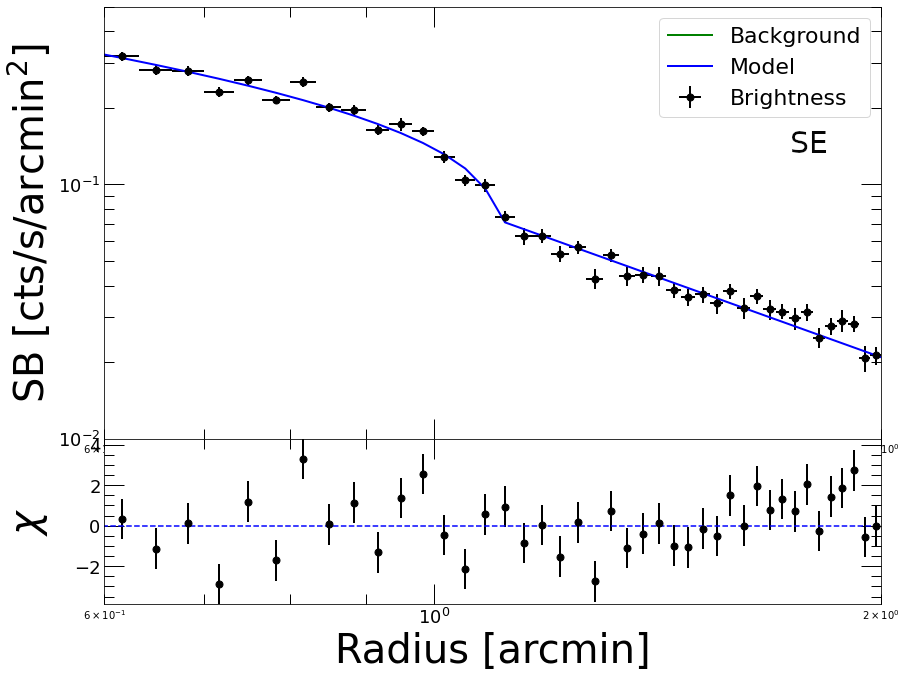}} \\
\subfloat{
\includegraphics[width=5.6cm]{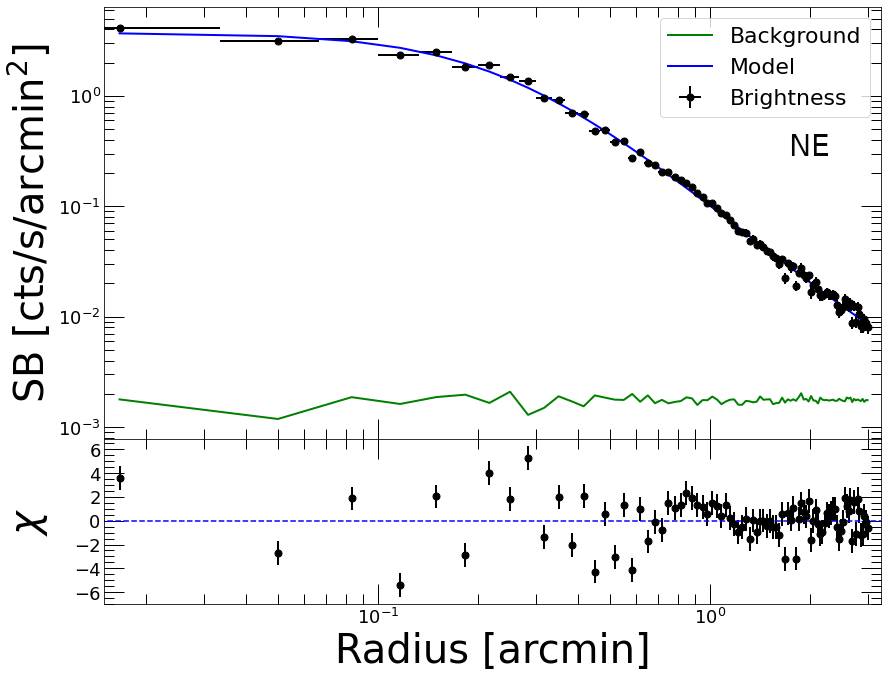} \
\includegraphics[width=5.6cm]{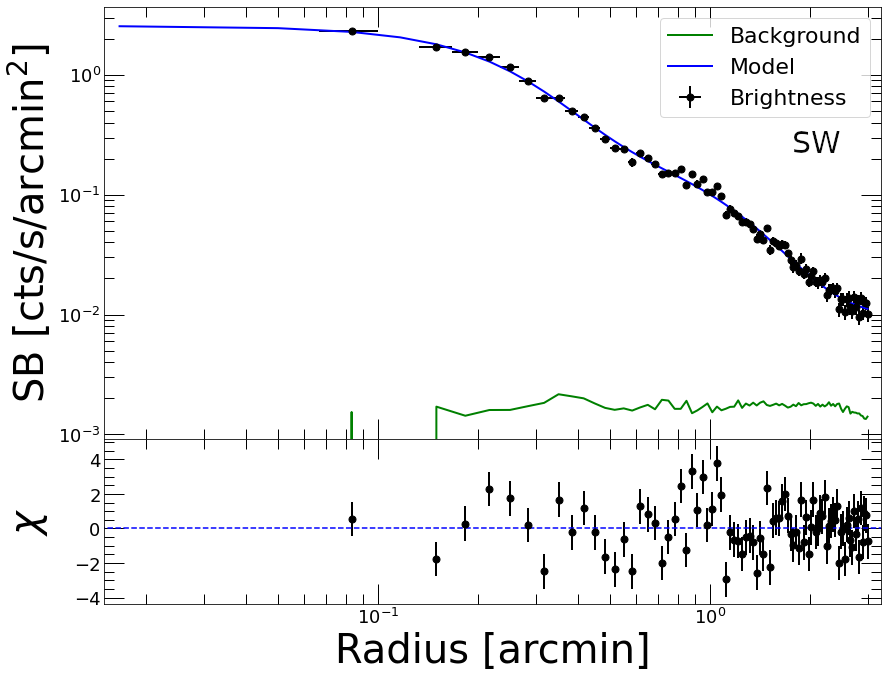}} 
\caption{X-ray analysis of RX J1720.1+2638. \emph{Upper left panel:} \emph{Chandra} 0.5–7.0 keV image, with represented a portion of the extraction regions. Black arcs indicate the position of cold fronts.  \emph{Upper central panel:} GGM filtered image. \emph{Upper right panel:} Residual image after the subtraction of a double $\beta$-model. \emph{Middle left panel:} SB profiles extracted along the directions depicted in the upper left panel. \emph{Middle central panel:} best-fitting broken power-law model (blue line) with associated residuals on the SB discontinuity in the NW profile. \emph{Middle right panel:} same as before for the SB discontinuity in the SE profile. \emph{Lower left panel:} best-fitting double $\beta$-model (blue line) with associated residuals on NE profile. \emph{Lower right panel:} same as before for the SW profile.}
\label{fig:RXJ1720-1_Xmaps}
\end{figure*}

\subsubsection*{RX J1720.1+2638}
We combined three \emph{Chandra} observations of the cluster RXJ1720.1 (ObsID 3224, 4361 and 1453), for a total exposure time of 57.3 ks.
The GGM image (Fig. \ref{fig:RXJ1720-1_Xmaps}, upper central panel) evidences a spiral structure, typical of sloshing clusters, characterised by an excess of surface gradient around the cluster core in the SE and NW direction and an arc shape in the SE direction at a larger distance. Similarly, the residual map (upper right panel) shows an excess of SB in the SE direction, corresponding to the tail of the spiral. We extracted radial profiles in four different directions and adjusted the opening angles to enhance the detection of SB discontinuities noticed in the NW and SE profiles.
In particular, the NW profile presents a SB jump of 1.7 at $r\sim25$\arcsec (lower central profile), while the SE profile has a SB jump of 1.6 at $r\sim65$\arcsec (lower right profile).
Those features were already detected by \cite{Mazzotta2008}, with a corresponding temperature jump characterised by a higher temperature at larger radii, typical of cold fronts.
We note that also the SW profile is not well-fitted by a double $\beta$-model, presenting a hint of SB jump at $r\sim65$\arcsec. This feature is an extension of the jump in the adjacent SE sector that has been adapted to maximise the SB jump. The magnitude of the feature in fact decreases changing the angular aperture of the sector.

\begin{figure*}
\centering
\subfloat{
\includegraphics[width=17cm]{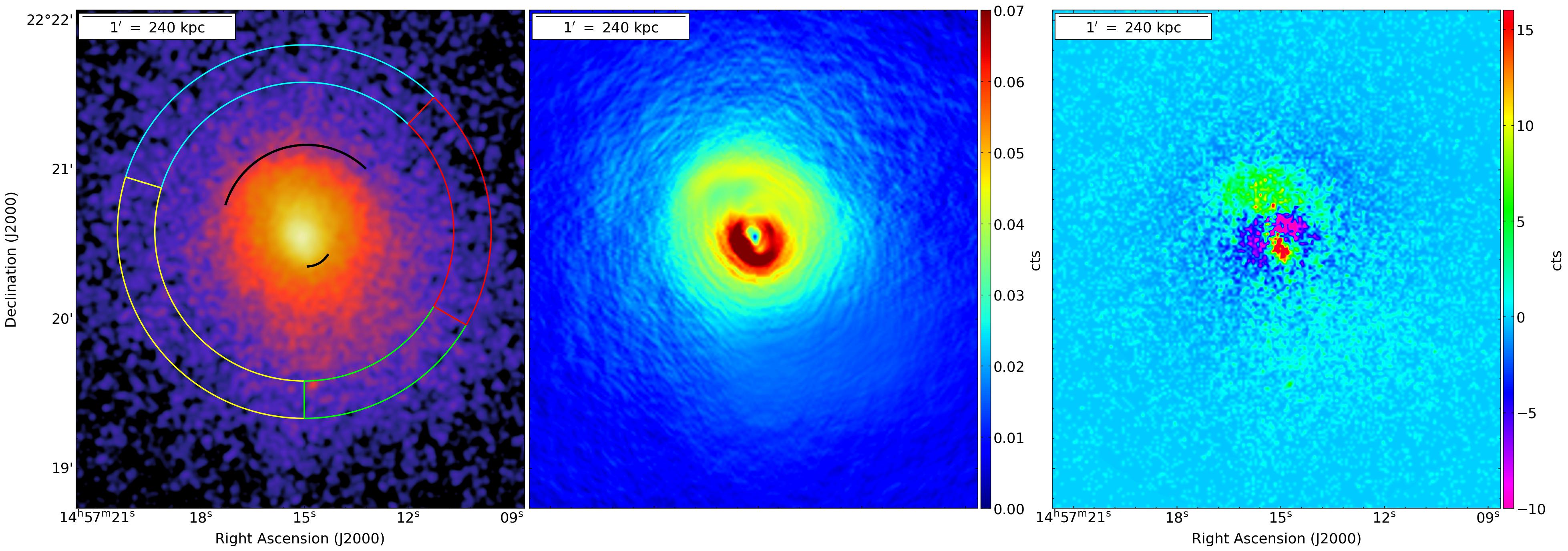}}\\
\subfloat{
\includegraphics[width=5.6cm]{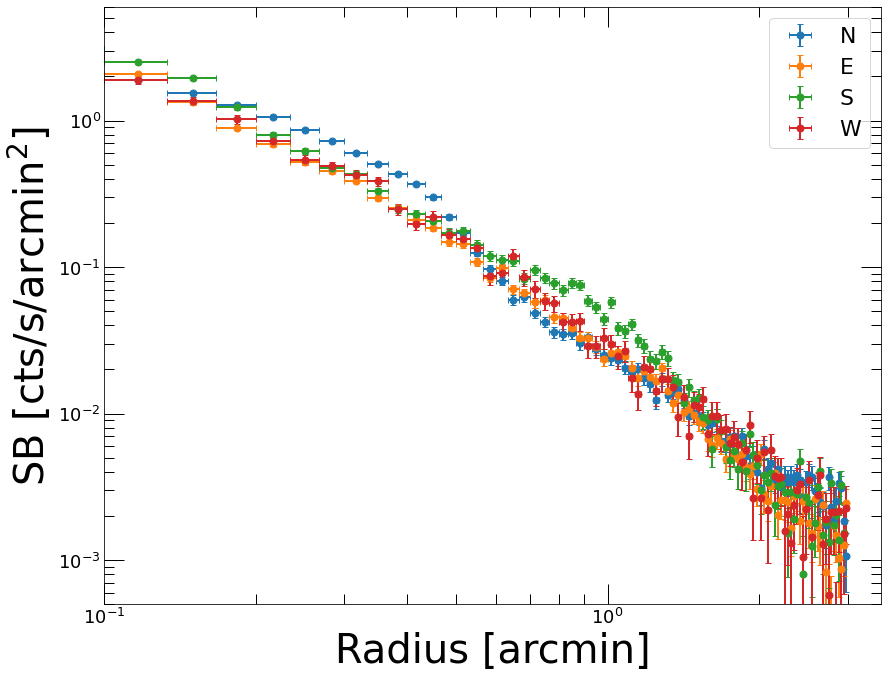} \
\includegraphics[width=5.6cm]{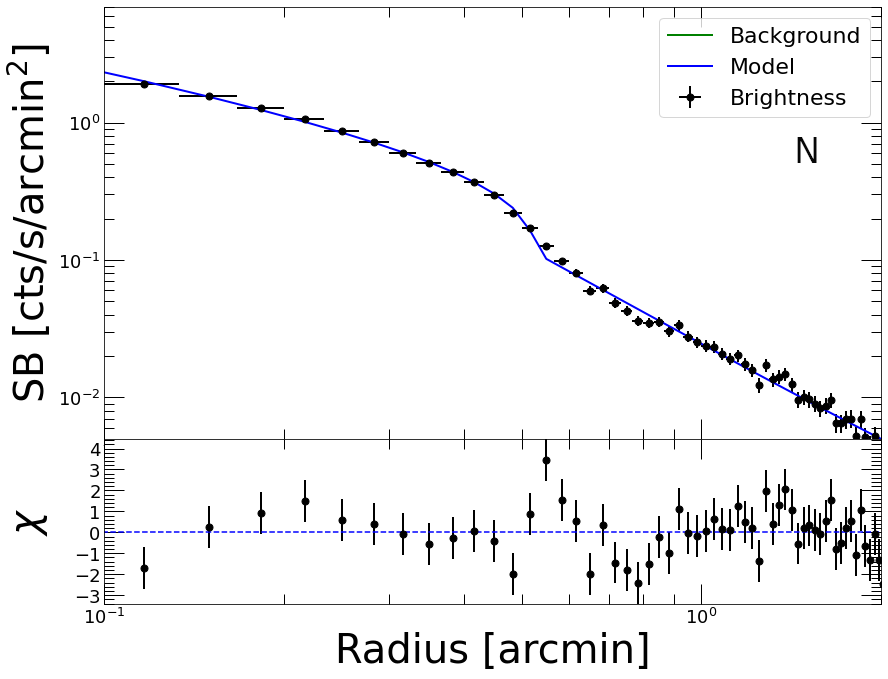} \ \includegraphics[width=5.6cm]{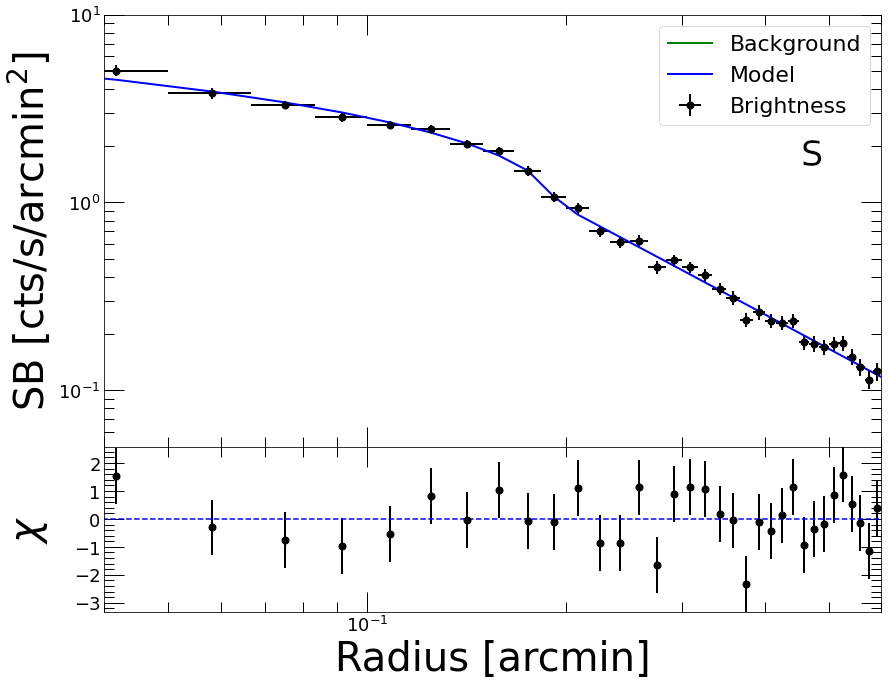}} \\
\subfloat{
\includegraphics[width=5.6cm]{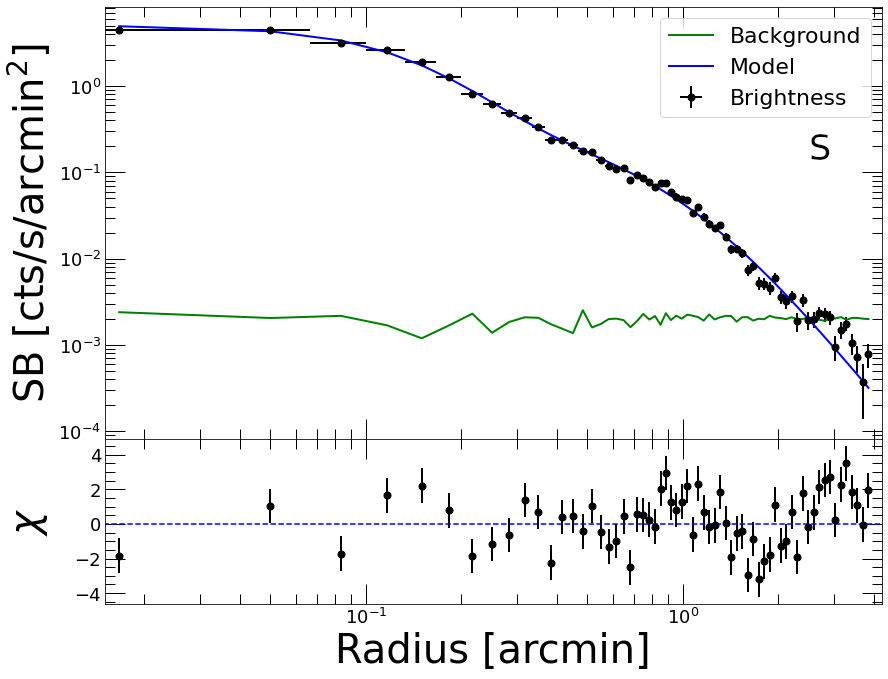}}
\caption{X-ray analysis of MS 1455.0+2232. \emph{Upper left panel:} \emph{Chandra} 0.5–7.0 keV image, with represented a portion of the extraction regions. Black arcs indicate the position of cold fronts. \emph{Upper central panel:} GGM filtered image. \emph{Upper right panel:} Residual image after the subtraction of a double $\beta$-model. \emph{Middle left panel:} SB profiles extracted along the directions depicted in the upper left panel. \emph{Middle central panel:} best-fitting broken power-law model (blue line) with associated residuals on the SB discontinuity in the N profile. \emph{Middle right panel:} same as before for the SB discontinuity in the S profile. \emph{Lower panel:} best-fitting double $\beta$-model on the S profile highlighting the SB fluctuation identified by \cite{Giacintucci2024} at r=1.9\arcmin.}
\label{fig:MS1455_Xmaps}
\end{figure*}

\subsubsection*{MS 1455.0+2232}
For the cluster MS1455, we re-analysed the \emph{Chandra} observation (ObsID 4192, exposure time 91.9 ks) presented in \cite{Mazzotta2008}.
The GGM map (Fig. \ref{fig:MS1455_Xmaps}, upper central panel) shows a spiral structure with an excess of surface gradient south of the cluster core and in the northern region at larger radii, corresponding to an excess of SB in the residual image.
We investigated radial profiles in the north and south directions, adjusting the sectors' aperture to increase the detection of SB discontinuities. We also inspected complementary sectors for comparison.
We detected a SB jump of 1.80 in the northern profile at $r\sim34\arcsec$, and a less prominent jump of 1.35 in the southern profile at $r\sim45\arcsec$.
Those SB discontinuities were previously identified by \cite{Mazzotta2008}, with a corresponding temperature jump with a lower temperature in the more central regions, classifying these features as cold fronts.
During the preparation of this article, a new X-ray study of the cluster has been presented in \cite{Giacintucci2024}, detecting a further CF southern of the cluster centre at $r\sim114\arcsec$. We then expanded the extraction of the southern radial profile to larger radii to check for this feature. We confirmed the presence of a SB jump in the residuals of the double $\beta$-model fit at the radial distance of the newly claimed CF. Our goal, however, was to reveal the presence of sloshing and we did it by focusing on the central regions of the clusters, where the features are overwhelming. Verifying the presence of large-scale sloshing and its connection with the halo extension is beyond the scope of this article and will be investigated in the future.

\begin{figure*}
\centering
\subfloat{
\includegraphics[width=17cm]{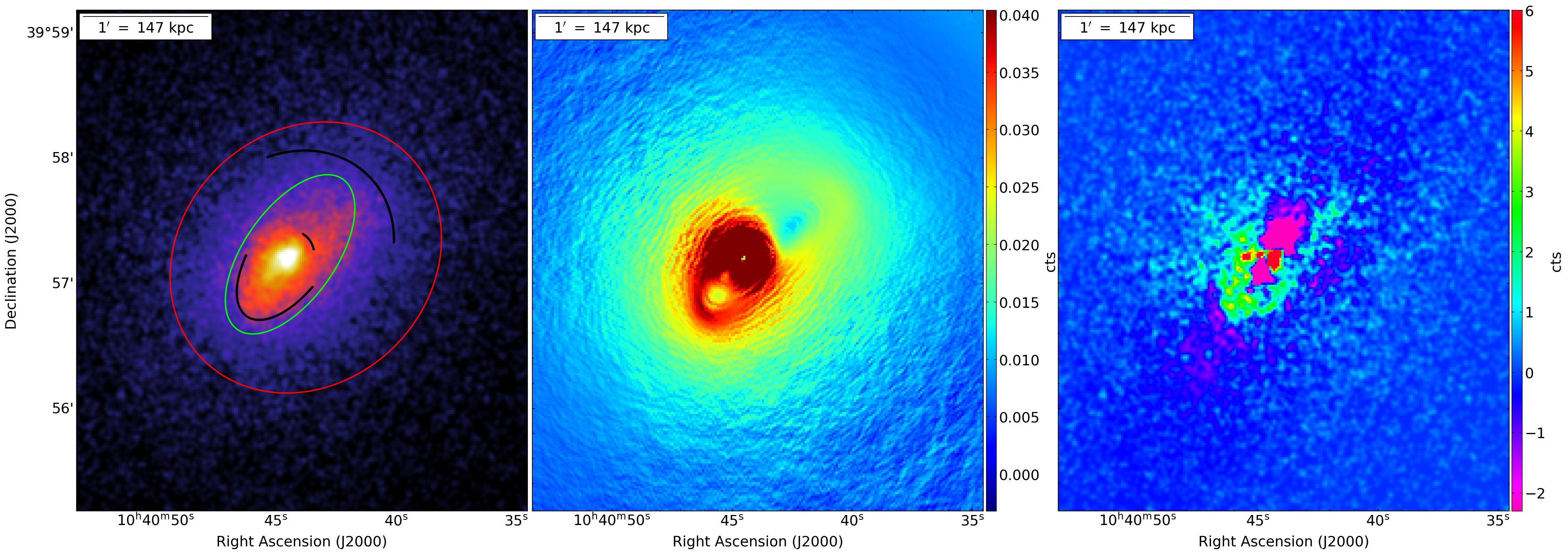}}\\
\subfloat{
\includegraphics[width=5.6cm]{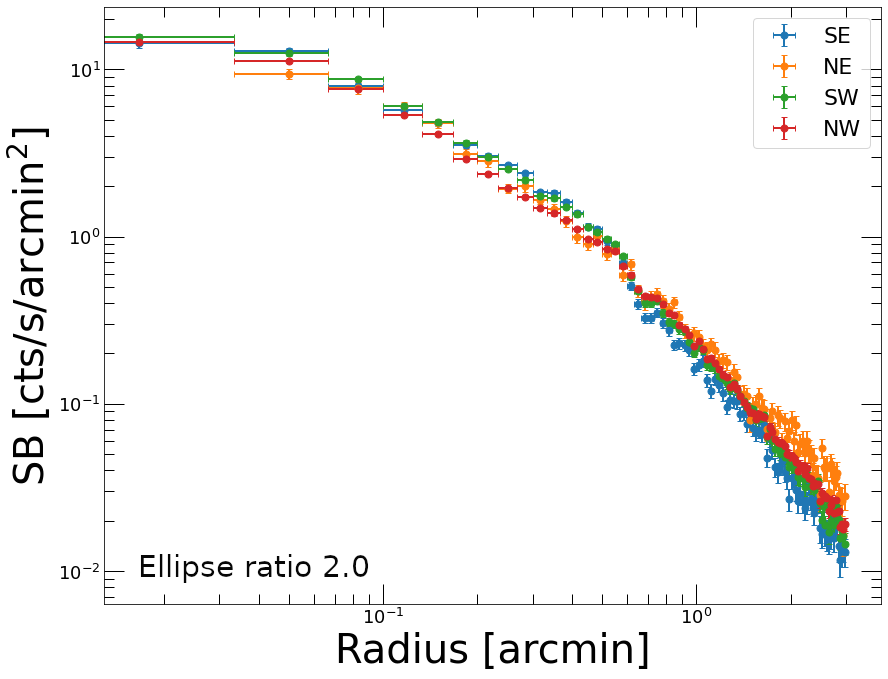} 
\includegraphics[width=5.6cm]{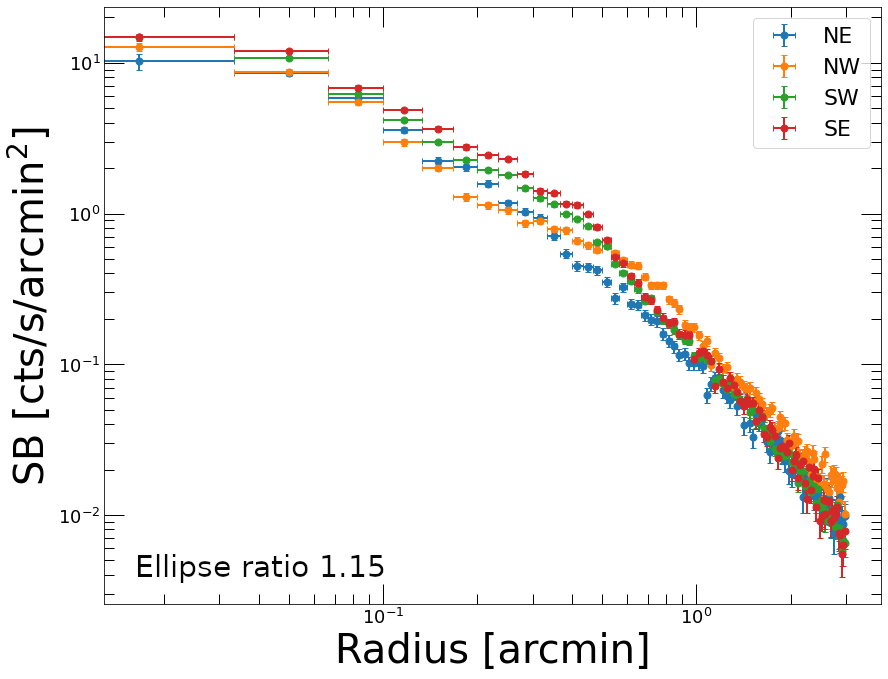}
\includegraphics[width=5.6cm]{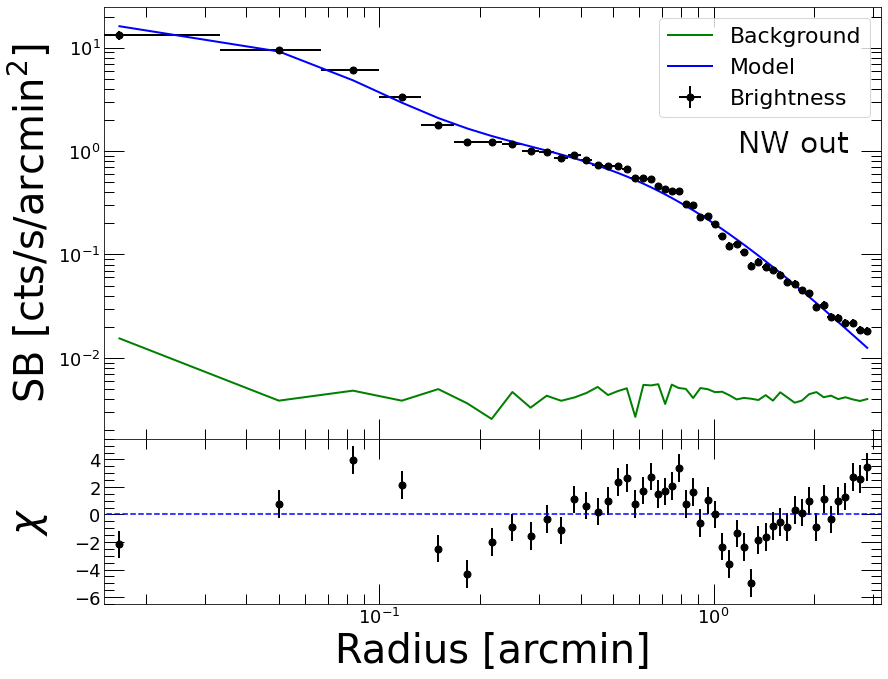}} \\
\subfloat{
\includegraphics[width=5.6cm]{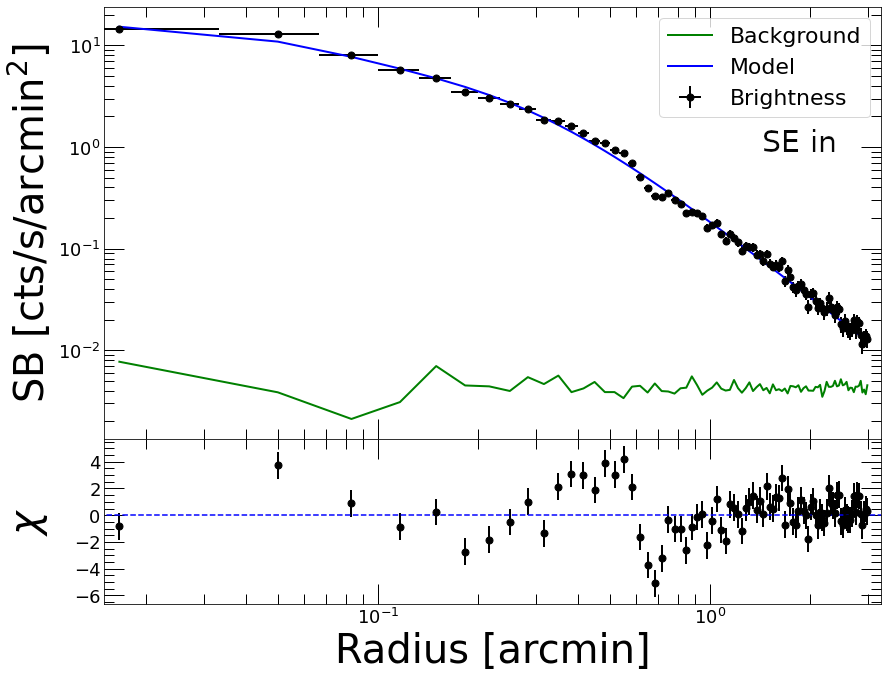} 
\includegraphics[width=5.6cm]{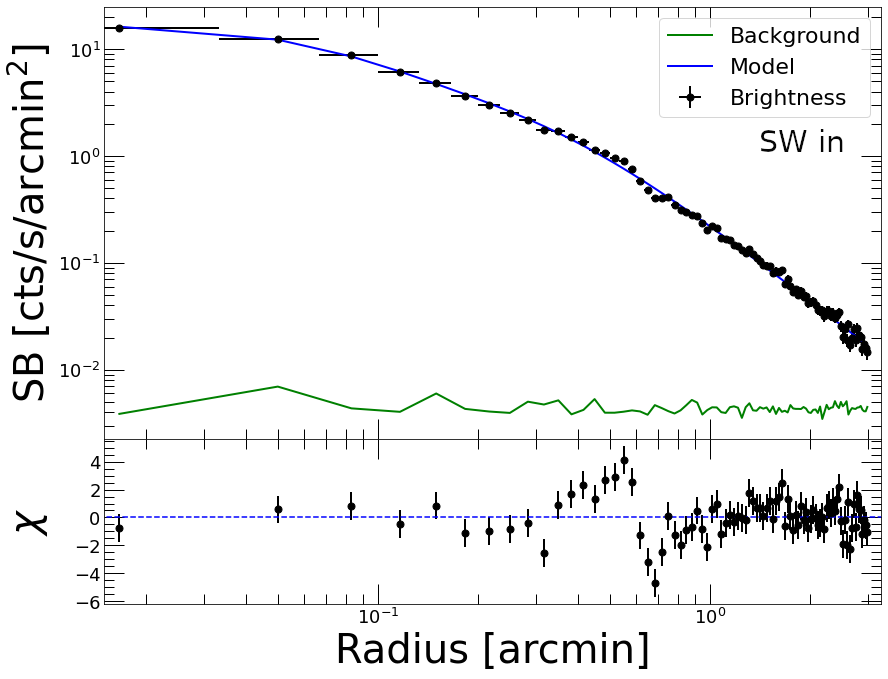}}
\caption{X-ray analysis of A1068. \emph{Upper left panel:} \emph{Chandra} 0.5–7.0 keV image. The black arcs indicate the position of cold fronts. The green and red ellipses represent the shape of the extraction regions. \emph{Upper central panel:} GGM filtered image. \emph{Upper right panel:} Residual image after the subtraction of a double $\beta$-model. \emph{Middle left panel:} SB profiles extracted using regions with ellipticity ratio of 2.0. \emph{Middle central panel:} SB profiles extracted using regions with ellipticity ratio of 1.15. \emph{Middle right panel:} best-fitting double $\beta$-model (blue line) with associated residuals of the NW profile (ellipticity ratio of 1.15). \emph{Lower left panel:} same as before for the SE profile (ellipticity ratio of 2.0). \emph{Lower right panel:} same as before for the SW profile (ellipticity ratio of 2.0).}
\label{fig:A1068_apx}
\end{figure*}

\subsubsection*{A1068}
The cluster A1068 was observed with \emph{Chandra} with a total exposure time of 26.8 ks.
Both the GGM and residual maps (Fig. \ref{fig:A1068_apx}, upper central and right panels) evidence an highly elliptical shape of the thermal emission in the central regions, elongated in the NW-SE direction, with an excess of SB in the SE direction. At larger scales, instead, the ellipticity of the emission is less accentuated. We therefore, extracted radial profiles along elliptical annuli with two different elliptical ratios: an elliptical ratio of 2.0 (represented by the green ellipse in the upper left panel) to investigate the inner features and an elliptical ratio of 1.15 (represented by the red ellipse in the upper left panel) to investigate the outer features.
Deviations from a double $\beta$-model are present along the NW, SE and SW profiles using both types of elliptical annuli, but the jumps are more accentuated using a proper region of extraction.
In particular, the profiles extracted using the highest elliptical ratio evidence a SB jump in the SE and SW profiles (lower left and central panels), at a similar distance from the cluster centre of $r\sim37.2\arcsec$ and $r\sim38.4\arcsec$, respectively, that could represent a single front.
In the NW direction, instead, we individuated two SB jumps, one internal at $r\sim11.4\arcsec$ and an external one at $r\sim53\arcsec$ (middle right panel).
Spectral analysis allows us to classify these fronts as cold fronts.
For more details see Section \ref{A1068}.

\begin{figure*}
\centering
\subfloat{
\includegraphics[width=10.5cm]{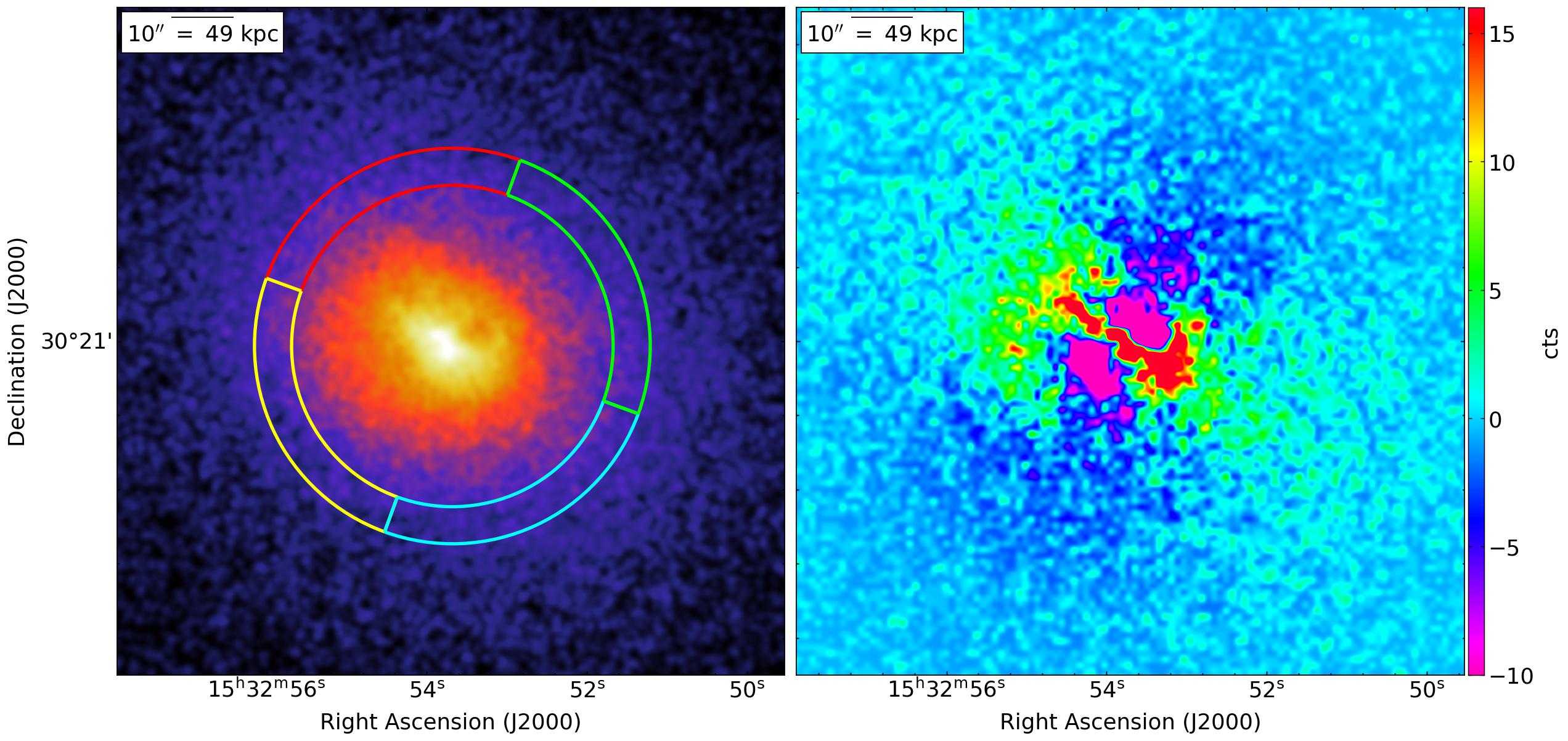}
\includegraphics[width=6.5cm]{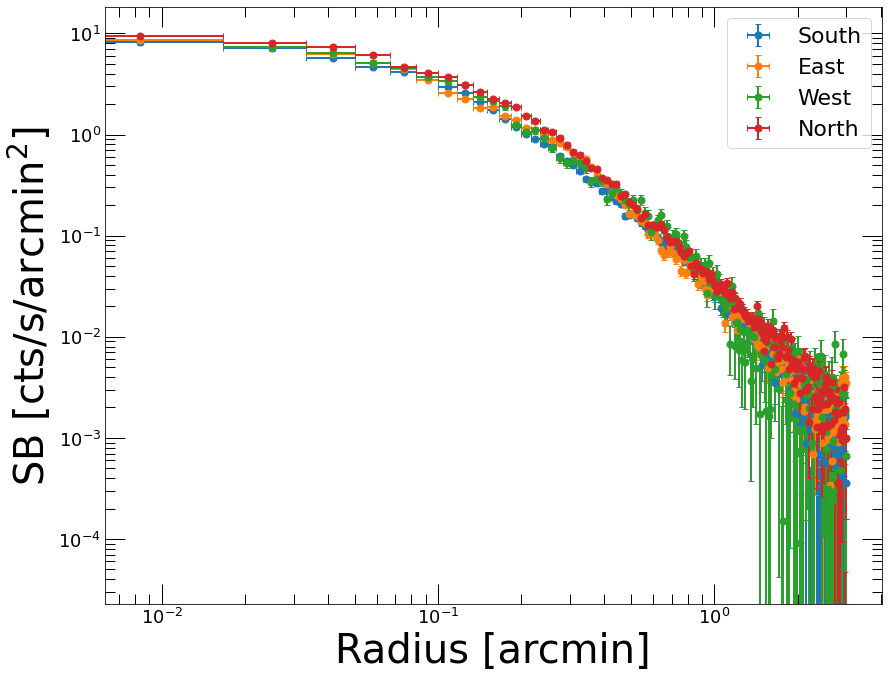}}\\
\subfloat{
\includegraphics[width=5.6cm]{figures/RXJ1532_W_bin1.png}
\includegraphics[width=5.6cm]{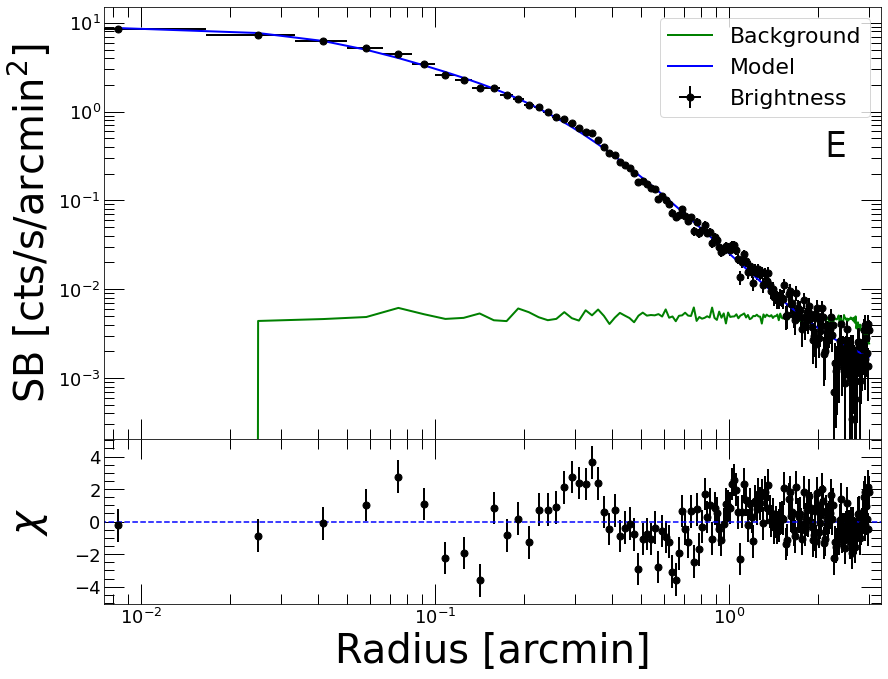}}
\caption{X-ray analysis of RX J1532.9+3021. \emph{Upper left panel:} \emph{Chandra} 0.5-7.0 keV image, with represented a portion of the extraction regions. \emph{Upper central panel:} Residual image after the subtraction of a double $\beta$-model. \emph{Upper Right panel:} Surface brightness profiles extracted along the direction of the regions depicted in the left panel. \emph{Lower Left panel:} Best-fitting double $\beta$-model (blue line) with associated residuals of the western profile. \emph{Lower right panel:} same as before for the eastern profile.}
\label{fig:RXJ1532_App}
\end{figure*}

\subsection{Non-detection of cold fronts}
\subsubsection*{RX J1532.9+3021}
For the cluster RXJ1532, we combined the three \emph{Chandra} observations (total exposure time of 108 ks) already analysed by \cite{Hlavacek2012}.
In the X-ray image (Fig. \ref{fig:RXJ1532_App}), two central X-ray cavities located west and east with respect to the cluster centre are visible. 
Correspondingly, there is a deficit of X-ray emission in the residual image. In the opposite directions, there is instead an excess of X-ray emission.
\cite{Hlavacek2012} reported the presence of a cold front in the western direction at a distance of 65 kpc $\equiv$ 0.22\arcmin.
We tried to reproduce the extraction regions used in \cite{Hlavacek2012}. The western sector is well fitted by a double $\beta$-model, no SB discontinuities are visible, and there is only a slight decrease of SB at $r\le0.1$\arcmin, corresponding to the X-ray cavity. The eastern sector, instead, deviates slightly from a double $\beta$-model, but this is due to the presence of the cavity which is not well modelled.
Therefore, we do not find statistical evidence for the presence of a cold front in this cluster.

\subsubsection*{MS 0735.6+7421}
\emph{Chandra} X-ray observations (6 observations, total exposure time of 550 ks) of the cluster MS0735 were studied in detail by \cite{Vantyghem2014}. 
The source presents two X-ray cavities located north and south of the cluster centre, enveloped by a cocoon, well visible both in the X-ray image and in the residual map (Fig. \ref{fig:X_noCF_1}, first row). 
The SB profiles clearly show a SB jump along all the directions considered, associated with the cocoon.
Spectral analysis performed by \cite{Vantyghem2014} indicates that this feature is a weak shock front.

\subsubsection*{MS 0839.8+2938}
The cluster MS0838 was observed with \emph{Chandra} (ObsID 2224) for an exposure time of 29.8 ks.
The X-ray emission (Fig. \ref{fig:X_noCF_1}, second row) is slightly elongated in the southern direction, as also shown by the residual image, presenting an X-ray excess south of the cluster core. The analysed radial profiles do not present SB discontinuities. 

\subsubsection*{Z2089}
\begin{figure*}
\centering
\subfloat{
\includegraphics[width=17cm]{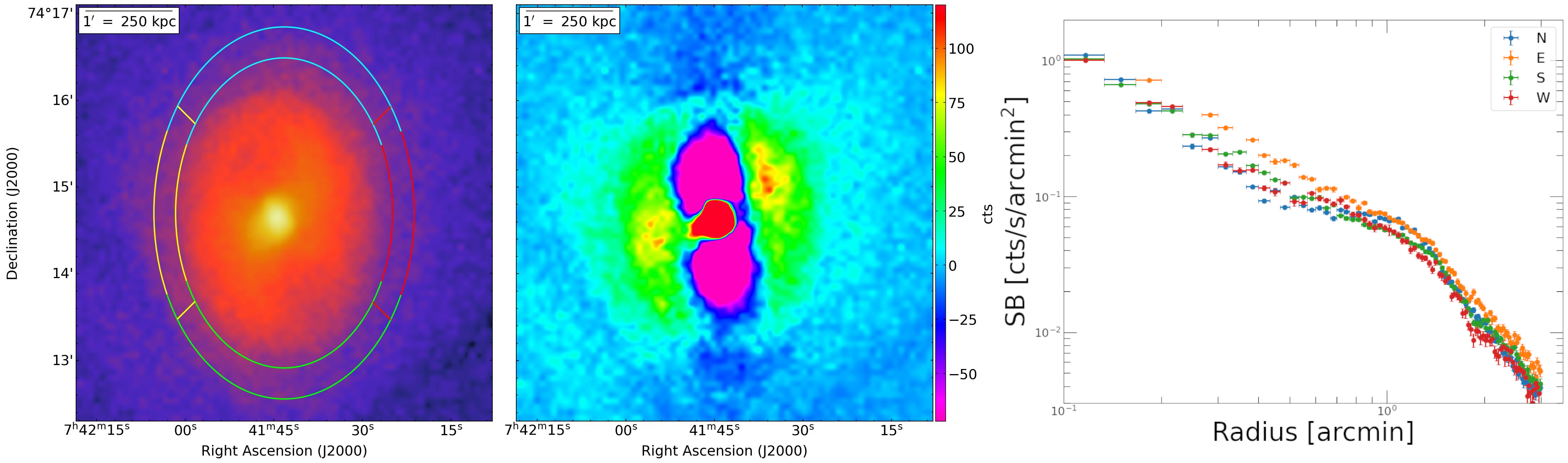}}\\
\subfloat{\includegraphics[width=17cm]{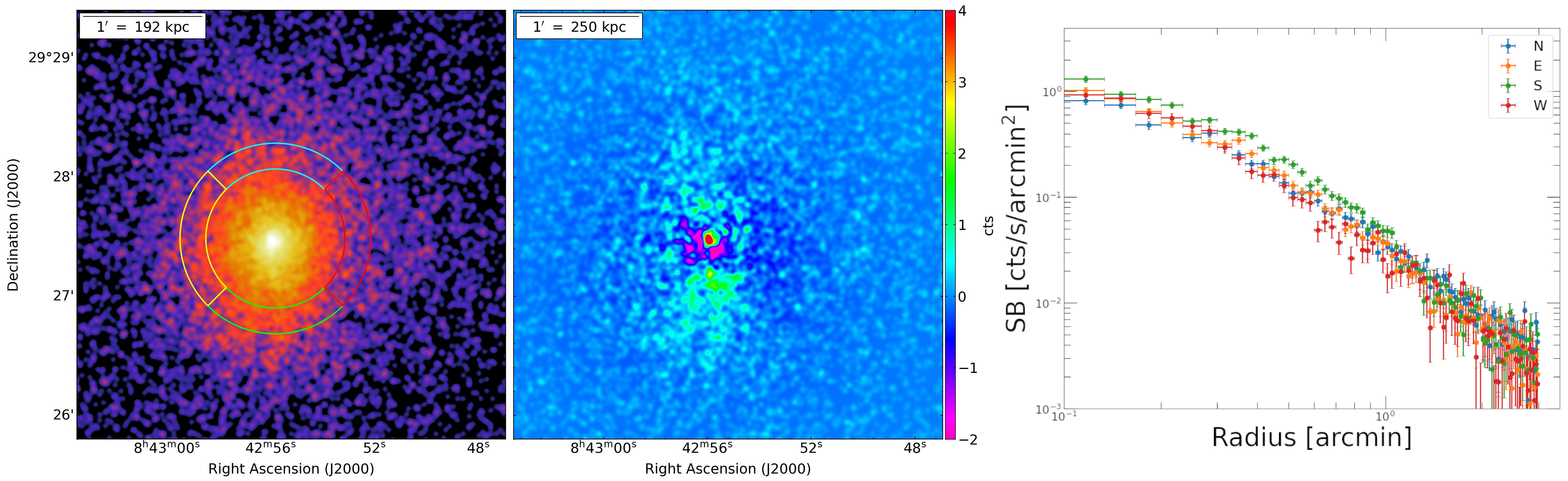}}\\
\subfloat{
\includegraphics[width=17cm]{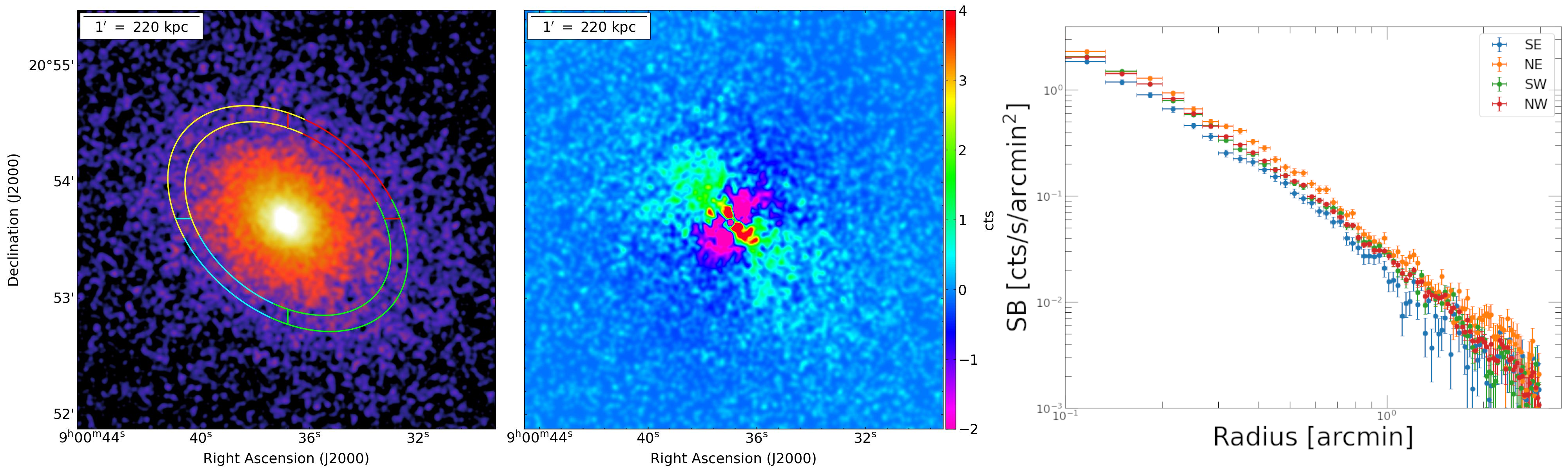}}
\caption{X-ray analysis of MS 0735.6+7421 (\emph{First row}), MS 0839.8+2938 (\emph{Second row}) and Z2089 (\emph{Last row}). \emph{Left panels:} \emph{Chandra} 0.5-7.0 keV images with represented a portion of the extraction regions. \emph{Central panels:} Residual images obtained subtracting an elliptical double $\beta$-model for the cluster MS 0735.6+7421 and a circular double $\beta$-model for the other clusters. \emph{Right panels:} Surface brightness profiles extracted along the direction of the regions depicted in the left panels.}
\label{fig:X_noCF_1}
\end{figure*}

The cluster Z2089 was observed with \emph{Chandra} (ObsID 10463) for an exposure time of 40.6 ks.
The X-ray emission of the cluster is elliptical, elongated in the NE-SW direction (Fig. \ref{fig:X_noCF_1}, last row). The residual image obtained after the subtraction of a circular double $\beta$-model shows an excess of X-ray emission in the direction of the major axis. We extracted radial profiles in the NE, SW, and complementary sectors, but we did not notice particular deviations or SB fluctuations in either direction.

\subsubsection*{RBS797}
We have re-analysed the \emph{Chandra} X-ray observation of RBS797 performed in 2007 with ObSID 7982 and exposure time of 38.3 ks.
The X-ray emission of the cluster (Fig. \ref{fig:X_noCF_2}, first row) is not circularly symmetric, but slightly elongated in the east-west direction. In the central region is clearly visible a pair of small but strong depressions in the NE-SW direction, firstly detected by \cite{Schindler2001}. 
Our radial analysis did not evidence SB discontinuities. We note that the W profile is not well fitted by a double $\beta$-model ($\chi^2_{red}=2.0$), but the goodness of fit improves if we restrict the profile to $r\ge0.1$\arcmin, where a SB depression, associated to a cavity, is present.  
However, deeper observations (total exposure time of 427 ks) revealed the presence of three series of shock fronts at different radial distances, which could be associated with the multiple phases of AGN activity \citep{Ubertosi2023}. Whereas, no cold fronts were found.

\subsubsection*{A1204}
We re-analysed the X-ray \emph{Chandra} observation of the cluster A1204 (ObsID 2205, exposure time 23.6 ks) presented in \cite{Cavagnolo2009}.
We noticed a pair of X-ray depressions in the east and west direction near the cluster core, also visible in the residual image (Fig. \ref{fig:X_noCF_2}, second row). The radial profiles analysed do not present SB discontinuities. 

\subsubsection*{MACS J1720.2+3536}
For the cluster MACSJ1720.2, we combined the three \emph{Chandra} observations (total exposure time of 61.7 ks) presented in \cite{Hlavacek2012}.
The thermal emission of the cluster is elongated in the south direction (Fig. \ref{fig:X_noCF_2}, third row), as also shown by the excess of thermal emission in the residual map in that direction.
Even the radial profiles show a higher SB in the south and west directions, compared to the northern and eastern, but no SB discontinuities are present in those profiles.

\subsubsection*{MACS J2245.0+2637}
\begin{figure*}
\centering
\subfloat{
\includegraphics[width=17cm]{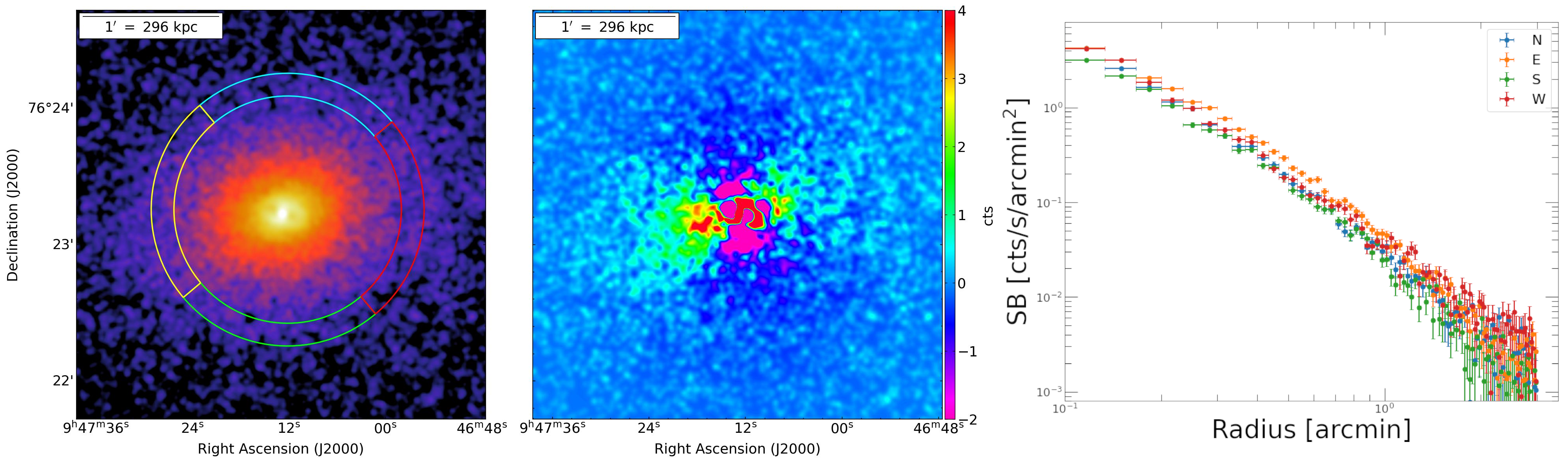}}\\
\subfloat{
\includegraphics[width=17cm]{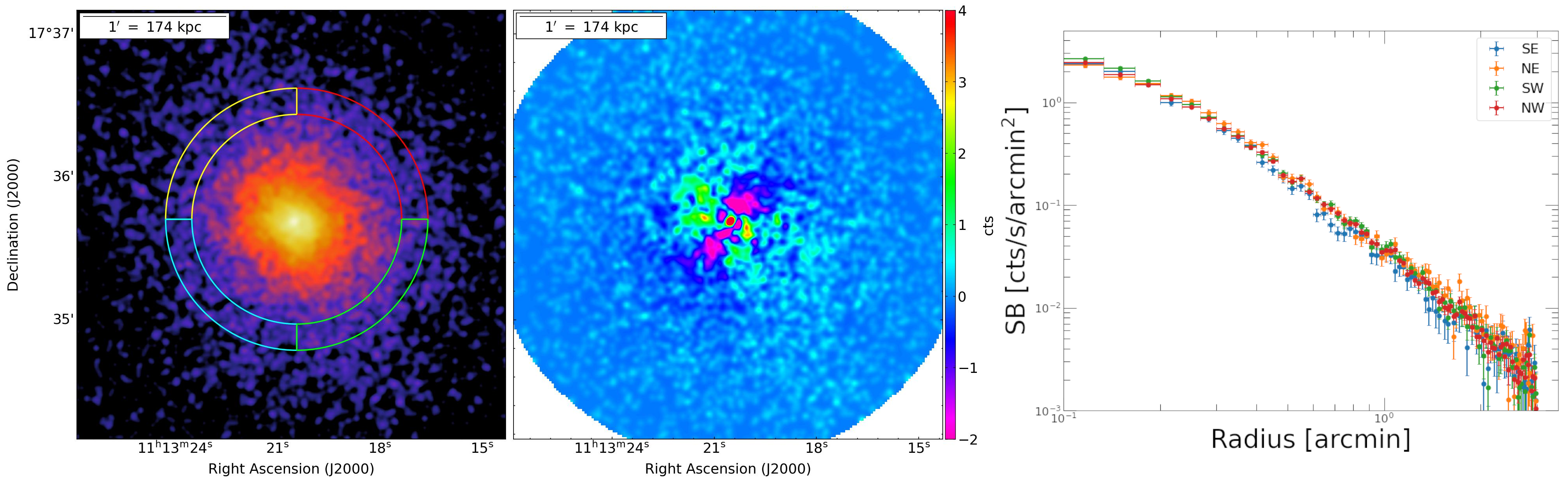}}\\
\subfloat{\includegraphics[width=17cm]{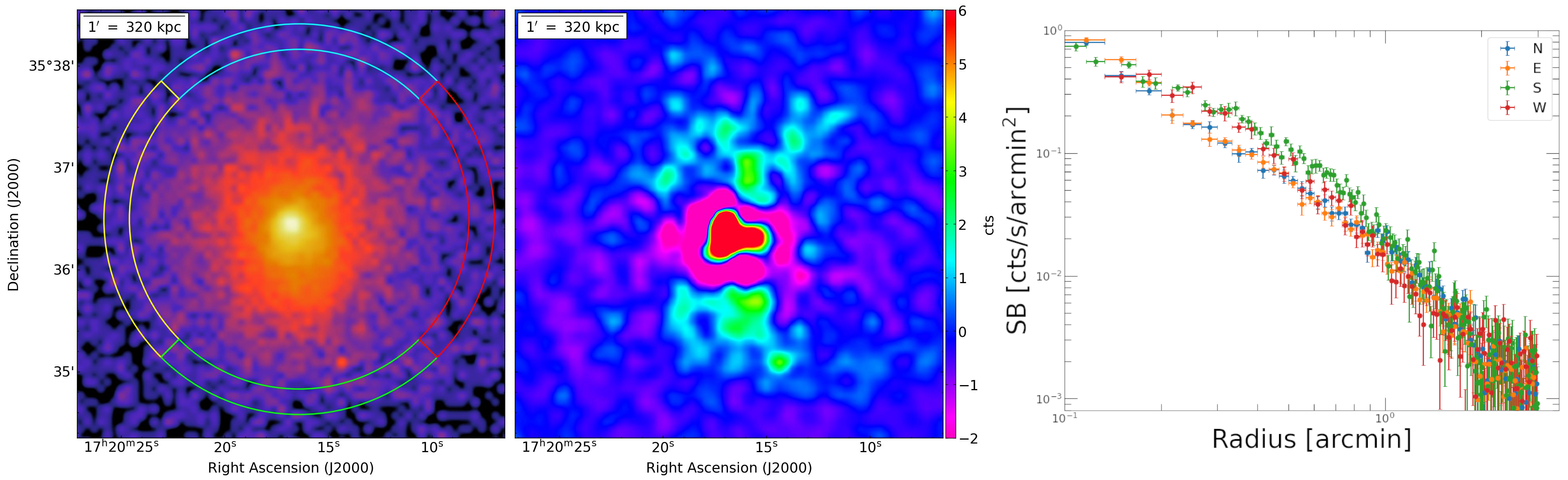}}\\
\subfloat{
\includegraphics[width=10.5cm]{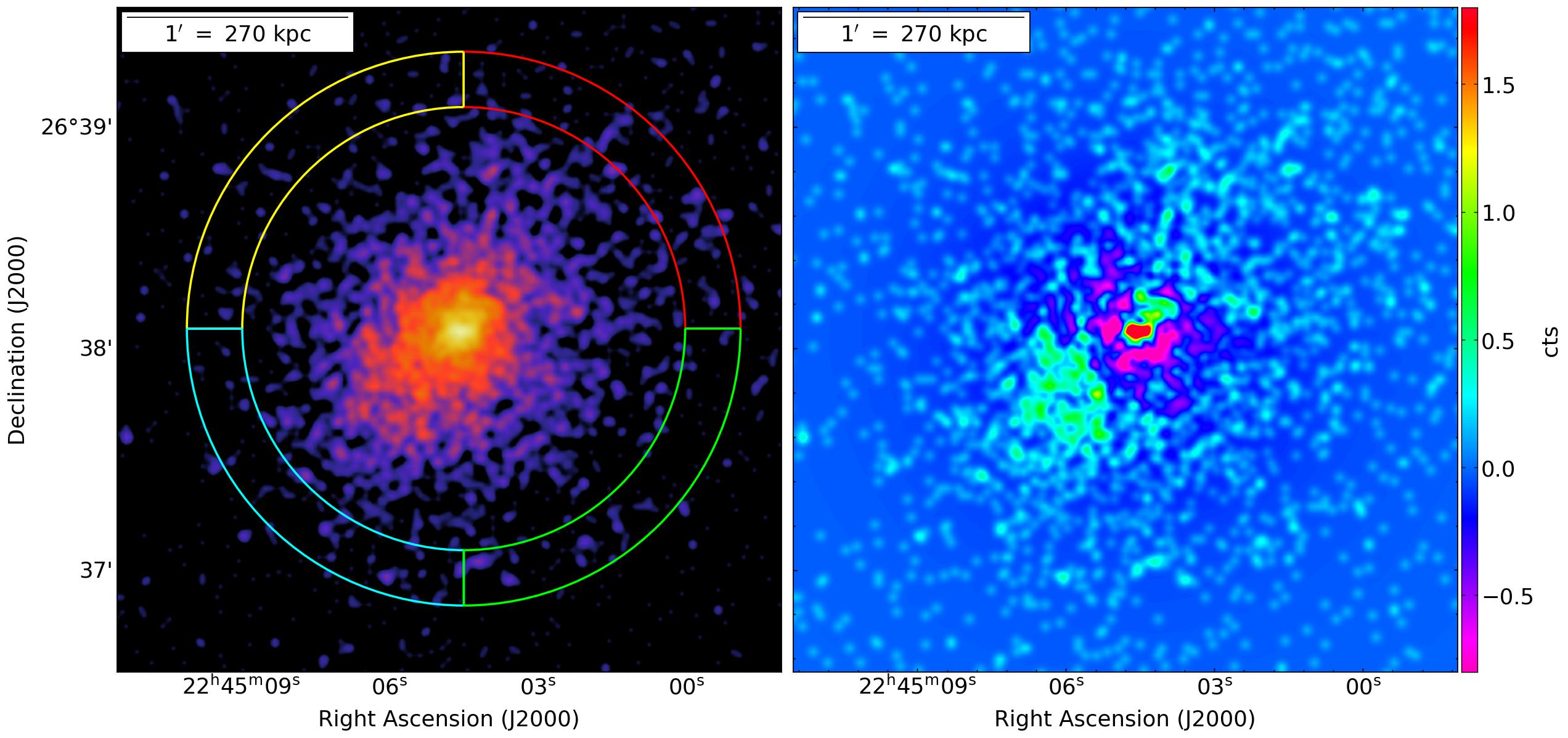}
\includegraphics[width=6.5cm]{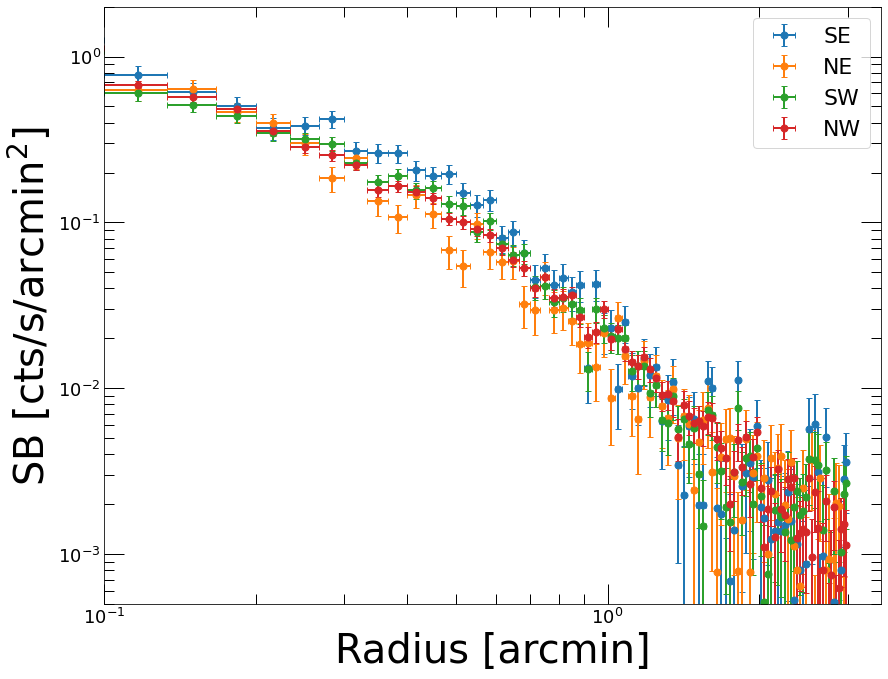}}
\caption{Same as Fig. \ref{fig:X_noCF_1}, for RBS 797 (\emph{First row}), A1204 (\emph{Second row}), MACS J1720.2+3536 (\emph{Third row}), and MACS J2245.0+2637 (\emph{Last row}). The residual images represented in the central panels are obtained subtracting a circular double $\beta$-model for RBS 797 and MACS J1720.2+3536, a circular $\beta$-model for A1204 and an elliptical double $\beta$-model for MACS J2245.0+2637. }
\label{fig:X_noCF_2}
\end{figure*}

The cluster MACSJ2245 was observed with \emph{Chandra} in a single observation (ObsID 3287, exposure time of 16.9 ks).
The thermal emission of the cluster is elongated towards the south-east (Fig. \ref{fig:X_noCF_2}, last row), as also indicated by the X-ray excess in the residual map. All the radio profiles analysed are well fitted with a double $\beta$-model, so no SB discontinuities are present.

\label{lastpage}
\end{document}